\documentclass[11pt]{article}
\usepackage{fullpage}
\usepackage[affil-it]{authblk}
\usepackage[margin=3.3cm]{geometry}
\usepackage{amsmath}
\usepackage{dsfont}
\usepackage[english]{babel}
\usepackage[applemac]{inputenc}
\usepackage[T1]{fontenc}
\usepackage[active]{srcltx}
\usepackage{amsfonts}
\usepackage{amssymb}
\usepackage{graphicx}
\usepackage{color}
\usepackage{float}
\usepackage{bm}
\usepackage{bbm}
\usepackage{authblk}
\usepackage{comment}
\usepackage{enumerate}
\usepackage[titletoc,title]{appendix}
\usepackage{ulem}
\usepackage{url}
\usepackage[dvipsnames]{xcolor}

\newcommand\seq[3]{{#1}_{#2}^{#3}}
\newcommand{\bd}{\begin{document}}
\newcommand{\ed}{\end{document}}
\newcommand{\beq}{\begin{equation}}
\newcommand{\eeq}{\end{equation}}
\newcommand{\bef}{\begin{figure}}
\newcommand{\enf}{\end{figure}}
\newcommand{\bea}{\begin{eqnarray}}
\newcommand{\eea}{\end{eqnarray}}
\newcommand{\baR}{\begin{array}}
\newcommand{\eaR}{\end{array}}
\newcommand{\bc}{\begin{center}}
\newcommand{\ec}{\end{center}}
\newcommand{\ben}{\begin{enumerate}}
\newcommand{\een}{\end{enumerate}}
\newcommand{\bit}{\begin{itemize}}
\newcommand{\eit}{\end{itemize}}
\newcommand{\su}{\section}
\newcommand{\ssu}{\subsection}
\newcommand{\sssu}{\subsubsection}
\newcommand\ssssu[1]{\textbf{#1}}
\newcommand{\nid}{\noindent}
\newcommand{\nnb}{\nonumber}
\newcommand\bloc[2]{{\omega}_{#1}^{#2}}
\newcommand\blocp[2]{{\omega'}_{#1}^{#2}}
\newcommand{\un}{\mathbbm{1}}
\newcommand{\setZ}{\mathbbm{Z}}
\newcommand{\setN}{\mathbbm{N}}
\newcommand{\setR}{\mathbbm{R}}
\newcommand{\pare}[1]{\left(#1 \, \right)}
\newcommand{\scal}[2]{\langle #1 \mid #2 \rangle}
\newcommand{\scalm}[2]{\langle #1 \mid #2 \rangle_{\musp}}
\newcommand{\qdf}[3]{\langle #1 \mid #2 \mid #3 \rangle}
\newcommand{\bra}[1]{\left[#1 \, \right]}
\newcommand{\brac}[2]{\left[\, #1 \, | \, #2 \right]}
\newcommand{\Set}[1]{\left\{\, #1 \, \right\}}
\newcommand{\Setc}[2]{\left\{\, #1 \, ; \, #2 \right\}}
\renewcommand{\P}[1]{\cP\bra{#1}}
\newcommand{\Prob}[1]{\mathds{P}\left[\, #1 \, \right]}
\newcommand{\PT}[2]{\pi^{(T)}_{#2}\left[\, #1 \, \right]}
\newcommand{\pT}{\pi^{(T)}}
\newcommand{\Probex}[1]{P^{(\ast)}\left[\, #1 \, \right]}
\newcommand{\probc}{\mathds{P}}
\newcommand{\Probc}[2]{\mathds{P}\bra{#1 \, \left| \, #2 \right.}}
\newcommand{\probcsp}{\mathds{P}^{(sp)}}
\newcommand{\Probcsp}[2]{\mathds{P}^{(sp)}\bra{#1 \, \left| \, #2 \right.}}
\newcommand{\Probca}[2]{\mathds{P}^{(1)}\bra{#1 \, \left| \, #2 \right.}}
\newcommand{\Pnc}[2]{\mathds{P}_n\bra{#1 \, \left| \, #2 \right.}}
\newcommand{\Probcex}[2]{P^{(\ast)}\bra{#1 \, \left| \, #2 \right.}}
\newcommand{\Probcp}[2]{P^{(ap)}\bra{#1 \, \left| \, #2 \right.}}
\newcommand{\abs}[1]{\left|\, #1 \, \right|}
\newcommand{\deq}{\stackrel {\rm def}{=}}
\newcommand{\sqsubsetneq}{\stackrel {\tiny\sqsubset}{\tiny \neq}}
\newcommand\moy[1]{\mu\bra{#1}}
\newcommand\dKL[2]{d_{\tiny{KL}}\bra{#1 \, \mid #2}}
\newcommand\cA{{\cal A}}
\newcommand\cB{{\cal B}}
\newcommand\cC{{\cal C}}
\newcommand\cD{{\cal D}}
\newcommand\cN{{\cal N}}
\newcommand\cX{{\cal X}}
\newcommand\cY{{\cal Y}}
\newcommand\Q{\mathcal{Q}}
\newcommand{\ProbC}[1]{\mathds{P}_\cC\bra{#1}}
\renewcommand\H[1]{{\cH^{(#1)}}}
\newcommand\noy[1]{\nu\bra{#1}}
\newcommand\s[1]{{\cS\bra{#1}}}
\newcommand\Xk[1]{X_k({#1})}
\newcommand\Ik[1]{I_k({#1})}
\newcommand\Skj[1]{S_{kj}({#1})}
\newcommand\X[2]{X_{#1}\pare{#2}}
\newcommand\et[2]{\eta_{#1}\pare{#2}}
\newcommand\tO[2]{\tau_{#1}\pare{#2}}
\newcommand\cF{{\cal F}}
\newcommand\cG{{\cal G}}
\newcommand\cI{{\cal I}}
\newcommand\cH{{\cal H}}
\newcommand\cK{{\cal K}}
\newcommand\cL{{\cal L}}
\newcommand\cM{{\cal M}}
\newcommand\cT{{\cal T}}
\newcommand\cS{{\cal S}}
\newcommand\cP{{\cal P}}
\newcommand\cO{{\cal O}}
\newcommand\cU{{\cal U}}
\newcommand\cV{{\cal V}}
\newcommand\cW{{\cal W}}
\renewcommand\bf{{\bm{f}}}
\newcommand\baf{{\bar{f}}}
\newcommand\bag{{\bar{g}}}
\newcommand\baF{{\bar{F}}}
\newcommand\baG{{\bar{G}}}
\newcommand\bap{{\bar{\psi}}}
\newcommand\baP{{\bar{\Psi}}}
\newcommand\bF{{\bm{F}}}
\newcommand\bo{{\bm{o}}}
\newcommand\blambda{{\bm{\lambda}}}
\newcommand\bbeta{{\bm{\beta}}}
\newcommand\gk[1]{g_k\pare{#1}}
\newcommand\gLk{g_{L,k}}
\newcommand\akj[1]{\alpha_{kj}(#1)}
\newcommand\sif[1]{\seq{\omega}{-\infty}{#1}}
\newcommand\XkD{\Xk{\bloc{0}{D-1}}}
\newcommand\Xkr{X_k^r}
\newcommand\tLk{\tau_{L,k}}
\newcommand\moyc[2]{\mu\bra{#1 \, | \, #2}}
\newcommand\vkspont{V_k^{(sp)}}
\newcommand\Vkspont[1]{V_k^{(sp)}({#1})}
\newcommand\VkL[1]{V_k^{(L)}({#1})}
\newcommand\vkdet{V_k^{(det)}}
\newcommand\Vkdet[1]{V_k^{(det)}({#1})}
\newcommand\Vksyn[1]{V_k^{(syn)}({#1})}
\newcommand\vkext{V_k^{(S)}}
\newcommand\Vkext[1]{V_k^{(S)}({#1})}
\newcommand\Vknoise[1]{V_k^{(noise)}({#1})}
\newcommand\sk[1]{\sigma_k({#1})}
\newcommand\sj[1]{\sigma_j({#1})}
\newcommand\skr{\sigma_k^r\pare{\bloc{0}{D-1}}}
\newcommand{\ent}[1]{\left[\, #1 \, \right]}
\newcommand\vm[1]{var_m\left[#1 \right]}
\newcommand\eg[1]{\stackrel {\rm #1}{=}}
\newcommand\tk[1]{\tau_k\pare{#1}}
\newcommand\tko{\tau_k(t,\omega)}
\newcommand\cBm[1]{\cB^{-1}\pare{#1}}
\newtheorem{theorem}{Theorem}
\newcommand{\bth}{\begin{theorem}}
\newcommand{\enth}{\end{theorem}}
\newcommand\omD[1]{\seq{\bra{\omega^{(#1)}}}{0}{D-1}}
\newcommand\omz[1]{\omega^{(#1)}(D)}
\newcommand\om[1]{\varpi^{(#1)}}
\newcommand\skjq{s_{kj;q}}
\newcommand\ikq{i_{k;q}}
\newcommand\skjqs{p_{kj;q_s}}
\newcommand\ikqs{i_{k;q_s}}
\newcommand\ikqsu{i_{k;q_{s_u}}}
\newcommand\ikqu{i_{k;q_u}}
\newcommand\xkq{x_{k;q}}
\newcommand\skjqv{s_{kj;q_{v}}}
\newcommand\tjr{t_j^{(r)}(\omega)}
\newcommand\dtt{\Phi(t,\omega)}
\newcommand\dts{\Phi(s,\omega)}
\newcommand\dto{e^{\Phi(t-t_0)}}
\newcommand\dtes{e^{\Phi(t-s)}}
\newcommand{\ket}[1]{\mid #1 \rangle}
\newcommand{\brak}[1]{\langle #1 \mid}
\newcommand{\Sum}[2]{\sum_{\tiny{\baR{ccc} #1 \eaR}}^{\tiny{\baR{ccc} #2 \eaR}}}
\newcommand{\zb}{\zeta_{\bbeta}}
\newcommand{\etab}{\eta_{\bbeta}}
\newcommand{\mex}{\mu^{(\ast)}}
\newcommand{\phex}{\phi^{(\ast)}}
\newcommand{\Phex}{\Phi^{(\ast)}}
\newcommand{\Hex}{\cH_{\bbeta'}^{(\ast)}}
\newcommand\p[1]{\cP\bra{#1}}
\newcommand\w[2]{w_{#1}^{(#2)}}
\newcommand\Gb{\cG_\bbeta}
\newcommand\blp[3]{\omega^{#3,(#1)}_{#2}}
\newcommand\blpb[3]{\overline{\omega^{#3,(#1)}_{#2}}}
\newcommand\moyex[1]{\mu^{(\ast)}\bra{#1}}
\newcommand\E[1]{{\cal E}_{#1}}
\newcommand\Gap{\overline{G}}
\newcommand\gap[1]{\overline{g_{#1}}}
\newcommand{\K}[2]{\cK^{(#1)}(#2)}
\newcommand{\Vr}{V_{reset}}
\newcommand{\bcoul}[2]{{\color{#1}{\textbf{#2}}}}
\newcommand{\bruno}[1]{\bcoul{red}{Bruno says: #1}}
\newcommand{\rod}[1]{\bcoul{blue}{Rodrigo says: #1}}
\newcommand{\tf}[2]{t_{#1}^{(#2)}}
\newcommand{\Tf}[1]{\mathcal{T}^{(#1)}(\omega)}
\newcommand{\Exp}[1]{\mathds{E}\left[\, #1 \, \right]}
\newcommand{\Expc}[2]{\mathds{E}\left[\, #1 \, | \, #2 \right]}
\newcommand{\Expsp}[1]{\mathds{E}^{(sp)}\left[\, #1 \, \right]}
\newcommand{\Expcsp}[2]{\mathds{E}^{(sp)}\left[\, #1 \, \mid \, #2 \right]}
\newcommand{\Cov}[1]{Cov\left[\, #1 \, \right]}
\newcommand{\bD}[2]{\underline{D}_{#1}(#2,\omega)}
\newcommand{\hk}{\hat{k}}
\renewcommand{\i}{\mathrm{i}}
\newcommand\rpf{R}
\newcommand\rpfc[1]{R\pare{#1}}
\newcommand\lpf{L}
\newcommand\lpfc[1]{L\pare{#1}}
\newcommand{\vect}[1]{
\left(\begin{array}{lll}
#1
\end{array} \right)
}

\newcommand{\frc}[1]{\left\{\, #1 \, \right\}}
\newcommand{\Cor}[2]{{\cC}^{(sp)}\left[\, #1, #2 \, \right]}
\newcommand{\Covc}[2]{Cov\left[\, #1 \,\, | \, #2 \right]}
\newcommand{\pr}[3]{p_{#1}\pare{#2, \, #3}}
\newcommand\pTo{{\pi_{\omega}^{(T)}}}
\newcommand\tj[2]{t_j^{(#1)}(#2)}
\newcommand\tjro{\tjr{\omega}}
\newcommand\tjk{t_j^{(k)}(\omega)}
\newcommand\Vk[1]{V_k({#1})}
\newcommand\Vlnoise[1]{V_l^{(noise)}({#1})}
\newcommand\Vnoise[1]{V^{(noise)}({#1})}
\newcommand\VkB[1]{V_k^{(B)}({#1})}
\newcommand\dXk[1]{\delta X_k\pare{#1}}
\newcommand\Repk[2]{\cH^{({#1})}_i\pare{#2}}
\newcommand\Xksp[1]{X^{(sp)}_k\pare{#1}}
\newcommand\sko[2]{\sigma_k({#1},{#2},\omega)}
\newcommand\sr{\sigma_R}
\newcommand\sdr{\sigma^2_R}
\newcommand\skd[1]{\sigma^2_k({#1})}
\newcommand\skop[1]{\sigma_k({#1},\omega')}
\newcommand\skopd[1]{\sigma^2_k({#1},\omega')}
\newcommand\ta[2]{\tau_{#1}({#2})}
\newcommand\tauk[1]{\ta{k}{#1}}
\newcommand\tkos{\tau_k(s,\omega)}
\newcommand\tkop{\tau_{k'}(t,\omega)}
\newcommand\tkoptp{\tau_{k'}(t',\omega)}
\newcommand\tlo{\tau_l(t,\omega)}
\newcommand\nko{\tau_k(n,\omega)}
\newcommand\nkrom{\tau_k(r-1,\omega)}
\newcommand\nkrm{\tau_k(r-1,.)}
\newcommand\nkop{\tau_k(n,\omega')}
\newcommand\Vkosyn[2]{V_k^{(syn)}({#1},{#2},\omega)}
\newcommand\Vkoext[2]{V_k^{(S)}({#1},{#2},\omega)}
\newcommand\Vkodet[2]{V_k^{(det)}({#1},{#2},\omega)}
\newcommand\Vkonoise[2]{V_k^{(noise)}({#1},{#2},\omega)}
\newcommand\VkoB[2]{V_k^{(B)}({#1},{#2},\omega)}
\newcommand\Vkto{\Vko{t}{t+1}}
\newcommand\Vksto{\Vko{s}{t}}
\newcommand\tmk{\tau_{M,k}}
\newcommand\gmk{g_{M,k}}
\newcommand\dl[1]{\delta_l\left[#1 \right]}
\newcommand\nega[1]{\stackrel {\rm #1}{\neq}}
\newcommand\moyt[1]{\overline{#1}}
\newcommand\Phisp[1]{\Phi^{(sp)}\pare{#1}}
\newcommand\phisp[1]{\phi^{(sp)}\pare{#1}}
\newcommand\phis{\phi^{(sp)}}
\newcommand\phim{\phi^{(Markov)}}
\newcommand\phispk[1]{\phi^{(sp)}_k\pare{#1}}
\newcommand\Zsp[1]{Z^{(sp)}\bra{#1}}
\newcommand\musp{\mu^{(sp)}}
\newcommand\Csp[2]{\mathcal{C}^{(sp)}\bra{#1, #2}}
\newcommand\Chi[2]{\chi_{#1}\pare{#2}}
\newcommand\CHi[2]{\chi_{#1}^{(#2)}}
\newcommand\moysp[1]{\mu^{(sp)}\bra{#1}}
\newcommand\moyspc[2]{\mu^{(sp)}\bra{#1 \, | \, #2}}
\newcommand\tmoy[1]{\tmu\bra{#1}}
\newcommand\tmoyc[2]{\tmu\bra{#1 \, | \, #2}}
\newcommand\dmu[1]{\delta^{(1)}\bra{#1}}
\newcommand\M[1]{\cM{(#1)}}
\newcommand\pM[2]{\cM^{(#1)#2}}
\newcommand\Mc[3]{\cM_{#2,\,#3}{(#1)}}
\newcommand\pc[2]{p_{#1}{(#2)}}
\newcommand\tphi[1]{\tilde{\phi}^{(#1)}}
\newcommand\tfi{\tilde{\phi}}
\newcommand\tphic[3]{\tilde{\phi}^{(#1)}_{#2,\,#3}}
\newcommand\tfc[2]{\tilde{f}_{#1}{(#2)}}
\newcommand\tmu{\tilde{\mu}}
\newcommand\cst[3]{#1_{#2}^{(#3)}}
\newcommand\spont[2]{#1_{(sp)}\bra{#2}}
\newcommand\spontc[3]{#1_{(sp)}\brac{#2}{#3}}
\newcommand\valp[2]{s_{#1}(#2)}
\newcommand\eqm{\stackrel {\rm \musp}{=}}
\newcommand\emp[3]{\pi_{#1}^{(#2)}\pare{#3}}
\newcommand\Int[2]{\bra{#1, \, #2}}
\newcommand\pder[3]{\left. \frac{\partial #1}{\partial #2}   \right|_{#3}}
\newcommand\cR{\mathcal{R}}
\newcommand\Minv{\mathcal{M}_{inv}}
\newcommand{\tbc}{\bruno{TBC \bigskip}}
\newcommand\bam{\bar{m}}
\newcommand\baM{\bar{M}}
\newcommand\baMc{\bar{M}^{(c)}}
\newcommand\baMcd{\bar{M}^{(c)\dag}}
\newcommand\baMe{\bar{M}^{(e)}}
\newcommand\baMed{\bar{M}^{(e)\dag}}
\newcommand{\Conv}[3]{\pare{#1 \ast #2}\bra{#3}}
    
\newcommand{\B}[1]{{\textcolor{blue}{[bruno: \textit{#1}]}}}
\definecolor{antiquefuchsia}{rgb}{0.57, 0.36, 0.51}
\definecolor{armygreen}{rgb}{0.29, 0.33, 0.13}
\definecolor{applegreen}{rgb}{0.55, 0.71, 0.0}\definecolor{ao(english)}{rgb}{0.2, 0.5, 0.2}\definecolor{arsenic}{rgb}{0.23, 0.27, 0.29}
\definecolor{britishracinggreen}{rgb}{0.0, 0.26, 0.15}
\newcommand{\new}[1]{{\textcolor{britishracinggreen}{[\\new \B{Rodrigo could you check ? \\} \textit{#1}]}}}
\newcommand{\nnew}[1]{{\textcolor{britishracinggreen}{[\\new: \\ \textit{#1}]}}}

\newcommand{\Expm}[2]{\mathds{E}_{#1}\left[\, #2 \, \right]}

\newtheorem{name}{Printed output}
\newenvironment{remark}[1][Remark:]{\begin{trivlist}
\item[\hskip \labelsep {\bfseries #1}]}{\end{trivlist}}
\newcommand{\TBC}{\B{To be continued\\ \pagebreak}}
\newcommand{\R}[1]{{\textcolor{red}{[Rodrigo: \textit{#1}]}}}

\begin{document}

\title{
Linear response for spiking neuronal networks with unbounded memory}
\author[1]{Bruno Cessac \thanks{Electronic address: \texttt{bruno.cessac@inria.fr}}}
\author[2]{Ignacio Ampuero \thanks{Electronic address: \texttt{ignacio.ampuero.13@sansano.usm.cl }}}
\author[3]{Rodrigo Cofre \thanks{Electronic address: \texttt{rodrigo.cofre@uv.cl}; Corresponding author}}
\affil[1]{Biovision team, INRIA and Neuromod Institute, Universit\'{e} C\^{o}te d'Azur, France.}
\affil[2]{Departamento de Inform\'{a}tica, Universidad T\'{e}cnica Federico Santa Mar\'{i}a, Valpara\'{i}so, Chile}
\affil[3]{CIMFAV, Facultad de Ingenier\'{i}a, Universidad de Valpara\'{i}so, Valpara\'{i}so, Chile}

\maketitle

\begin{abstract}
We establish a general linear response relation for spiking neuronal networks, based
on chains with unbounded memory. This relation allow us to predict the influence of a weak amplitude time dependent external stimuli on spatio-temporal spike correlations, from the spontaneous statistics (without stimulus) in a general
context where the memory in spike dynamics can extend arbitrarily far in the past. Using this approach, we show how linear response is explicitly related to neuronal dynamics with an example, the gIF model, introduced by M. Rudolph and A. Destexhe.
This example illustrates the collective effect of the stimuli, intrinsic neuronal dynamics, and network connectivity on spike statistics. We illustrate our results with numerical simulations.
\end{abstract}


\su{Introduction}

Neurons communicate by short-lasting electrical signals called action potentials or ``spikes'', allowing the rapid propagation of information throughout the nervous system, with a minimal energy dissipation \cite{abbott-dayan:99}. The spike shape is remarkably constant for a given neuron, and it is a contemporary view to consider spikes as quanta (bits) of information \cite{rieke-etal:96}. As a result, information is presumably encoded in the spike timing \cite{masquelier:08}. 

\begin{figure}[h!]
  \centering
    \includegraphics[width=0.9\textwidth]{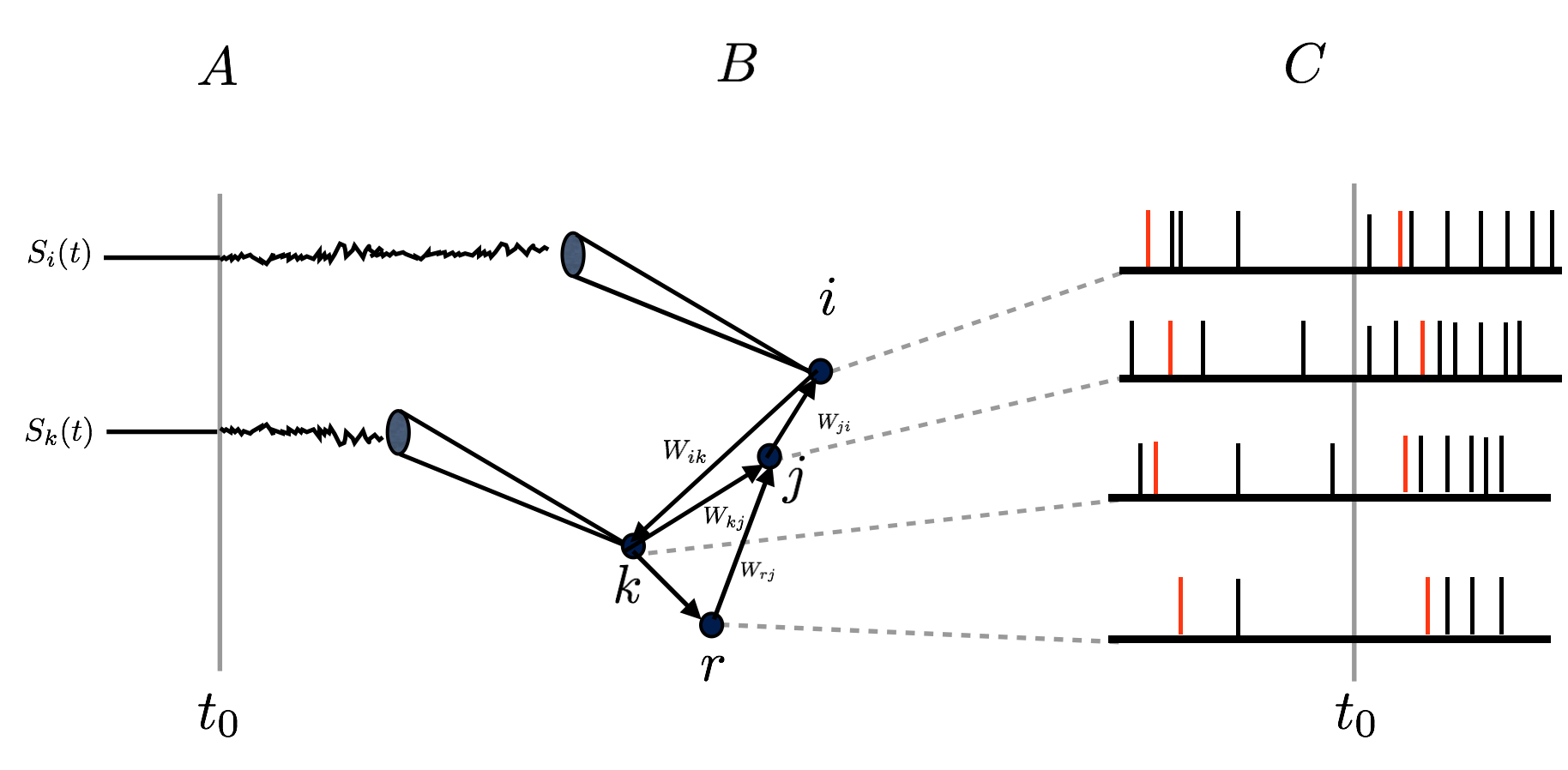}
    \caption{ A time-dependent stimulus ($A$) is applied at time $t_0$  to some neurons of a spiking neuronal network ($B$). As a result, the spiking activity is modified as well as spike correlations between neurons ($C$), even for neurons \textit{not directly stimulated}, because of interactions.}
\label{Fig:StimulusResponse}
\end{figure}

The simplest quantitative way to characterize the spiking activity of a neuron is its firing rate $r(t)$, where $r(t) dt$ is the probability that this neuron spikes during a small interval $[t,t+dt]$. Under the influence of an external stimulus the firing rate changes. A classical ansatz, coming from the Volterra expansions \cite{rieke-etal:96}  is to write the variation in the firing rate of a neuron as the convolution form:
\beq\label{eq:LinearResponseRate}
\delta^{(1)}[r(t)] = (K*S)[t].
\eeq
\noindent
where the exponent $^{(1)}$ recalls that we consider a first-order effect of the stimulus $S$, that is, the stimulus is weak enough so that higher order terms in the Volterra expansion can be neglected. This is an example of linear response: the variation in the rate is proportional to the stimulus. Here, $K$ is a convolution kernel constrained by the underlying network dynamics. For example, in sensory neurons, where $S$ and $K$ are functions of space and time, the convolution \eqref{eq:LinearResponseRate} takes the explicit form:
\begin{equation} \label{eq:KernelRF}
(K*S)[t] = \int_{x=-\infty}^{+\infty} \int_{y=-\infty}^{+\infty} \int_{\tau=-\infty}^t K(x,y,t-\tau) S(x,y,\tau) d\tau dx dy,
\end{equation}
where $K$ decays sufficiently fast at infinity (in space and time) to ensure that the integral is well defined. $K$ mimics the receptive field (RF) of the neuron.
In general, the response of spiking neuronal networks to a time dependent stimulus does not only affect rates, it also has an impact on higher order correlations between neurons, because neurons are connected. This situation is sketched in Fig. \ref{Fig:StimulusResponse}. 

In particular, sensory neurons collectively convey to the brain information about external stimuli using correlated spike patterns resulting from the conjunction of stimulus influence, intrinsic neurons dynamics and neurons interactions via synapses \cite{schneidman-berry-etal:06,tkacik-marre-etal:13,pillow-paninski-etal:05, vasquez-palacios-etal:12,cofre-cessac:14}. 
This correlated firing has been linked to stimulus encoding \cite{deCharms:96}, stimulus discrimination \cite{alonso:96,marre-etal:15} and to intrinsic properties of the network which remain in absence of stimulus \cite{tsodyks-etal:99}. However, disentangling the biophysical origins of the correlations observed in spiking data is still a central and difficult problem in neuroscience \cite{dayan-abbott:01,gerstner-kistler:02,rieke-etal:96}. As a consequence, correlations in spiking neuronal networks have attracted a considerable amount of attention in the last years, from experimental data analysis perspectives \cite{schneidman-berry-etal:06, nirenberg-latham:03,pillow-paninski-etal:05,vasquez-palacios-etal:12} as well as from the theoretical modeling viewpoint \cite{abbott-dayan:99, sompolinsky-etal:01, brunel-hakim:99,trousdale:12, lindner:05, ostojic-etal:09b}.

On one hand, novel experimental recording techniques (Multi-Electrode Arrays, MEA) permit the measurement of the collective spiking activity of larger and larger populations of interacting neurons responding to external stimuli \cite{marre-etal:12,ferrea-etal:12}. These recordings allow, in particular, for better characterization of the link between stimuli and the correlated responses of a living neuronal network, paving the way to better understand ``the neural-code'' \cite{rieke-etal:96}. 
Yet, neuronal responses are highly variable \cite{dayan-abbott:01}. Even at the single neuron level, when presenting repetitions of the same stimulus under controlled experimental conditions, the neural activity changes from trial to trial \cite{croner:93, shadlen-etal:98}. Thus, researchers are seeking statistical regularities in order to unveil a probabilistic, causal, relationship between stimuli and spiking responses \cite{schneidman-berry-etal:06,tkacik-marre-etal:13,vasquez-palacios-etal:12,ferrari:16,botella:18,nghiem:18}.

On the other hand, mathematical models of spiking neuronal networks offer a complementary approach to biological experiments \cite{dayan-abbott:01,ermentrout-terman:10,gerstner-kistler:02,izhikevich:07}. Based on biophysically plausible mechanisms controlling the dynamics of neurons, mathematical modeling provides a framework to characterize the population spike train statistics in terms of biophysical parameters, synaptic connectivity, history of previous spikes and stimuli. The hope is that understanding these aspects in a model will allow for better processing and extraction of information from real data. 
Yet, there is a large gap between what is learned from a model and experiments on real neurons. From this theoretical and computational point of view, characterizing the response of a neuronal network model to a (time-dependent) stimulus and relating this correlated response to spike trains measured from MEA recordings involves several  modeling steps. 

\begin{enumerate}[(i)]

\item \textbf{Modeling spontaneous activity.} The goal here is to characterize the collective spiking activity in the absence of external stimuli.  In this situation, spike correlations are, by assumption, only due to neuronal dynamics and interactions. 

\item \textbf{Modeling the response to stimuli.} Assume that the spiking neuronal network receives  a time-dependent stimulus $S(t)$ from time $t_0$ to time $t_1$, as in Fig. \ref{Fig:StimulusResponse}. Even if the stimulus is applied to a subset of neurons in the network, its influence will eventually propagate to other neurons, directly or indirectly connected. The stimulus will act on spikes timing, modifying spike correlations.  In particular, the relationship \eqref{eq:LinearResponseRate} ought to extend to more general statistical indicators than rates. That is, for a statistical indicator  $f$ - rate, correlation, or more generally, an observable as defined in section \ref{Sec:SpikesObs} - one expects from Volterra expansion that the time-dependent variation of $f$ under the action of the stimulus would take the form $\delta^{(1)}[f(t)] = (K_f*S)[t]$, where the convolution kernel $K_f$ depends on $f$ as well as on the network dynamics (including synaptic interactions). Determining the explicit mathematical form of $K_f$ is difficult in general (see \cite{cessac-sepulchre:04,cessac-sepulchre:06} for an example based on Ruelle's linear response theory \cite{ruelle:99} applied to the Amari-Wilson-Cowan model \cite{amari:77,wilson-cowan:73}).
 
\item \textbf{Experimental characterization of spontaneous activity and response to a stimulus.} In MEA experiments the underlying neuronal network is not known. One only has access to spike trains. Therefore, at this level, one has to define an efficient, operational, way to characterize spike statistics from MEA, in the spontaneous regime, as well as  in the stimulated regime. This last point is particularly tricky because it implies a non-stationary response, whereas some prominent standard statistical methods like Maximum Entropy or Boltzmann machines heavily rely on a stationarity assumption \cite{jaynes:57}. In addition, the processes generating spikes trains are causal, with memory, suggesting a Markovian dynamics. In fact, the memory depth can be dependent upon neurons, and dynamics, and variable length Markov dynamics could be more realistic \cite{cessac:11b,galves:13,vidybida:15}.   
\end{enumerate}

We don't know about any work addressing these three points simultaneously and the road toward this achievement seems to be long. This paper is one step further in that direction. In contrast, there is a relatively large amount of literature considering the links between two of these points.

\paragraph{Modeling the collective neural response to a stimulus ((i) $\to$ (ii)).}

There is a large body of theoretical work linking the spike responses of neuronal network models to their structural properties in the presence of stimuli. 
For example, the relationship between stimuli and the firing rate as a function of the parameters of the model can be obtained in a network of homogeneous Leaky Integrate-and-Fire neurons, considered in the mean-field limit \cite{johannesma:68,knight:72, brunel-hakim:99,gerstner-kistler:02}. Computing the firing rate for time-varying stimuli is, however, a much more difficult problem \cite{brunel-hakim:99,lindner:05}. Beyond firing rates, there are also results concerning correlations. In \cite{paninski:04}, Toyoizumi et al develop mean-field methods for approximating the stimulus-driven firing rates (both in the time-varying and steady-state case), auto- and cross-correlations, and stimulus-dependent filtering properties of spiking neuronal networks with Markov
refractoriness. In \cite{trousdale:12}, the authors obtain formulas for cross-correlations with an arbitrary delay in terms of ``motifs'' in the neuronal connectivity, while, in \cite{pernice:11,reynaud:13}, the correlation structure in networks of interacting Hawkes processes is investigated.

In general, neuronal networks dynamics involves interactions between neurons with time delays and strongly depends on the network history. 
 The statistics of general spike events involving distinct neurons spiking at different times ought to be affected by stimuli and could shed light on the coding process. Obviously, a thorough characterization of such events gets rapidly out of reach in experimental data as the number of neurons involved in this event and time delays increases.  On the other hand, this characterization can be achieved in neuronal network models using analytic expressions, as we show.

\paragraph{Statistical models of MEA recordings ((i) $\to$ (iii)).}
There is a wide variety of literature about the modeling and inference approaches that have been recently developed to
describe the correlated spiking activity of populations of neurons (see \cite{gardella:18} for a recent review, rather complete, although not mentioning the mathematical literature on the subject). These approaches cover a variety of models describing correlations between pairs of neurons as well as between larger groups,
synchronous or delayed in time, with or without the explicit influence of the stimulus,
and including or not hidden, latent variables. One can distinguish two main trends. The first one, inspired from statistical physics seeks to characterize the spike correlations by a restricted form of Gibbs distribution - i.e. derived from the Maximum Entropy Principle - of the form \eqref{eq:GibbsStatPhys} below.  
We use the term ``restricted'' because the Maximum Entropy Principle requires stationarity and because most models used in the literature do not consider spike time correlations; for instance in the Ising model and extensions to higher order spatial correlations \cite{schneidman-berry-etal:06}, successive times are independent. In fact, Gibbs distributions can get rid of these limitations  as discussed in detail in section \ref{Sec:LinRepGen}. The second trend consists of building stochastic processes reproducing the spike correlations, based on a family of transition probabilities taking into account causality and history. 
Prominent examples are the Generalized Linear Model (GLM) model \cite{paninski:04,pillow-paninski-etal:05} or Hawkes processes \cite{reynaud:13,reynaud:14,ferrari:18,galves:20}. As discussed in \cite{cofre-cessac:14,cessac-cofre:13,cofre-cessac:13}  and reviewed in the next section, the probability distributions generated this way are as well Gibbs distributions, considered in a more general setting than in the standard statistical physics courses. 

\paragraph{Time-dependent, non-stationary, spike population response to a stimulus ((i-iii) $\to$ (ii)).} 
Assume that we have a reasonable characterization of spike train statistics, e.g. with the techniques described in the previous paragraphs,  for spontaneous activity (assumed to be stationary). Can we predict how the spike correlations will be modified under the influence of a time-dependent stimulus, with a weak enough amplitude so that we may neglect higher order corrections? As we show in this paper (section \ref{Sec:LinRepGen}) one can derive a quite general linear response theory, somewhat linking (i-iii) to (ii) extending the notion of linear response theory used in non-equilibrium statistical physics and ergodic theory, to spiking processes (chains) with infinite memory. In particular, it generalizes \eqref{eq:LinearResponseRate} to general observables.
 However, this theory stays at a formal level without a concrete example where the convolution kernel is explicitly computed from neurons dynamics. As we show here this can be achieved in the generalized Integrate and Fire model (gIF) introduced by M. Rudolph and A. Destexhe in 2006 \cite{rudolph-destexhe:06}. \\

The theory we develop has its roots in non-equilibrium statistical physics (briefly reviewed in section \ref{Sec:LinRepStatPhys}) and ergodic theory (section \ref{Sec:LinRepSD}).  
The linear response determines how the expectation value of an observable of a dynamical system changes upon weakly perturbing the dynamics. A seminal result in this context is the \textit{fluctuation-dissipation theorem} \cite{kubo:66}, where the linear response only depends on the correlation functions of the unperturbed system. Our approach proceeds along similar lines, meaning that the linear response can be predicted from the spikes correlations of the unperturbed system.

The paper is organized as follows. In section \ref{Sec:LinRepGen}, we briefly review linear response in statistical physics and ergodic theory allowing us to make a link between neuronal networks, considered as dynamical systems, and the statistics of spikes. In section \ref{Sec:LinRepInfMem} we introduce the formalism of chains with unbounded memory  (which are, as we explain, equivalent to left-sided one-dimensional Gibbs distributions), allowing the handling of non-stationary spike distribution with unbounded memory. In this context, we derive the general linear response formula \eqref{eq:dmu} used throughout the paper. This equation expresses  the time-dependent variation in the average of an observable $f$ as a time series of
specific correlation functions  computed with respect to spontaneous activity (without stimulus). This result, reminiscent of the fluctuation-dissipation theorem in statistical physics \cite{kubo:57,ruelle:99}, is applied here to spike statistics. 

In section \ref{Sec:LinRepgIF}  we introduce a spiking neuronal network model to instantiate our analysis. This model has been presented in \cite{rudolph-destexhe:06}. We associate the spiking activity to a discrete stochastic process defined from transition probabilities where memory is unbounded. These probabilities are written as a function of the parameters of the model. From this, we are able to explicitly write a discrete time form of the convolution kernel \eqref{eq:LinearResponseRate} as an explicit function of the model parameters, especially synaptic weights. The expression relies on a Markovian approximation of the chain and on a decomposition theorem of spike observables, introduced in  a more general context by Hammersley and Clifford in 1971 \cite{hammersley-clifford:71} (see section \ref{Sec:SpikesObs}).  
What is the main result of linear response theory? The response of a system, originally at equilibrium, to a time-dependent stimulus is proportional to the stimulus, with coefficients obtained via correlations functions computed at equilibrium. We derive a result of this type in the gIF model here addressing a central question: which correlations matter and how they are related  to synaptic interactions?

Although, the main content of this paper is mathematical we illustrate, in section \ref{sec:Numerics} a few examples where our linear response theory is applied. In particular, we show how one can predict the variation in the firing rate and in the delayed pairwise correlation between two neurons from the mere knowledge of the stimulus and relevant spontaneous correlations. 

\section{Linear response, Gibbs distributions and probabilistic chains with unbounded memory} \label{Sec:LinRepGen}

Neuronal networks can be considered either as dynamical systems (when the dynamics is known) or as spike generating processes characterized by transition probabilities computed from spike train observations. In the first case, it is natural to seek a linear response from dynamics itself, using possible approximations (e.g. mean field \cite{paninski:04}). In the second case, one has to define a probability distribution on the spike trains in order to investigate the effect of a perturbation. In this section, we show how these 2 approaches are related, making a link between the classical statistical physics approach of linear response, dynamical systems and ergodic theory, and neuronal networks. We introduce then the general formalism of chains with unbounded memory allowing the handling of non-equilibrium linear response for spiking neuronal networks. All of the material in this section is known in different domains, statistical physics, ergodic theory, stochastic processes, neuronal networks, and is presented here for a better understanding of the next sections.

\subsection{Linear response in statistical physics} \label{Sec:LinRepStatPhys}

For simplicity, we consider in this introductory section a dynamical system taking a finite number of ``states", where a state is denoted by $\omega$.  In statistical physics, the linear response theory can be addressed in these terms. In a system at thermodynamic equilibrium,
the probability of observing a state $\omega$ is given by the Boltzmann-Gibbs distribution:
\begin{equation}\label{eq:GibbsStatPhys}
\moy{\omega}=\frac{1}{Z}e^{-\frac{H\pare{\omega}}{k_B T} },
\end{equation}
where $Z=\sum_{\omega} e^{-\frac{H\pare{\omega}}{k_B T}}$ is called the partition function, with $k_B$, the Boltzmann constant and $T$ the temperature. The function of state:
\begin{equation} \label{eq:HStatPhys}
H\pare{\omega} =  \sum_{\alpha} \lambda_\alpha X_\alpha\pare{\omega},
\end{equation}
is called the energy of the state $\omega$. The functions $X_\alpha$ are extensive quantities (proportional to the number of particles) such as energy, electric charge, volume, number of particles, magnetic field, $\dots$ 
The conjugated parameters $\lambda_\alpha$ 
correspond to intensive quantities (not proportional to the number of particles), like temperature, electric potential, pressure, chemical potential, magnetic susceptibility, $\dots$. In general, they depend on the location in the physical space (e.g. the temperature depends on the position in a fluid). At equilibrium, they are uniform in space though. 
\textit{The form of $H$, i.e. the choice of the $\lambda_\alpha$ and $X_\alpha$ is constrained by the physical properties of the system. It is also constrained by boundary conditions.}

In standard statistical physics courses, the Gibbs distribution form \eqref{eq:GibbsStatPhys} is obtained as a consequence of a principle, the Maximum Entropy Principle \cite{jaynes:57}. For a 
probability measure $P$ on the set of states, the statistical entropy is:
\begin{equation}\label{Eq:EntropyPhyStat}
S\bra{P} = - k_B \sum_\omega \log P\bra{\omega} \log P\bra{\omega}.
\end{equation}
Denote $\Expm{P}{}$ the expectation with respect to $P$. The Maximum Entropy Principle seeks a probability distribution maximizing the statistical entropy under the constraint that the average energy is constant, i.e. $\Expm{P}{H}=C$ for any probability measure $P$ on the set of states. This probability exists and is unique when the set of states is finite; this is \eqref{eq:GibbsStatPhys}. 
When this set is infinite (e.g. thermodynamic limit) additional summability conditions are required on $H$ to ensure existence and uniqueness \cite{ruelle:67,georgii:88}.

A non-equilibrium situation arises when the $\lambda_\alpha$s are not uniform in space, generating gradients $\vec{\nabla} \lambda_\alpha$ (temperature gradient, electric potential gradient ...). These gradients result in currents $\vec{j}_\alpha$ of $X_\alpha$ (e.g. a temperature gradient induces a heat current). In general, the currents are nonlinear functions of gradients. The Onsager linear response theory assumes that currents are linear combinations of gradients (i.e. gradients are weak enough so that non-linear terms can be neglected). Known examples are Ohm's law where the electric current is proportional to the gradient of the electric potential,  Fourier's law where the heat flux is proportional to the temperature gradient, Fick's law etc. Several gradients can be simultaneously involved like in Peltier effect. The proportionality coefficients are called Onsager coefficients \cite{bellac:04}.

Now, the property that interests us is that Onsager coefficients are obtained as correlations functions computed \textit{at equilibrium} (Kubo relations, \cite{kubo:57}). Thus, the knowledge of correlations
at equilibrium allows the inference of the non-equilibrium response of the system to (weak) perturbations.

\subsection{Linear response in dynamical systems} \label{Sec:LinRepSD}

The Maximum Entropy Principle is a powerful tool as it allows for the establishment of a link between the description of a system in terms of its states $\omega$ and thermodynamics, characterized by macroscopic averages. However, many scientists \cite{gaspard:04,dorfman:99,hoover:12,gallavotti:95,ruelle:97} 
starting with Boltzmann himself \cite{boltzmann:02}, tried to construct equilibrium and non-equilibrium statistical physics from the microscopic dynamics of the system. At equilibrium, the problem can be formally stated this way. ``Given an autonomous dynamical system in a compact phase space $\cM$, is there a natural probability measure characterizing how trajectories sample this phase space? Under which conditions does this probability take  the exponential form \eqref{eq:GibbsStatPhys}?".
Autonomous means here that the vector field of the dynamical system is independent of time, what we refer to as an equilibrium situation.
Natural means ``for typical initial conditions", i.e. selected with respect to the Lebesgue measure on $\cM$.  

The natural context to address this question is ergodic theory. For flows preserving the volume in the phase space, the natural measure is the Liouville measure \cite{gaspard:98}. For dissipative systems, the natural measure is the so-called Sinai-Ruelle-Bowen measure (SRB) \cite{young:02} whose existence is guaranteed in a specific class of dynamical systems (e.g. axiom A, uniformly hyperbolic \cite{bowen:75,walters:75}). It is given by the weak-limit of the Lebesgue measure under the flow of the dynamics \cite{ruelle:99}. In these systems, there exists in addition a remarkable finite partition of the phase space, called Markov partition. This allows us to map the trajectories of the dynamical system to the trajectories of a Markov chain (symbolic coding).
The SRB measure is then the invariant measure of this Markov chain. This is a Gibbs distribution with a potential determined by the Jacobian of the flow \cite{ruelle:67,bowen:75,georgii:88}. It obeys a Maximum Entropy Principle as well.

The SRB measure extends to time-dependent flow \cite{ruelle:99} in uniformly hyperbolic systems \cite{bowen:75,walters:75}. In this context, David Ruelle provided a proof of the linear response, as a consequence of the differentiability of the SRB measures with respect to smooth perturbations of the flow \cite{ruelle:97,ruelle:08}. He established a linear response formula and Onsager coefficients, depending on the Jacobian of the flow. There are Kubo relations where Onsager  coefficients are obtained from correlations at equilibrium. In non-uniformly hyperbolic systems there may not exist a linear response  \cite{baladi:14} like in the Henon map. However, this violation can be detected and quantified numerically \cite{cessac:07}.

\subsection{Linear response in neuronal networks} \label{Sec:LinRepRN}

Neuronal networks are modeled by dynamical systems (possibly stochastic). Therefore, the linear response theory can be addressed using the tools briefly presented in the previous section. This has for example been done for a discrete time in the Amari-Wilson-Cowan \cite{amari:77,wilson-cowan:73} model where the convolution kernel $K$ appearing in \eqref{eq:KernelRF} can be explicitly computed \cite{cessac-sepulchre:04,cessac-sepulchre:06}. Notably, one can compute the response of a neuron to a weak harmonic perturbation of another neuron, exhibiting specific resonances and a functional connectivity distinct from the synaptic graph. 

When dealing with spiking models such as Integrate and Fire, the dynamics are not differentiable anymore (because of the mechanism of reset at threshold). Still, the Markov chain formalism (and its extension to infinite memory) can be used, as developed below.
In particular, we exhibit an example where transition probabilities can be explicitly computed and directly related to dynamics. 

More generally, the formalism presented in section \ref{Sec:LinRepInfMem} extends to a situation where the dynamics are unknown and only spike trains are observed: one only has access to spike correlations, and the idea, inherent to linear response theory is to use these correlations in the absence of stimulus to infer the linear response to a time-dependent stimulus of weak amplitude. This leads to specific questions intrinsic to neuronal dynamics:

\begin{itemize} 
\item \textbf{Are there Kubo-like relations in spiking neural networks?} How are linear response coefficients related to spontaneous spiking dynamics? From the statistical physics wisdom, we expect the linear response to be given by spikes correlations functions. This paper aims at providing a systematic way to express the linear response in terms of spike correlations. 
\item \textbf{What is the natural basis to expand the linear response?} This question is related to the previous one. There are many possible spike correlations in space and time, pairs, triplets, and so on. Clearly, one expects high correlations to play a less important role than, say, pairs correlations, but is it possible to quantify how it depends on neurons dynamics. As we show in this paper, the dependence lies in a transfer matrix constructed from transition probabilities, depending themselves, upon dynamics. An important consequence derived from this formalism is exponential correlation decay. Under fairly general conditions the spectrum of the transfer matrix has a gap between the largest eigenvalue and the remainder of the spectrum (this results from the Perron-Frobenius theorem \cite{seneta:06,gantmacher:98}). This implies exponential correlations decay, and ensures the existence of a linear response. It also implies the existence of resonances in the power spectrum with strong consequences on the linear response to harmonic signals. This is further developed in section \ref{Sec:CorrDecay} and \ref{Sec:MonDecomp} below. 
\item \textbf{Synaptic versus effective connectivity.} Modeling the spike train statistics from experimental data without knowing the underlying dynamics leads to a partial knowledge of the causality in the spike trains induced by dynamics and synaptic connectivity. This approach makes notions appear such as ``effective connectivity'' build from stationary correlations between neurons (think of the ``Ising'' model build from instantaneous pairwise correlations \cite{schneidman-berry-etal:06}). How is this effective connectivity related to the synaptic connectivity? Another notion of effective connectivity arises from the neuronal response. Excite a single neuron with a stimulus and check whether other neurons respond. This provides a third notion of connectivity. It has been shown, for a discrete time, a version of Amari-Wilson-Cowan model, that these three notions of connectivity lead to completely different graphs of connectivity \cite{amari:77,wilson-cowan:73}. What is the situation for spiking neurons? A partial answer to this question can be found in the paper \cite{cocco:09} where the mapping from the Integrate and Fire model to Ising has been considered with statistical physics methods. However, the Ising model considers instantaneous pairwise spike events whereas spike interactions involve delays. Thus, a better-adapted notion of effective connectivity should include such delays. We provide a general formalism to do this in this paper. We show in particular that linear response connectivity differs in general from pairwise correlations connectivity, except in simplified models like Ising where the fluctuation-dissipation theorem establishes, in this case, the proportionality between the susceptibility and the instantaneous pairwise correlations (this does not hold for spike train statistics including memory and delays). 
\item \textbf{Plasticity.} Another advantage of the linear response theory is to handle the effect of synaptic plasticity and learning. Assume that a neuronal network, submitted to a stimulus, has its synaptic weights evolving in time according to a learning rule (Hebbian, STDP) involving spike pairwise correlations (with a time delay). The response to the stimulus depends on these correlations, and these correlations depend upon the stimulus, in a feedback loop. Although this aspect is not addressed in this paper, our results open up the possibility of describing this feedback loop. 
\item \textbf{Data.} An important application of linear response is that correlations are computed with respect to the statistics in spontaneous activity, which can be approximated from experimental recordings of a neuronal network spiking, in absence of external stimuli, using the Maximum Entropy Principle. Thus a linear response formula could be used to anticipate the response of a biological neuronal network to a weak stimulus, knowing the structure of its spontaneous spike correlations.
\end{itemize}

\ssu{Spike trains and observables}\label{Sec:SpikesObs}

Neuron variables such as membrane potential or ionic currents are described by continuous-time equations. In contrast, spikes resulting from the experimental observation are discrete events, binned with a certain time resolution $\delta$. In this paper, we will jointly consider these two time descriptions.

We consider a network of $N$ neurons, labeled by the index $k=1 \dots N$. We define a spike variable $\omega_{k}(n)=1$ if neuron $k$ has emitted a spike in the time interval $[n \delta, (n+1)\delta[$, and $\omega_k(n)=0$ otherwise. 
We denote by $\omega(n) := \bra{\omega_{k}(n)}_{k=1}^{N}$ the spike-state of the entire network at time $n$, which we call a \textit{spiking pattern}. We denote by $\cA=\Set{0,1}^N$, the state space of spiking patterns in a  network of $N$ neurons;
a \textit{spike block} denoted by $\bloc{m}{n}$, $n \geq m$, is the sequence of spike patterns $\omega(m), \omega(m+1), \dots, \omega(n)$; blocks are  elements of the product set $\cA^{n-m}$ also denoted $\cA^{n}_{m}$ in the text. We use this last notation because we  will consider processes with infinite memory ($m \to -\infty$) and we want to have an explicit notation $\cA^{n}_{-\infty}$ for the corresponding set of events.
 The \textit{time-range} (or ``range'') of a block $\bloc{m}{n}$ is $n-m+1$, the number of time steps from $m$ to $n$. We call a spike train an infinite sequence of spikes both in the past and in the future.  The set of spike trains is  thus $\Omega \equiv \cA^\setZ$. To simplify notations we note a spike train $\omega \in \Omega$.
The shift operator  $\sigma: \Omega \to \Omega$ is $\sigma \omega = \omega'$, with $\omega'(n) = \omega(n+1)$. This allow us to go one step forward in time along the spike train $\omega$.

We note $\mathcal{F}_{\leq n}$ the set of measurable events (filtration) before time $n$ and 
$\mathcal{F}$ the filtration on $\Omega$. $\mathcal{P}(\Omega,\mathcal{F})$ is the set of probability measures on $\Omega,\mathcal{F}$.\\

We will use the notion of (spike) \textit{observable}.
This is a function $f : \Omega \rightarrow \mathds{R}$  that associates a real number to a spike-train.
We say that the observable $f:\Omega \rightarrow \mathds{R}$ has range $R=D+1$ if $f(\omega) \equiv f(\bloc{0}{D})$. It follows from the Hammersley-Clifford theorem \cite{hammersley-clifford:71,moussouris:74}  that any range-$R$ observable can be written in the form:
\beq\label{eq:monomial_decomp}
f(\omega)=\sum_l f_l \, m_l(\omega),
\eeq
where $f_l$ are real numbers, the coefficients of the decomposition of $f$ in the finite space of range $R$-observables. The functions $m_l$ spanning this space are called monomials \cite{cofre-cessac:14}. They have the form:
\begin{equation}\label{eq:monomial_def}
m_l(\omega)=\prod_{k=1}^n \omega_{i_k}(t_k).
\end{equation}
where $i_k = 1 \dots N$ is a neuron index, and $t_k=0 \dots D$. Thus,  $m_l(\omega)=1$ if and only if, in the spike train $\omega$, neuron $i_1$ spikes at time $t_1$, $\dots$, neuron $i_k$ spikes at time $t_k$. Otherwise, $m_l(\omega)=0$. The number $n$ is the \textit{degree} of the monomial; degree one monomials have the form $\omega_{i_1}(t_1)$, degree two monomials have the form $\omega_{i_1}(t_1)\omega_{i_2}(t_2)$, and so on. Thus, monomials are similar to what physicists call (spike) interactions; in our case these interactions involve a time delay between spikes. 
There are $L=2^{NR}$ monomials of $N$ neurons and range $R$ and one can index
each of them by an integer $l$ in one-to-one correspondence with the
set of pairs $(i_k,t_k)$ (see eq. \eqref{eq:DefBlockIndex}). The advantage of monomial representation is to focus  on spike events, which is natural for spiking neuronal dynamics. 

The decomposition \eqref{eq:monomial_decomp} is straightforward. However, we would like to give a simple, yet illustrative example.  Let us consider $1$ neuron at times $0$ and $1$. The spiking patterns are $\omega=(\omega_1(0) \, \omega_1(1))$=$\pare{0,0}$, $\pare{1,0}$, $\pare{0,1}$, $\pare{1,1}$, and, a function of these patterns takes four values: $f\pare{0,0}=F_0$, $f\pare{1,0}=F_1$, $f\pare{0,1}=F_2$, $f\pare{1,1}=F_3$. We have:
$$
f(\omega)=F_0 \, \pare{1-\omega_1(0)}\pare{1-\omega_1(1)}
+ F_1 \, \omega_1(0)\pare{1-\omega_1(1)}$$
$$
+ F_2  \, \pare{1-\omega_1(0)}\omega_1(1)
+ F_3  \, \omega_1(0)\omega_1(1),
$$
that we may rewrite in the form \eqref{eq:monomial_decomp} :
\begin{equation*}
\begin{split}
f(\omega) &= f_0 + f_1 \, \omega_1(0) + f_2 \, \omega_1(1) + \, f_3 \,  \omega_1(0)\omega_1(1);\\
f_0&=F_0; \, f_1=F_1-F_0; \, f_2 = F_2-F_0; f_3 = F_0-F_1-F_2+F_3\\
\end{split}
\end{equation*}
More generally the transformation from the ``function'' representation (the vector $\bF$ of $F_i$s in the example) to the monomial representation  (the vector $\bf$, of  $f_l$s) is given by the linear transformation:
\begin{equation}\label{eq:Hammersley-Clifford}
\bf = Q\cdot\bF,
\end{equation}
where $Q$ is a triangular matrix given by \cite{moussouris:74}:
\begin{equation*}
Q_{ll'} = \left\{
\begin{array}{lll}
\pare{-1}^{d(l)-d(l')}; \quad l' \sqsubseteq l \\
0; \quad otherwise
\end{array}
\right.
\end{equation*}
Here the notation $l' \sqsubseteq l$ means the following: To each spike block $\bloc{0}{D}$, one can associate a
unique integer:
\beq\label{eq:DefBlockIndex}
l =\sum_{k=1}^N \sum_{n=0}^D 2^{n\,N+k-1} \, \omega_k(n),
\eeq
called the \textit{index} of the block. 
We define the block inclusion $\sqsubseteq$ on $\Omega_{N,R}$:
$\bloc{0}{D} \sqsubseteq \seq{\omega\,'}{-D}{0}$ if $\omega_k(n)=1
\Rightarrow \omega'_k(n)=1$ (all bits '1' in $\bloc{0}{D}$ are bits
'1' in $\seq{\omega \,'}{-D}{0}$), with the convention
that the block of degree $0$ is included in all blocks. By extension $l' \sqsubseteq l$ means that all bits '1' in the block corresponding to the integer  $l$ are included in the block corresponding to $l'$ \cite{cofre-cessac:14}.\\
The result \eqref{eq:Hammersley-Clifford} shows that the coefficient of a monomial is the linear combination of function values where the number of terms in the combination increases with the monomial degree. \\

We now introduce \textit{time dependent observables}. These are functions $f(t,\omega)$ depending on time $t$ (continuous or discrete) and on the spike train $\omega$. The notation $f(t,\omega)$ stands here for $f(t,\bloc{-\infty}{\ent{t}})$ where $\ent{t}$ is the integer part of $t$: the function depends on the spike train $\omega$ via spikes \textit{preceding} the current time $t$. This is an implementation of causality. A  range-$R$
time dependent observable is a function $f(t,\omega) \equiv f(t,\bloc{\ent{t}-D}{\ent{t}})$. The decomposition \eqref{eq:monomial_decomp} also holds for a time dependent range-$R$ observable 
\beq\label{eq:monomial_decomp_t}
f(t,\omega)=\sum_l f_l(t) \, m_l(\sigma^{\ent{t}} \omega),
\eeq
where $f_l(t)$ are now functions of time, and $\sigma^{\ent{t}}$ is the $\ent{t}$-iterate of the time shift operator $\sigma$.

\subsection{Homogeneous Markov chains and Gibbs distributions}\label{Sec:HomogeneousMarkov}

A ``natural'' way to characterize the statistics of observed spike trains is to associate them to a Markov chain with transition probabilities $\Pnc{\omega(n)}{\bloc{n-D}{n-1}}$, where the index $n$ in $\mathds{P}_n$ indicates that the transition probabilities depend on time $n$. 
This approach is ``natural'' because it captures causality by conditioning on the past spikes. We call $D$ the memory depth of the chain and set $R=D+1$. 

\subsubsection{Invariant probability} 

Let us start the discussion when transition probabilities are independent of time (homogeneous Markov chain). In this case, we drop the index $n$ in the transition probabilities,  $\Probc{\omega(n)}{\bloc{n-D}{n-1}}$. Assuming that all transition probabilities are strictly positive, it follows from the Perron-Frobenius theorem \cite{seneta:06,gantmacher:98} that the Markov chain has a unique invariant probability $p$ on $\cA^{D}$. From the Chapman-Kolmogorov equation \cite{seneta:06} one constructs, from $p$ and transition probabilities, a probability measure $\mu$ on $\mathcal{P}(\Omega,\mathcal{F})$.
where:
\begin{equation}\label{eq:Chapman-Kolmogorov}
\moy{\bloc{m}{n}} 
= \prod_{l=m+D}^{n} \Probc{\omega(l)}{\bloc{l-D}{l-1}} \, p \bra{\bloc{m}{m+D-1}}, \quad  \forall m< n \in \setZ.
\end{equation}
As we will now discuss, there is a natural correspondence between $\mu$ and exponential distributions of the form \eqref{eq:HStatPhys} (Gibbs distributions). 

\subsubsection{Transfer matrix}
Let us indeed now consider a range-$R$ observable:
\begin{equation}\label{eq:H_decomp}
H\pare{\omega} = \sum_l h_l \, m_l(\omega),
\end{equation}
 where $h_l > C > -\infty$. 
  Any block $\bloc{0}{D}$ of range $R=D+1$ can be viewed as a transition from a block $\om{u}=\bloc{0}{D-1}$ to the block 
$\om{u'}=\bloc{1}{D}$. We write $\bloc{0}{D} \sim \om{u}\om{u'}$. 
By extension, for two blocks $\om{u},\om{u'}$ of range $D\geq 1$ we say that the transition $\om{u} \to \om{u'}$ is \textit{legal} if there is a block $\bloc{0}{D} \sim \om{u}\om{u'}$. 
On this basis, one can construct a transfer matrix with positive entries:
\begin{equation}\label{eq:PF}
\cL_{\om{u},\om{u'}}=   
\left\{
\begin{array}{lll}
 e^{H(\bloc{0}{D})},
\quad &\mbox{if }  \bloc{0}{D} \sim \om{u}\om{u'};  \\
0, \quad &\mbox{otherwise}.
\end{array}
\right. 
\end{equation}

It follows from the Perron-Frobenius theorem \cite{seneta:06,gantmacher:98} that $\cL$ has a unique real positive eigenvalue $s$, strictly larger in modulus than the other eigenvalues, and with positive right eigenvector $\cL\rpf=s\rpf$, and left eigenvector $\lpf \cL=s \lpf$. Moreover, the range-$R$ observable:
%
$$\phi(\bloc{0}{D})=H(\bloc{0}{D}) - \log \rpfc{\bloc{0}{D-1}} + \log \rpfc{\bloc{1}{D}}-\log s$$
%
%
defines an homogeneous Markov chain \cite{walters:75} with transition probabilities $\Probc{\omega(D)}{\bloc{0}{D-1}} = e^{\phi(\bloc{0}{D})}$.  

\subsubsection{Invariant probability and Gibbs distribution}

The unique invariant probability of this Markov chain is:
\beq\label{eq:p}
p(\bloc{0}{D-1})=\rpfc{\bloc{0}{D-1}}\lpfc{\bloc{0}{D-1}}.
\eeq
Using the Chapman-Kolmogorov equation \eqref{eq:Chapman-Kolmogorov} one extends $p$ to a probability $\mu$ on $\Omega$, where, for $m+n > D$ :

\beq\label{eq:CondProbMarkov}
\moyc{\bloc{m}{n}}{\bloc{m-D-1}{m-1}} = \frac{e^{\sum_{l=m-D}^{n-D} H\pare{\bloc{l}{l+D}}}\rpfc{\bloc{n-D+1}{n}}\lpfc{\bloc{m-D}{m-1}}}{s^{n-m+1}}.
\eeq
This emphasizes the Markovian nature of the process since the conditioning has a finite time horizon of depth $D$. 

It follows therefore that the probability of observing a spike block $\bloc{m}{n}$, given a certain past $\bloc{m-D-1}{m-1}$ 
is proportional to $e^{\sum_{l=m-D}^{n-D} H\pare{\bloc{l}{l+D}}}$. If $H$ is formally interpreted as an energy\footnote{Note that, in contrast to \eqref{eq:HStatPhys} we have removed the $-$ sign which has no reason to be present in this context, and is a source of nuisance when doing computations.} then $\sum_{l=m-D}^{n-D} H\pare{\bloc{l}{l+D}}$ is the energy of the block $\bloc{m}{n}$.

This establishes a first relationship with Gibbs distributions\footnote{More precisely one has, $\exists \, A, B >0$ such that, for any block $\bloc{0}{n}$, $$A \leq \frac{\moy{\bloc{0}{n}}}{e^{-(n-D+1)\cP(\mathcal{H})} e^{-\sum_{k=0}^{n-D}\mathcal{H}\pare{\bloc{k}{k+D}}}}\leq B,$$ which defines, in ergodic theory, a Gibbs measure in the sense of Bowen \cite{bowen:75}.} of the form \eqref{eq:GibbsStatPhys}, with a strong difference though. Whereas we assumed \eqref{eq:GibbsStatPhys} to hold on a finite set of states characterizing the system at a given time, here $\omega$ is a trajectory of the system describing its time evolution. In addition, the probability of the block $\bloc{m}{n}$ is conditioned upon the past, which, in statistical physics would correspond to determine the probability of a block of binary variables (say spins) $\bloc{m}{n}$ with left boundary conditions $\bloc{m-D-1}{m-1}$. This analogy is further developed in the next section.

In the case of a Markov chain, the entropy \eqref{Eq:EntropyPhyStat} extends to:
\begin{equation*}
S(\mu)=-\sum_{\bloc{0}{D}} p\pare{\bloc{0}{D-1}}  \Probc{\omega(D)}{\bloc{0}{D-1}} \log \Probc{\omega(D)}{\bloc{0}{D-1}},
\end{equation*}
where we have dropped the Boltzmann constant as it plays no role here. Then, it can be shown that $\mu$ satisfies a variational principle \cite{bowen:75, walters:75}; it maximizes $S\bra{\nu} \, + \, \Expm{\nu}{H}$, where $\nu$ is an invariant probability on $\mathcal{P}(\Omega,\mathcal{F})$. If $\Expm{\nu}{H}$ is fixed this amounts to maximizing the entropy under the constraint that the average energy $\Expm{\nu}{H}$ is fixed.
Finally, the supremum $\cF=S\bra{\nu} \, + \, \Expm{\nu}{H}$ corresponds to the free energy; the generating function of cumulants.

We have therefore shown that a potential of the form \eqref{eq:H_decomp} is associated to a homogeneous Markov chain where the invariant probability, extends the notion of Gibbs distribution, introduced in section \ref{Sec:LinRepStatPhys}, to systems with memory where the probability to be in a state depends on a finite history. The extension to infinite history is made in the next section.
As an important remark, \textit{the detailed balance condition not required to define Gibbs distributions from Markov chains.\\}

Reciprocally, one can associate to a Markov chain with strictly positive transition probabilities a function of the form \eqref{eq:H_decomp}. In fact, there are infinitely many such functions (related through cohomology relations), but there is one having a minimal number of monomials \cite{cofre-cessac:14}.  
This actually raises a serious problem when dealing with so-called Maximum Entropy models to handle neuronal spike trains. \textit{The energy is not known a priori}.
In contrast to thermodynamics, where the $\lambda_\alpha$ and $X_\alpha$ entering in the definition of the energy \eqref{eq:HStatPhys}
are known from first principles, here we have no such principle to guide us in order to reduce the number of terms in \eqref{eq:H_decomp}. There exist methods to reduce this complexity \cite{cofre-cessac:14,herzog:18} but still there remain a large number of terms. For example, one can exactly map a Generalized Linear Model, depending on $O(N^2)$ parameters, to a Maximum Entropy model, but the number of terms in this model is generically exponential in $N$ \cite{cofre-cessac:14}.

At this stage, still dealing with time-translation invariant systems, we have therefore two possible representations to handle a spike train statistics : (i) the Maximum Entropy approach with a generic potential of the form \eqref{eq:H_decomp}; (ii) the Markov chain approach. While the two approaches are equivalent, the first one is agnostic about the underlying model, but has many redundant and irrelevant terms, and is hard to interpret. The second requires the the knowledge of the transition probabilities, either inferring them from data - which is in general difficult because many blocks do not appear in the sample so that one cannot reliably estimate the conditional probabilities - or by guessing their form from ad hoc models, like LN or GLM \cite{paninski:04,pillow-paninski-etal:05}. However, it is possible to establish this form analytically in some IF models, as developed in section \ref{Sec:LinRepgIF}.

\sssu{Correlation decay}\label{Sec:CorrDecay}

Replacing $H$ by $\phi$ in \eqref{eq:PF} one constructs a matrix, $\cL_\phi$, which, from the Perron-Frobenius theorem, has a largest eigenvalue of $1$ with eigenvector \eqref{eq:p}. The rest of the spectrum is bounded away from $1$ (spectral gap). A very important property resulting from this, used throughout the paper, is the exponential correlation decay. For two observables $f, g$, and integer times $m,n$ we define the correlation:  
\begin{equation}\label{eq:CorrDef}
\cC_\mu\bra{f(m,.),g(n,.)} \deq 
\mathds{E}_\mu\left[\, f(m,.)g(n,.) \, \right] -  \mathds{E}_\mu\left[\,f(m,.)\, \right] \mathds{E}_\mu\left[\,g(n,.) \right]
\end{equation}
In the present case, where $\mu$ is time translation invariant, $\cC_\mu\bra{f(m,.),g(n,.)}$ depends only on $m-n$ so that we write, for simplicity $\cC_\mu\bra{f(m,.),g(n,.)} \equiv \cC_{f,g}(m-n)$. 
Using the fact that $\cL_\phi$ is the adjoint (with respect to $\mu$) of the time shift operator $\sigma$, and
using the spectral decomposition theorem one obtains that:
\begin{equation}\label{eq:CorrDecay}
\cC_{f,g}(r) = \sum_{k=2}^{L}  \lambda_k^r \, \Gamma_{k,f,g},
\end{equation}
where $\Gamma_{k,f,g}$ are complex numbers. Note that, as $\cL_\phi$ is real, the eigenvalues are complex-conjugated. The coefficients $\Gamma_{k,f,g}$ combines together to produce real correlations functions with an exponentially decaying part (the modulus of $\lambda_k$s) and an oscillatory part (the phase of  $\lambda_k$s). 
Note that the sum starts at $k=2$ because the term corresponding to $\lambda_1=1$ (the first, largest eigenvalue corresponding to the invariant probability) is removed from the definition \eqref{eq:CorrDef} of the correlations. It follows from the spectrum of $\cL_\phi$ that $\abs{\lambda_k} < 1-\epsilon < 1$. Therefore, correlations decay exponentially fast with a characteristic rate $-\log\pare{\abs{\lambda_2}}$.

\subsection{Chains with infinite memory and Gibbs distributions}\label{Sec:ChainsGibbs}

In the previous section we have made two important assumptions: (i) memory is bounded (finite memory depth $D$); (ii) the correspondence between Markov chains and Gibbs form was established for homogeneous Markov chains. 

However, when considering neural networks, the memory is not necessarily, neither constant nor bounded: consider for example an Integrate and Fire model where the memory goes back to the last time in the past when the neuron has fired; in general, it is not possible to bound this time. So, the most general formalism is to consider chains with unbounded memory \cite{galves:13,reynaud:14,vidybida:15}. Of course, as we discuss below, Markovian approximations are possible and useful. Still, one needs to properly control these approximations. In addition, we want to consider here the case of a system submitted to a time-dependent stimulus, where the dynamic is not time-translation invariant.

Thus, we are now considering a family of transition probabilities of the form $\Pnc{\omega(n)}{\sif{n-1}}$, which represent the probability that at time $n$ one observes the spiking pattern $\omega(n)$ given the (unbounded) network spike history. Such a non-Markovian stochastic process is known as a ``chain with complete connections'' or a ``chain with unbounded memory'' (\cite{onicescu:35}) defined in more detail here. This section follows very closely from \cite{fernandez-maillard:05}.\\

\noindent
\textbf{Definition }
A system of transition probabilities is a family $\{\probc_n\}_{n \in \mathds{Z}}$ of functions with $\Pnc{\cdot}{\cdot}: \mathcal{A}\times \mathcal{A}^{n-1}_{-\infty} \rightarrow [0,1]$ such that the following conditions hold for every $n \in \mathds{Z}$:\\

\noindent
\textbf{(a)} For every $\omega(n) \in \mathcal{A}$ the function $\Pnc{\omega(n)}{\cdot}$ is measurable with respect to $\mathcal{F}_{\leq n-1}$.\\

\noindent
\textbf{(b)}  For every $\sif{n-1} \in \mathcal{A}^{n-1}_{-\infty}$,

$$\sum_{\omega(n) \in \mathcal{A}} \Pnc{\omega(n)}{\sif{n-1}}=1.$$

\noindent
\textbf{Definition }
A probability measure $\mu$ in $\mathcal{P}(\Omega,\mathcal{F})$
is \textbf{consistent} with a system of transition probabilities $\{\mathds{P}_n\}_{n \in \mathds{Z}}$ if:

\beq\label{eq:def_consistent}
\int h\left(\seq{\omega}{-\infty}{n}\right) \mu(d\omega) = \int \sum_{\omega(n) \in \cA} h\left(\sif{n-1} \omega(n) \right)\Pnc{\omega(n)}{\sif{n-1}} \mu(d\omega).
\eeq

\noindent
for all $n \in \mathds{Z}$ and all $\mathcal{F}_{\leq n}$-measurable functions $h$. The probability measure $\mu$, when it exists, is called a \textit{chain with complete connections} consistent with the system of transition probabilities $\{\mathds{P}_n\}_{n \in \mathds{Z}}.$ It is possible that multiple measures are consistent with the same system of transition probabilities.\\

\noindent
We now give conditions ensuring the existence of a probability measure consistent with the system of transition probabilities \cite{fernandez-maillard:05}.\\

\noindent
\textbf{Definition }
A system of transition probabilities is \textbf{non-null} on $\Omega$  if, for all $n \in \mathds{Z}$ and all $\sif{n} \in  \mathcal{A}^{n}_{-\infty}$:

$$\Probc{\omega(n)}{\sif{n-1}}>0$$

\noindent
We note, for $n \in \mathds{Z}$, $m \geq 0$, and $r$ integer:

$$\omega \stackrel{m,n}{=} \omega', \,  \mbox{ if } \omega(r) = \omega'(r), \forall r \in \{n-m, . . . , n\}.$$

\noindent
\textbf{Definition}
Let $m$ be a positive integer. The $m$-variation of  $\Pnc{\omega(n)}{\cdot}$ is:

\begin{equation}\label{eq:Variation}
var_m[\Pnc{\omega(n)}{\cdot}]=\sup\left\lbrace \mid \Pnc{\omega(n)}{\sif{n-1}}-\mathds{P}_n[\omega(n)\mid \omega'^{n-1}_{-\infty}] \mid : \omega \stackrel{m,n}{=} \omega'\right\rbrace
\end{equation}

\noindent

\noindent
\textbf{Definition}
The function $\Pnc{\omega(n)}{\cdot}$ is \textbf{continuous }if $var_m[\Pnc{\omega(n)}{\cdot}] \rightarrow 0$ as $m \rightarrow + \infty$.\\

The intuitive meaning of continuity is the following. The quantity $var_m[\Pnc{\omega(n)}{\cdot}]$ corresponds to the maximum variation one can observe on the probability of the spike state at time $n$, \textit{given that the history is fixed up to time $n-m$}. Thus, continuity implies that this variation tends to zero as $m$ tends to infinity: the further in the past that the spike sequence is fixed, the smaller the probability that the past influence the present. \\

The following result holds (see \cite{fernandez-maillard:05}):\\

\noindent
\textbf{Theorem } A system of continuous transition probabilities on a compact space has at least one probability measure consistent with it. \\

Uniqueness requires additional technical assumptions \cite{fernandez-maillard:05}. These conditions hold in the gIF model \cite{cessac:11b} considered in section \ref{Sec:LinRepgIF}.\\

Let us now elaborate on the link with Gibbs distributions. First, we define $\phi\pare{n,\omega}: \mathbb{Z} \times \Omega \rightarrow \mathbb{R}$ by:
\beq\label{eq:phi_def}
\phi\pare{n,\omega} \equiv \log \Probc{\omega(n)}{\sif{n-1}},
\eeq 
and:
\beq\label{eq:pCondphi}
\Phi(m,n,\omega)=\sum_{r=m}^n \phi\pare{r,\omega}.
\eeq
Then:
\begin{equation}\label{eq:CondProbBlocChain}
\Probc{\bloc{m}{n}}{\sif{m-1}}=e^{\Phi(m,n,\omega)} = 
e^{\sum_{r=m}^n \phi\pare{r,\omega}}
\end{equation}
and:
\beq\label{eq:MuBlockChain}
\mu[\bloc{m}{n}] =  \int_{\seq{\cA}{-\infty}{m-1}} 
 e^{\Phi(m,n,\omega)} \mu(d\omega).
\eeq

These equations emphasize the connection with Gibbs distributions in statistical physics where $\phi$ acts as an ``energy'' \cite{kozlov:74,fernandez-maillard:05}. From now on we will instead use the term ``potential". The correspondence in  our case is to consider ``time'' as a 1-dimensional ``lattice'' and the ``boundary conditions'' as the past $\sif{m-1}$ of the stochastic process. In contrast to statistical physics, and because the potential is defined via transition probabilities, the normalization factor (partition function) is equal to $1$. For this reason we call
$\phi$  \textit{a normalized Gibbs potential}. 

Equations \eqref{eq:CondProbBlocChain} and \eqref{eq:MuBlockChain} are similar to \eqref{eq:CondProbMarkov} with an essential difference: the memory is now infinite, and the potential $\phi$ has an infinite range. As it is well known in statistical physics \cite{ruelle:67,georgii:88}, infinite range potentials require specific conditions to be associated with a unique Gibbs distribution.
There is a mathematically well founded correspondence between chains with complete connections and Gibbs distributions \cite{fernandez-maillard:05,georgii:88,ruelle:67}. However, while chains with complete connections define probability transitions where the present is conditioned upon the past, Gibbs distributions allows conditioning ``upon the future'' as well
\footnote{More generally, Gibbs distributions in statistical physics extend to probability distributions on $\setZ^d$ where the probability \eqref{eq:GibbsStatPhys} to observe a certain configuration of spins in a restricted region of space is constrained by the configuration at the boundaries of this region. They are therefore defined in terms of specifications \cite{ruelle:67,georgii:88}, which determine  finite-volume conditional probabilities when the exterior of the volume is known. In one spatial dimension ($d=1$), identifying $\setZ$ with a time axis, this corresponds to conditioning both in the past and in the future. In contrast, families of transition probabilities with an exponential continuity rate define so-called left-interval specifications (LIS) \cite{fernandez-maillard:05,leny:08}.}. This leads to different notions of ``Gibbsianness", not equivalent \cite{fernandez-gallo:11}. We shall not develop on these distinctions here and will call Gibbs distribution a chain with complete connection.

\section{Linear response for neuronal networks with unbounded memory} \label{Sec:LinRepInfMem}

We consider a neural system where spike statistics is characterized by a time-translation invariant Gibbs distribution (chain with unbounded memory) $\musp$ where ``sp'' stands for ``spontaneous". That is, we suppose that, in the absence of a stimulus, the spontaneous dynamics is stationary. 
We assume that a stimulus $S(t)$ is applied from time $t=t_0$, and that conditions of existence and uniqueness of a  chain with complete connection $\mu$ are fulfilled in the presence of the stimulus (an example is given in the next section). We note $n_0 = \ent{t_0}$. For times anterior to $n_0$, $\mu$ identifies with $\musp$, that is, for any $m < n \leq n_0$, for any block $\bloc{m}{n}$, $\moy{\bloc{m}{n}}=\moysp{\bloc{m}{n}}$. In contrast, for $n>n_0$ spike statistics is modified. Consider a range-$R$ observable $f(t,\omega)$ then $\Expm{\mu}{f(t,.)} \deq \Expm{\musp}{f(t,.)} + \delta \bra{f(t,.)}$ where $\delta \bra{f(t,.)} =0$ for $t<t_0$ and $\delta \bra{f(t,.)} \neq 0$ for $t \geq t_0$. 

The goal is to establish an explicit (formal) equation for $\delta \bra{f(t,.)}$, as a function of the stimulus. This is done via a Volterra-like expansion in powers of the stimulus, cut to the first order so as to obtain a linear response in terms of a convolution between the stimulus and a convolution kernel $K_f$, depending on $f$, $\dmu{f(t,.)} = \bra{K_f \ast S}(t)$. This way, we obtain a relationship between the proportionality coefficient $K_f$ in the linear response, and specific correlations functions computed at equilibrium (spontaneous activity). This provides a Kubo relation holding in the case of neuronal networks with unbounded memory and for an arbitrary range-$R$ observable $f$. In contrast to Volterra expansion, our formalism allows to explicit the dependence of $K_f$ in the neuronal network characteristics (parameters fixing the individual neurons dynamics and connectivity-synaptic weights). An example is provided in the next section.

\ssu{First order expansion} \label{Sec:FirstOrder}

We assume that the statistics of spikes is described by time-dependent chains with unbounded memory, with potential $\phi(n,\omega)$. We note $\delta \phi(n,\omega)= \phi(n,\omega)-\phisp{\omega}$. We define likewise $\Phi(n,\omega)= \Phisp{\omega} + \delta \Phi(n,\omega)$ using \eqref{eq:pCondphi}. 

From the definition \eqref{eq:phi_def}, $e^{\phi^{(sp)}}$ corresponds to the family of transition probabilities $\Set{\probcsp}$ defining the Gibbs distribution $\musp$ in the spontaneous regime, whereas $e^{\phi}$ corresponds to the family of transition probabilities $\Set{\probc}$ defining the Gibbs distribution $\mu$ in time-dependent stimuli-evoked regime.
For $n  > n_0$, we have:
\begin{equation}\label{eq:ExpansionExpPhi}
e^{ \Phi\pare{n_0+1,n,\omega}}   = e^{\Phi^{(sp)}\pare{n_0+1,n,\omega} + \delta  \Phi \pare{n_0+1,n,\omega}} = e^{\Phi^{(sp)}\pare{n_0+1,n,\omega} } \,
\bra{1+ \sum_{p=1}^{+\infty} \frac{\delta \Phi(n_0+1,n,\omega)^p}{p!}},
\end{equation}
which, from equation (\ref{eq:pCondphi}), gives: 
$$
\Probc{\bloc{n_0+1}{n}}{\sif{n_0}} =\Probcsp{\bloc{n_0+1}{n}}{\sif{n_0}} \,
\bra{1+ \sum_{p=1}^{+\infty} \frac{\delta \Phi(n_0+1,n,\omega)^p}{p!}}.
$$

Taking the first order approximation of the exponential, we obtain:
$$
e^{ \Phi\pare{n_0+1,n,\omega}}  \sim e^{\Phi^{(sp)}\pare{n_0+1,n,\omega} }\,
\bra{1 \, + \, \delta \Phi(n_0+1,n,\omega)}.
$$

However, while $\Phi\pare{n_0+1,n,\omega}$ and $\Phi^{(sp)}\pare{n_0+1,n,\omega} $ are normalized potentials, i.e., the log of a conditional probability, the first order approximation of $e^{\Phi^{(sp)}\pare{n_0+1,n,\omega} }\,
\bra{1 \, + \, \delta \Phi(n_0+1,n,\omega)}$ is not. Normalization is obtained formally by introducing  the partition function:
$$
Z\bra{\sif{n_0}}  =\sum_{\bloc{n_0+1}{n}} e^{\Phisp{n_0+1,n,\omega}} [1 + \delta \Phi(n_0+1,n,\omega)],
$$
constrained by the past sequence $\sif{n_0}$, so that the quantity
$$
\Probca{\bloc{n_0+1}{n}}{\sif{n_0}}  \equiv  \frac{e^{\Phi^{(sp)}\pare{n_0+1,n,\omega}}}{Z\bra{\sif{n_0}}} \,
\bra{1 \, + \, \delta \Phi(n_0+1,n,\omega)},
$$
is the first order approximation of $\Probc{\bloc{n_0+1}{n}}{\sif{n_0}}$.

Setting:
$$
\Zsp{\sif{n_0}} = \sum_{\bloc{n_0+1}{n}} e^{\Phisp{n_0+1,n,\omega}},
$$
we have, to first order:
$$
\frac{1}{Z\bra{\sif{n_0}}}  = \frac{1}{\Zsp{\sif{n_0}}}  \, \bra{1 \, - \, \sum_{\bloc{n_0+1}{n}}  \, \frac{e^{\Phisp{n_0+1,n,\omega}} }{\Zsp{\sif{n_0}}} \, \delta \Phi(n_0+1,n,\omega)}.
$$

But, as $\Phi^{(sp)}$ is the log of a conditional probability, $\Zsp{\sif{n_0}}=1$. So,
finally, we obtain, to first order:
\begin{equation*}
\Probca{\bloc{n_0+1}{n}}{\sif{n_0}}  \sim \Probcsp{\bloc{n_0+1}{n}}{\sif{n_0}} \,
\bra{1 \, + \, \delta \Phi(n_0+1,n,\omega) \,- \, \Expcsp{\delta \Phi(n_0+1,n,.)}{\sif{n_0}}},
\end{equation*}
where $\Expsp{}$ denotes the expectation with respect to $\musp$. We use $\Expsp{}$ instead of $\Expm{\mu^{(sp)}}{}$ to alleviate notations.

\ssu{Time dependent average of an observable}\label{Sec:TimeDepAv}

We now consider a time-dependent observable $f$ with finite range $R \equiv R_f$.
We assume $t-t_0 > R_f$.
We set $R_f=D_f+1$. Setting $n = \ent{t}$ we note $\Expm{n}{f(t,.)} = \int f(t,\omega) \mu(d\omega)$.
Here, a note of explanation is necessary. Functions $f(t,\omega)$ are random functions, where the randomness comes from $\omega$. So, the law of  $f(t,\omega)$ is determined by the probability $\mu$.
$\Expm{n}{f(t,.)}$ is the average of the continuous time dependent observable $f(t,\omega)$, averaged over the discrete time spike train $\omega$,
up to the discrete time $n = \ent{t}$ (by definition $f(t,.)$ does not depend on spike events occurring at times posterior to $n$). Note that this average cannot be defined by an ergodic time average procedure as, here, the probability is non stationary (see section \ref{sec:Numerics} for a numerical implementation).

Because $f$ has finite range $R_f$ we may write:
$$
\Expm{n}{f(t,.)} = 
\sum_{\bloc{n-D_f}{n}}  
f\pare{t,\bloc{n-D_f}{n}} \mu[\bloc{n-D_f}{n}]
= \sum_{\bloc{n_0+1}{n}} f\pare{t,\bloc{n-D_f}{n}} \mu[\bloc{n_0+1}{n}].
$$
\noindent
The last equality holds because $f\pare{t,\omega}$ is independent of $\sif{n_0}$. Thus, using \eqref{eq:MuBlockChain}:
\begin{eqnarray*} 
\Expm{n}{f(t,.)} &=& \sum_{\bloc{n_0+1}{n}}  
f\pare{t,\bloc{n-D_f}{n}} \, \int_{\seq{\cA}{-\infty}{n_0}}  \,  \Probc{\bloc{n_0+1}{n}}{\sif{n_0}} \,  \mu(d \omega)\\
  &=&  \sum_{\bloc{n_0+1}{n}}  
f\pare{t,\bloc{n-D_f}{n}} \, \int_{\seq{\cA}{-\infty}{n_0}}  \,  \Probc{\bloc{n_0+1}{n}}{\sif{n_0}} \,  \musp(d \omega),
\end{eqnarray*} 
where the last equation holds because, on $\mathcal{F}_{\leq n_0}$, $\mu=\musp$.\\

\noindent
Thus, replacing $\Probc{\bloc{n_0+1}{n}}{\sif{n_0}}$ by $\Probca{\bloc{n_0+1}{n}}{\sif{n_0}}$, we obtain, up to first order:

$$
\Expm{n}{f(t,.)} \sim \sum_{\bloc{n_0+1}{n}} f\pare{t,\bloc{n-D_f}{n}} \, \int_{\seq{\cA}{-\infty}{n_0}}  \,  \Probcsp{\bloc{n_0+1}{n}}{\sif{n_0}} \, \musp(d \omega)
$$
$$
+\sum_{\bloc{n_0+1}{n}} f\pare{t,\bloc{n-D_f}{n}} \, \int_{\seq{\cA}{-\infty}{n_0}}  \,  \Probcsp{\bloc{n_0+1}{n}}{\sif{n_0}} \,\delta \Phi(n_0+1,n,\omega) \, \musp(d \omega)
$$
$$
-\sum_{\bloc{n_0+1}{n}} f\pare{t,\bloc{n-D_f}{n}} \, \int_{\seq{\cA}{-\infty}{n_0}}  \,  \Probcsp{\bloc{n_0+1}{n}}{\sif{n_0}} \,
 \Expcsp{\delta \Phi(n_0+1,n,.)}{\sif{n_0}} \, \musp(d \omega).
$$

The first term is $\Expsp{f(t,.)}$ from \eqref{eq:MuBlockChain}. The second term is:
$$
\sum_{\bloc{n_0+1}{n}}  \int_{\seq{\cA}{-\infty}{n_0}}  \,f\pare{t,\omega} \,\,\delta \Phi(n_0+1,n,\omega) \,  \Probc{\bloc{n_0+1}{n}}{\sif{n_0}} \,   \musp(d \omega) = \Expsp{f(t,.) \, \delta \Phi(n_0+1,n,.)}$$  
from the consistency property \eqref{eq:def_consistent}, and because by assumption ($n-n_0 > R_f$), $f\pare{t,\omega}$ does not depend on $\sif{n_0}$.\\

\noindent
For the third term:
 $$
 \sum_{\bloc{n_0+1}{n}}  \, f\pare{t,\bloc{n-D_f}{n}}  \, \int_{\seq{\cA}{-\infty}{n_0}}  \Expcsp{\delta \Phi(n_0+1,n,\omega)}{\sif{n_0}} \,  \Probcsp{\bloc{n_0+1}{n}}{\sif{n_0}} \,
    \, \musp(d \omega)
 $$
 $$
 =
 \Expsp{f(t,.) \, \Expcsp{\delta \Phi(n_0+1,n,\omega)}{\sif{n_0}}}.
$$
But, by assumption $f(t,\bloc{n-D_f}{n})$ does not depend on $\sif{n_0}$ ($n-n_0>D_f$), whereas
by definition of the conditional expectation $\Expcsp{\delta \Phi(n_0+1,n,\omega)}{\sif{n_0}}$ is the projection on the sigma-algebra $\mathcal{F}_{\leq n_0}$.
As a consequence, we have:

\begin{eqnarray*} 
\Expsp{f(t,.) \, \Expcsp{\delta \Phi(n_0+1,n,.)}{\sif{n_0}}}   &=& 
\Expsp{f(t,.)} \, \Expsp{\Expcsp{\delta \Phi(n_0+1,n,.)}{\sif{n_0}}}\\
  &=&  \Expsp{f(t,.)} \, \Expsp{\delta \Phi(n_0+1,n,.)}.
\end{eqnarray*} 

\noindent

Summing up, we have, using \eqref{eq:pCondphi}:
$$\Expm{n}{f(t,.)}
%
=\Expsp{f(t,.)}+\sum_{r=n_0+1}^{n=\ent{t}} \pare{\Expsp{f(t,.) \, \delta \phi(r,.)}-\Expsp{f(t,.)} \, \Expsp{\delta \phi(r,.)}},
$$
Using the correlation function:
$$\Csp{f(t,.)}{g(t',.)} \deq \Expsp{f(t,.)g(t',.)} - \Expsp{f(t,.)}\Expsp{g(t',.)}$$
we obtain 

\begin{equation}\label{eq:dmu}
\dmu{f(t,.)} 
  = \sum_{r=n_0+1}^{n=\ent{t}} \Csp{f(t,.)}{\delta \phi(r,.)}.
\end{equation} 

This equation expresses that the time-dependent variation in the average of an observable $f$ is expressed, to the first order, as a time series of
correlation functions, between $f$ and the time-dependent variation of the normalized potential, computed with respect to the equilibrium distribution. This is our main result. 

It is similar to the \textit{fluctuation-dissipation theorem} in statistical physics \cite{kubo:57,ruelle:99}. Here, it holds for Gibbs distributions with infinite range potential $\phi(t,\omega)$. A crucial point is the convergence of the series when the initial time of perturbation, $n_0$ tends to $-\infty$. This holds if correlations $\Csp{f(t,.)}{\delta \phi(r,.)}$ decay sufficiently fast, typically, exponentially. We come back to this point in the next section.

One of the advantages of this relationship is that averages are taken with respect to $\musp$. In the case of experimental data, these averages can be approximated by \textit{empirical averages} on spontaneous activity. Still, $\delta \phi$ is unknown. We now show how to get rid of this constraint. 

\ssu{Monomials decomposition}\label{Sec:MonDecomp}

\sssu{Linear response in the finite-range potential approximation} \label{Sec:LinRepFiniteRange}

Although  $\phi$, $\phis$ have infinite range, their memory dependence can decay fast, typically exponentially. This property is independent of the application of a stimulus: it holds for $\phi$ and $\phis$ as well, hence for $\delta \phi$. In this case, the infinite range potentials can be approximated by a finite-range one. We use here a classical result in ergodic theory: a potential $\phi$  with an exponentially decaying variation \eqref{eq:Variation} (more generally, so-called regular potential) can also be approximated by a finite-range potential $\phi^{(R)}$ in the sup norm where $  \|\phi-\phi^{(R)}\| \leq  C \Theta^R$, for some $0 < \Theta < 1$ \cite{ledrappier:74}. Equivalently the chain with infinite memory can be replaced by a Markov chain of memory depth $D$, $D=R+1$. Therefore, $\delta \phi$ can be approximated by a range-$R$ potential $\delta \phi^{(R)}$. Using the monomial decomposition \eqref{eq:monomial_decomp}, we have:

$$
\delta \phi(r,\omega) \sim \sum_{l'} \delta \phi_l(r) m_l(\bloc{r-D}{r}).
$$

In this setting, \eqref{eq:dmu} becomes: 
$$
\dmu{f(t)}=\sum_{r=n_0+1}^{n=\ent{t}} \sum_{l,l'} f_l(t) \, \delta \phi_{l'}(r) \,  \Csp{m_l(\bloc{n-D}{n})}{m_{l'}(\bloc{r-D}{r})}
.$$
But, $\musp$ is stationary by assumption. Hence the correlation $\Csp{m_l(\bloc{n-D}{n})}{m_{l'}(\bloc{r-D}{r})}$ only depends on the two monomials $m_l,m_{l'}$ and on the time lag $n-r$, i.e. $\Csp{m_l(\bloc{n-D}{n})}{m_{l'}(\bloc{r-D}{r})}=\Csp{m_l \circ \sigma^{n-r}}{m_{l'}}$, that we write $C_{l,l'}(n-r)$ to alleviate notations.
Thus:
$$
\dmu{f(t)}  \, = \, \sum_{r=-\infty}^{n=\ent{t}} \sum_{l,l'} f_l(t)  \,  C_{l,l'}(n-r) \, \delta \phi_{l'}(r),
$$
so that the linear response is a linear decomposition of monomial correlations computed \textit{at equilibrium}. We can write this in a more compact form introducing the $L$-dimension vectors $\bf(t)=
\vect{f_l(t)}_{l=1}^L$, $\bm{\delta \phi}(r) =
\vect{\delta \phi_l(r)}_{l=1}^L$ and the matrix $\cC(n-r)=\vect{C_{l,l'}(n-r)}_{l,l'=1}^L$. We write $\dmu{f(t)}$ in the form: \\
\beq\label{eq:LinRepCorrMonom}
\dmu{f(t)}  \, = \, \sum_{r=-\infty}^{n=\ent{t}} \scal{\bf(t)}{\cC(n-r) \cdot \bm{\delta \phi}(r)},
\eeq
where  $\scal{\,}{\,}$ denotes the standard scalar product.

This has three important consequences.

\sssu{Convergence of the linear response series.}
Adapting the notations it follows from section \ref{Sec:CorrDecay}, eq \eqref{eq:CorrDecay} that:
$$
C_{l,l'}(n-r) = \sum_{k=2}^{L}  \lambda_k^{n-r}  \Gamma_{k,l,l'},
$$
where $\lambda_k$ are the eigenvalues of the matrix $\cL_\phi$, introduced in section \ref{Sec:CorrDecay}. The coefficient $\Gamma_{k,l,l'}$ depends on the projection of $m_l$, $m_{l'}$ on the eigenvector $k$.  It follows from the exponential correlation decay that the series $\sum_{r=-\infty}^{n=\ent{t}} C_{l,l'}(n-r)$ converges, with an exponential rate controlled by the second eigenvalue (spectral gap). This ensures the convergence of the linear response equation \eqref{eq:LinRepCorrMonom}. In addition, this series can be truncated keeping only a time lag $n-r$ of the order of $-\log \abs{\lambda_2}$, or considering the projection of monomials on the highest eigenvectors.
Also, the linear response equation \eqref{eq:LinRepCorrMonom} involves monomials $m_l$, $m_{l'}$ whose probability decreases very fast with their degree. Consequently, one may truncate the sums on $l,l'$ to low degree monomials (e.g. $1$ and $2$-pairwise terms).
Note that the correlation of two monomials of degree two involves their product, which is already a monomial of degree four. These monomials have a low probability in general. 

\sssu{Resonances} \label{Sec:resonances}
Denote $\Pi$ the matrix which diagonalize the matrix $\cL_\phi$. For simplicity we consider here a time-independent observable $f(\omega)$.   We note $\baF=\Pi^{-1} \bf$. 
Assume that the vector $\delta \phi(r) \equiv \vect{\delta \phi_l(r)}_{l=1}^L$ is harmonic , i.e. it has the form $\delta \phi_l(r) = A_l e^{i  2 \pi \nu r}$, where $\nu$ is a real frequency. This happens for example when the stimulus itself is harmonic and weak enough so that we can compute $\delta \phi$ from a first order expansion of $\phisp{\omega}$ (see section \ref{Sec:LinRepgIF} for an example). Dealing with linear response in the stimulus one can take for $A_l$
a complex function (where the response $\dmu{a+ib}=\dmu{a} \, + \, i \, \dmu{b}$). 
It follows from \eqref{eq:LinRepCorrMonom} that:
\begin{eqnarray*}
\dmu{f(n)}  \, &=& \, \sum_{r=-\infty}^{n=\ent{t}} \scal{\bf}{\cC(n-r)\cdot\bm{\delta \phi}(r)} \\
 &=&  \sum_{i=2}^L  \, \baF_i(t) \, \sum_{r=-\infty}^{n=\ent{t}}   \lambda_k^{n-r} \,    \sum_{j=2}^L A_j e^{ \i 2 \pi \nu r} \, \Pi_{ij}\\
&=& e^{\i 2 \pi \nu n}\,  \sum_{i,j=2}  \, \baF_i \, \frac{\Pi_{ij}}{1-\lambda_i e^{-\i 2 \pi \nu}}   \, A_j 
\end{eqnarray*}

Thus, the linear response to an harmonic perturbation is harmonic. In particular, the response exhibits
resonance in the complex frequency domain for $\lambda_i e^{-\i 2 \pi \nu} = 1$. Therefore, the resonance observed in the dynamics is directly related to the eigenvalues of $\cL_\phi$ (in dynamical systems, these resonances are called Ruelle-Pollicott resonances \cite{gaspard:98}.

\sssu{The convolution kernel}\label{Sec:LinRepConv} 
For a time independent observable $f(\omega)$ we introduce $K_f(m)$ the $L \times L$ matrix with entries:
\begin{equation*}
K_{f;l,l'}(m) = f_l\,  \cC_{l,l'}(m),
\end{equation*}
so that:
\begin{equation*}
\dmu{f(t)}  \, = \, \sum_{r=-\infty}^{n=\ent{t}} \sum_{l,l'} \, K_{f;l,l'}(n-r) \delta \phi_{l'}(r),
 \end{equation*}
which is a discrete convolution. It is close to the form \eqref{eq:LinearResponseRate} with the difference that the time is discrete, coming from our spike trains discretization. The stimulus, explicit in \eqref{eq:LinearResponseRate} is here hidden in $\delta \phi(r)$. We give an illustration of this in the next section. But the fundamental result is that the convolution kernel defined this way is a linear combination of monomial correlations functions. This has, therefore, the form of a Kubo equation where the monomials play the role of the physical quantities introduced in section \ref{Sec:LinRepStatPhys}. As mentioned above, the main and  essential difference is that, in contrast to Physics, we have no a priori idea which monomials are the most important (except straightforward arguments on the decaying probability of high order monomials). \textit{There is no known principle to guide the choice.}

\section{An example: Linear response in a conductance based Integrate and Fire model} \label{Sec:LinRepgIF}
    
As an example of application of our results, we consider here the so-called generalized Integrate-and-Fire (gIF) model 
introduced by M. Rudolph and A. Destexhe \cite{rudolph-destexhe:06} and analyzed in \cite{cessac-vieville:08,cessac:11b,cofre-cessac:13}. We consider this model, despite its complex dynamics depending on spike history, because it allows an analytic treatment giving access to the collective spike statistics. Additionally, the potential $\phi$ can be computed. This is a key step toward applying the linear response theory introduced in the previous section and understanding its consequences.

The gIF model has a rather complex dynamics where the conductances and current depend on the spike history. So, for the sake of clarity, we introduce it through the leaky Integrate-and-Fire model. We then briefly recall the main results on gIF necessary to apply our formalism, up to the characterization of transition probabilities (section \ref{Sec:TransProbfgIF}). Note therefore that the material up to section \ref{Sec:TransProbfgIF} has been published elsewhere, \cite{cessac-vieville:08,cessac:11b,cofre-cessac:13}.

\subsection{Leaky Integrate-and-Fire model and spike events} \label{Sec:LIF}

We consider a neuron $k$ (reducing to a point), with membrane potential $V_k$, membrane capacity $C_k$, resistance $R$, submitted to a stimulus (current) $S_k(t)$. We call $\theta$ the spiking threshold. We define the sub-threshold dynamics as:

\beq\label{eq:LIF}
C_k \, \frac{dV_k}{dt}+\, \frac{1}{R} V_k=S_k(t), \quad \mbox{if  } V_k(t)< \theta.
\eeq
If there is a time $t_k$ such that the membrane potential of neuron $k$ reaches the firing threshold, $V_k(t_k) \geq \theta$, the neuron $k$ fires an action potential, i.e., it emits a spike and the membrane potential of neuron $k$ is reset to a fixed reset value $V_{res}$ instantaneously. Without loss of generality we set $V_{res}=0$.  The neuron's membrane potential remains at this value during a time denoted by $\Delta$ called ``refractory period'', i.e., $V_k(t')=V_{res}, \, t'\in [t_k,t_k +\Delta]$. Equation \eqref{eq:LIF}, with the reset condition, defines the LIF model first introduced by Lapique in 1907 \cite{lapicque:07}. %

\begin{figure}[h!]
  \centering
    \includegraphics[width=0.6\textwidth]{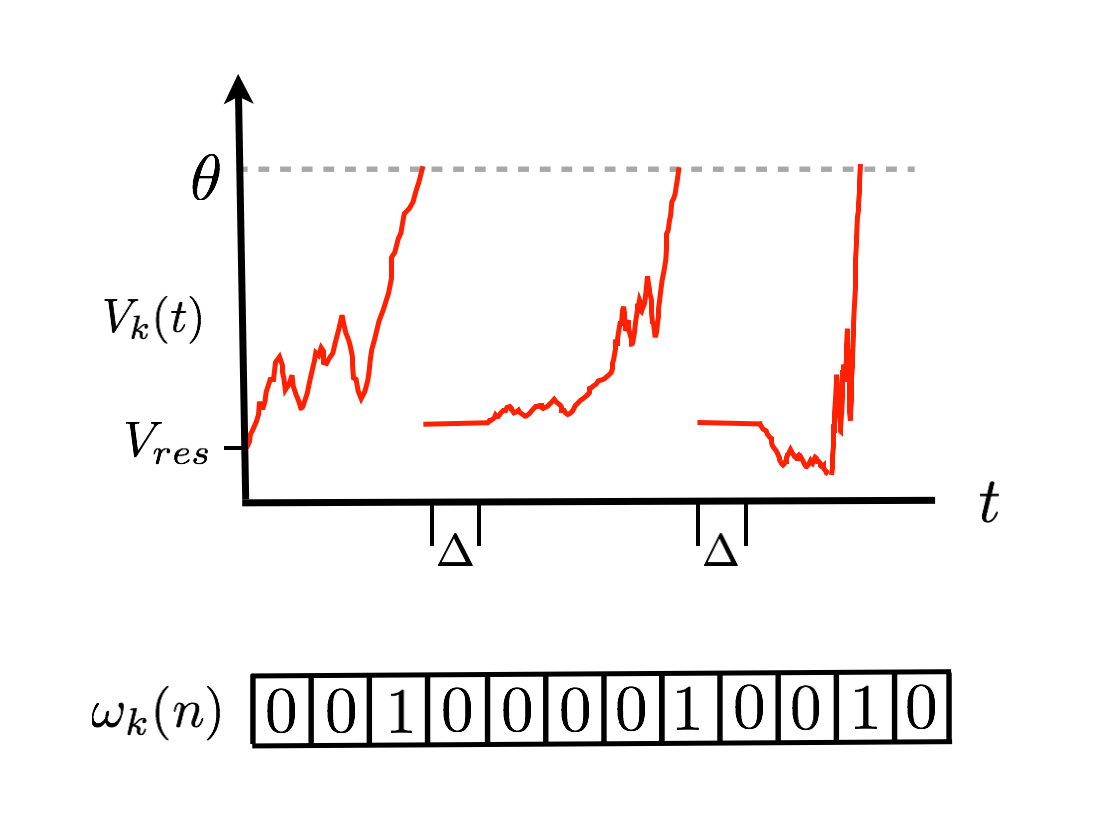}
    \caption{A sample of the time trajectory of the membrane potential of an IF neuron $k$ is plotted in continuous time. When the membrane potential reaches a fixed threshold $\theta$, it is reset to a fixed value $V_{res}$ and a spike is recorded in discrete time $\omega_k(n)=1$, otherwise $\omega_k(n)=0$. After the reset, the membrane potential remains reset for a period of time corresponding to the refractory period $\Delta$. }
    \label{Fig:LIF_Spikes}
\end{figure}

In the LIF, spikes are instantaneous, with no duration. So, if we denote $t_k^{(r)}$, the $r$-th spike emitted by neuron $k$, $t_k^{(r)}$ is a continuous variable. In agreement with section \ref{Sec:SpikesObs} we now assume that the spike train is binned with a time scale $\delta$, and we define a spike variable $\omega_{k}(n)$ so that $\omega_k(n)=1$ if $t_k^{(r)} \in [(n-1) \delta, n\delta[$, otherwise  $\omega_k(n)=0$. 
This is summarized in Fig. \ref{Fig:LIF_Spikes}.

\subsection{The gIF Conductance Based Model}\label{Sec:gIF}

In the gIF model the synaptic conductance $g_{kj}$ between the pre-synaptic neuron $j$ and the post-synaptic neuron $k$ depends on spike history as $g_{kj} (t) = G_{kj} \sum_{r \geq 0} \alpha_{kj} (t - t^{(r)}_j)$ where the sum holds on pre-synaptic spike-times  $t^{(r)}_j<t$. We use the convention $t^{(0)}_j= -\infty$. Here, $G_{kj} \geq 0$ is the maximal conductance between $j$ and $k$. It is zero when there is no synaptic connection between neurons $j$ and $k$.  $\alpha_{kj}$ is the so-called $\alpha$-profile that mimics the curse of a post-synaptic potential after a pre-synaptic spike \cite{rudolph-destexhe:06}. This function somewhat summarizes the complex dynamical process underlying the generation of a post-synaptic potential after the emission of a pre-synaptic spike. We may take:
\beq\label{eq:alpha}
\alpha_{kj}(t) = e^{-\frac{ t}{\tau_{kj}}} H(t),
\eeq
where $H(t)$ is the Heaviside function. Note we could also take $\alpha_{kj} (t) = P(t) e^{-\frac{ t}{\tau_{kj}}} H(t)$
where $P(t)$ is a polynomial in time. What indeed matters on mathematical grounds is the exponential tail of $\alpha_{kj}(t)$ \cite{cessac:11b}.

Using the spike binning where $t^{(r)}_j$ is replaced by $n \delta$ for some $n$, and setting $\delta=1$ we approximate $\alpha_{kj} (t - t^{(r)}_j)$ by $\alpha_{kj} (t - n)\omega_j(n) $ and  the conductance by :

\begin{equation}\label{eq:gkj}
g_{kj}(t,\omega) = G_{kj} \, \alpha_{kj}(t,\omega) ,
\end{equation}
where: 

\begin{equation}\label{eq:def_alpha_t_omega}
\alpha_{kj}(t,\omega) =\sum_{n=-\infty}^{\ent{t}} \alpha_{kj} (t - n) \, \omega_j(n).
\end{equation}

We recall that the notation $g_{kj}(t,\omega)$ means that function $g_{kj}$
depends on spikes occurring before time $t$.

Now, the LIF dynamics \eqref{eq:LIF} is generalized as \cite{cessac-vieville:08,cessac:11b,cofre-cessac:13}:
$$
C_k \frac{dV_k}{dt} +g_L(V_k-E_{L}) + \sum_j g_{kj}(t,\omega)(V_k-E_j)= S_k(t) + \sigma_B \xi_k (t), \quad \mbox{if  } V_k(t)< \theta,
$$
where $g_L, E_L$ are respectively the leak conductance and the leak reversal potential, $E_j$ the reversal potential characterizing the synaptic transmission between $j$ and $k$ \cite{destexhe:98}. Finally, $\xi_k (t)$ is a white noise, introducing stochasticity in dynamics. Its intensity is $\sigma_B$.

Setting:
\begin{equation*}
W_{kj} =G_{kj} E_j, 
\end{equation*}
\begin{equation*}
i_k(t,\omega) = g_L E_L + \sum_j W_{kj} \alpha_{kj}(t,\omega) + S_k(t) + \sigma_B \xi_k (t),
\end{equation*}
and:
\begin{equation*}
g_{k} (t,\omega) = g_L + \sum_{j=1}^N g_{kj}(t,\omega),
\end{equation*}
it is more convenient to write the gIF dynamics in the form:
\beq\label{eq:DyngIF}
C_k \, \frac{dV_k}{dt}+\gk{t,\omega} V_k=i_k(t,\omega), \quad \mbox{if  } V_k(t)< \theta,
\eeq
where $i_k(t,\omega)$ depends on the network spike history via $\alpha_{kj}(t,\omega)$, and contains a stochastic term.
As the reversal potential $E_j$ can be positive or negative, the 
synaptic weights $W_{kj}$ define an oriented and signed graph, whose vertices are the neurons.

\subsection{Solution of the sub-threshold equation} \label{Sec:SolgIF}

As in this model the conductances do not depend explicitly on the membrane potential $V_k$, one can explicitly integrate the sub-threshold dynamics. For fixed $\omega$, equation  \eqref{eq:DyngIF} is a linear equation in $V_k$ with flow $e^{-\frac{1}{C_k}\int_{t_1}^{t_2}\gk{u,\omega} \, du}$. This flow characterizes the membrane potential evolution of neuron $k$ below threshold, i.e., when
neuron $k$ does not spike in the time interval $[t_1,t_2]$. One can easily extend the definition to a flow including reset \cite{cofre-cessac:13} by denoting $\tko$ the last time before $t$ when neuron $k$ fired in the spike-train $\omega$:
\beq\label{eq:fgm}
\Gamma_k(t_1,t,\omega)=
\left\{
\baR{lll}
e^{-\frac{1}{C_k}\int_{t_1}^{t}\gk{u,\omega} \, du}, \, &\mbox{if}& t \geq  t_1  \geq \tko;\\
0, &\mbox{otherwise} \,  .
\eaR
\right.
\eeq

We then integrate the flow to obtain the voltage at time $t$, given the spike history $\omega$. We can split the solution of (\ref{eq:DyngIF}) as follows:
\beq\label{eq:Vk}
\Vk{t,\omega}= \Vkspont{t,\omega} +  \Vkext{t,\omega}+\Vknoise{t,\omega}.
\eeq
where the first term on the r.h.s corresponds to the ``spontaneous'' dynamics, which is independent of the external stimulus perturbation and noise. The spontaneous contribution can be divided again into a part $\Vksyn{t,\omega}$ corresponding to network effects (pre-synaptic neurons influence) and a part corresponding to the leak, $\VkL{t,\omega}$:
\beq\label{eq:Vkspont}
\Vkspont{t,\omega}  =  \Vksyn{t,\omega} +\VkL{t,\omega},
\eeq
with:
\beq\label{eq:Vksyn}
\Vksyn{t,\omega} \, = \, 
 \frac{1}{C_k} \sum_{j=1}^N   W_{kj} \, \int_{\tko}^{t} \Gamma_k(t_1,t,\omega)  \akj{t_1,\omega} dt_1,
\eeq
and:
$$
\VkL{t,\omega} \, = \,
\frac{E_L}{\tLk} \int_{\tko}^{t} \Gamma_k(t_1,t,\omega)   dt_1,
$$
where:
$$
\tLk \deq \frac{C_k}{g_L}.
$$
The second term in \eqref{eq:Vk} corresponds to the contribution due to the external stimulus:
\beq\label{eq:Vkext}
\Vkext{t,\omega} \, = \,
 \frac{1}{C_k} \, \int_{\tko}^{t} S_k(t_1) \Gamma_k(t_1,t,\omega)  dt_1.
\eeq
The last term is the stochastic part of the membrane potential:
$$
\Vknoise{t,\omega} = \frac{\sigma_B}{C_k} \int_{\tko}^{t}  \Gamma_k(t_1,t,\omega)  dB_k(t_1),
$$
where $B_k(t_1)$ is a Brownian process, thus Gaussian. 

\paragraph{Remarks.} 

\begin{itemize}
\item The voltage of the gIF model depends on the history in two ways. (i) The time integrals start at the time $\tko$, the last time in the past anterior to $t$ where neuron $k$ spiked. This time is unbounded and can go arbitrarily far in the past. (ii) The second history dependence is in the flow $\Gamma_k(t_1,t,\omega)$ constrained by the spike train history $\omega$ via the conductance $\gk{u,\omega}$. While the voltage is reset when a neuron spikes, the history dependence of the conductance is not. Thus, the memory of the gIF model is infinite. However, thanks to the exponential decay of the synaptic response \eqref{eq:alpha}, the memory dependence decays exponentially fast, ensuring the exponential continuity of transition probabilities, necessary to ensure the existence and uniqueness of a chain with unbounded memory \cite{cessac:11b,cofre-cessac:13}.
\item To our best knowledge, the literature on linear response in Integrate-and-Fire neurons relies on mean-field assumptions, in which previous spikes are averaged over. Here, in contrast, we do not average previous spikes, but condition on them. Nevertheless, a mean-field approximation can be used here as well, as developed in section \ref{Sec:Mean-Field}.
\end{itemize} 

\ssu{Transition Probabilities of the gIF Model} \label{Sec:TransProbfgIF}

The gIF model allows one to approximate, in the limit of small $\sigma_B$, the family of transition probabilities $\Pnc{\omega(n)}{\sif{n-1}}$
explicitly in terms of the parameters of the spiking neuronal network model. 
One can show that the gIF model as presented here\footnote{A more complete version of this model also includes electric synapses \cite{cofre-cessac:13}. In that case, the conditional independence is lost.} is conditionally independent i.e., it factorizes over neurons once the spike history has been fixed \cite{cessac:11b}.

%
$$\Pnc{\omega(n)}{\sif{n-1}} = \prod_{k=1}^N
\Pnc{\omega_k(n)}{\sif{n-1}},$$
%
where:
\beq\label{eq:PMarg}
\Pnc{\omega_k(n)}{\sif{n-1}}=
\omega_{k}(n) \,  \Pi\pare{\Xk{n-1,\omega}} \, + \,
\pare{1-\omega_{k}(n)} \,
\pare{1-\Pi\pare{\Xk{n-1,\omega}}},
\eeq
with:
$$
\Pi(x)=\frac{1}{\sqrt{2\pi}}\int_x^{+\infty} e^{-\frac{u^2}{2}}du,
$$
and:
\beq\label{eq:Xk}
\Xk{n-1,\omega} \deq \frac{\theta-\Vk{n-1,\omega}}{\sk{n-1,\omega}},
\eeq%
where,
$$
\skd{n-1,\omega}
= \left(\frac{\sigma_B}{C_k}\right)^2 \,
\int_{\tau_k(n-1,\omega)}^{n-1} \Gamma^2_k(t_1,n-1,\omega)  \, dt_1,
$$
corresponding to the  variance of the noise integrated along the flow up to time $n-1$.

In equation \eqref{eq:PMarg} there are two terms. The first term, corresponds to $\omega_{k}(n)=1$ which requires that $\Vk{t,\omega} >  \theta$ for some $t \in [(n-1)\delta,n \delta[$ (i.e. $\Vknoise{t,\omega} > \theta -\Vkspont{t,\omega} - \Vkext{t,\omega}$). We have assumed that the binning time $\delta$ is small enough that one can replace $t$ by $(n-1)\delta$ in the computation of the spiking condition which becomes $\Vknoise{n-1,\omega} > \theta -\Vkspont{n-1,\omega} - \Vkext{n-1,\omega}$.
Likewise, the second term corresponds to $\omega_{k}(n)=0$. \\

From (\ref{eq:phi_def}), we obtain the normalized potential for the gIF model.

$$\phi\pare{n,\omega} = \sum_{k=1}^N \phi_k\pare{n,\omega},$$
\noindent
where

\beq\label{eq:Phi_GIF}
\phi_k\pare{n,\omega} =
\omega_{k}(n) \, \log \Pi\pare{\Xk{n-1,\omega}} \, + \,
\pare{1-\omega_{k}(n)} \,
\log\pare{1-\Pi\pare{\Xk{n-1,\omega}}},
\eeq
\\
which depends on all of the parameters of the network via the variable $\Xk{n-1,\omega}$ (\ref{eq:Xk}). \\


\textbf{Remark.} The function $\Pi(x)$ is a sigmoid which tends to $1$ when $x \rightarrow -\infty$ and   tends to $ 0$ when $x \rightarrow \infty$. This has two consequences:
\begin{enumerate}
\item When $\Xk{n-1,\omega} \to +\infty$ (which arises when $\Vkspont{n-1,\omega} \to -\infty$) $, \Pi\pare{\Xk{n-1,\omega}} \to 0$) so that $\phi_k\pare{n,\omega} \to -\infty$. This expresses that the probability of having a spike at time $n$ when $\Xk{n-1,\omega}$ becomes large (neuron strongly hyper-polarized)  tends to $0$. The same argument holds \textit{mutatis mutandis} for the limit $\Xk{n-1,\omega} \to -\infty$.

\item When $\mid \Xk{n-1,\omega}\mid$ is large (neuron either strongly hyper-polarized or strongly depolarized) the effect of a variation of the membrane potential on the firing probability is negligible. Thus, we will study the effect of a perturbation in bounded range for $\Xk{n-1,\omega}$:

\begin{equation}\label{eq:bounds}
0 < \epsilon < \Pi\pare{\Xk{n-1,\omega}} < 1- \epsilon < 1,
\end{equation}
uniformly in $n,\omega$. This is ensured by natural assumptions on synaptic weights and on $\sigma_B$, the mean-square deviation of the noise, which has to be bounded away from $0$.

\end{enumerate}

\ssu{Expansion of the normalized potential} \label{Sec:ExpPot}

 The normalized potential of the gIF model  can be separated into: (i) a ``spontaneous'' part $\phi^{(sp)}(\omega)$, which before time $t_0$ is independent of the stimuli and time and; (ii) a ``perturbation'' part $\delta \phi(n,\omega)$ depending on a  time-dependent stimuli, which is non-zero from time $t_0$. 
Mathematically this is achieved by adding an extra term to the spontaneous potential after time $t_0$. 

$$\phi(n,\omega)=
\left\{
\baR{lll}
 \phi^{(sp)}(\omega) \, &\mbox{if}& n < \ent{t_0};\\
 \phi^{(sp)}(\omega) + \delta \phi(n,\omega)  \, &\mbox{if}&  n \geq \ent{t_0} \,  .
\eaR
\right.$$

Note that, at this stage, this is just a definition of $ \delta \phi(n,\omega)  = \phi(n,\omega)- \phi^{(sp)}(\omega)$.\\ 

From section \ref{Sec:LinRepInfMem} this perturbation induces a time-dependent variations on the average of an observable $f$:

$$
\mu_t[f(t,\cdot)] = \Expsp{f(t,\cdot)} + \dmu{f(t,\cdot)},
$$

If $t \leq t_0, \, \dmu{f(t,\cdot)} =0 , \forall f(t,\cdot)$ as $\mu_t=\musp$. Thus, the term $ \Expsp{f(t,\cdot)}$ refers to an average with respect to the unperturbed system and $\dmu{f(t,\cdot)}$. As we show, the variation $\delta^{(1)}$ can be explicitly written in terms of the variation on the normalized potential induced by the introduction of the stimulus.

Note that if the external stimuli is switched on at time $t_0$, spike statistics are still constrained by the previous spontaneous activity, since transition probabilities have memory. This effect is especially salient in the gIF model which has an unbounded memory.\\

We rewrite \eqref{eq:Xk} in the form:
%
$$\Xk{n-1,\omega}=\Xksp{n-1,\omega} \,+ \,\dXk{n-1,\omega},$$
%
where
\beq\label{eq:Xsp}
\Xksp{n-1,\omega}=\frac{\theta-\Vkspont{n-1,\omega}}{\sk{n-1,\omega}}
\eeq
is independent of the stimulus, and:
\begin{equation*}
\dXk{n-1,\omega}=- \frac{\Vkext{n-1,\omega}}{\sk{n-1,\omega}}=- \frac{1}{C_k \, \sk{n-1,\omega}} 
\int_{\max(t_0,\tau_k(n-1,\omega))}^{n-1} S_k(t_1) \Gamma_k(t_1,n-1,\omega)  dt_1,
\end{equation*}
where the last equality holds because $S(t_1)=0$ for $t_1 <t_0$. \\

In the next computation we write $X^{(sp)}_k, \delta X_k$ instead of $\Xksp{n-1,\omega}, \dXk{n-1,\omega}$ to alleviate notations. We make a series expansion of $\phi_k\pare{n,\omega}$
at $X^{(sp)}_k$, under the conditions of \eqref{eq:bounds}. We have:

\bea
\log\Pi(X^{(sp)}_k+\delta X_k)=
\log\Pi(X^{(sp)}_k)+\sum_{u=1}^{+\infty} \frac{a^{(u)}(X^{(sp)}_k)}{u!} \, \pare{\delta X_k}^u, \nonumber\\
\log\pare{1-\Pi(X^{(sp)}_k+\delta X_k)}=
\log\pare{1-\Pi(X^{(sp)}_k)}+\sum_{u=1}^{+\infty} \frac{b^{(u)}(X^{(sp)}_k)}{u!} \, \pare{\delta X_k}^u \nonumber
\eea
where $a^{(u)}$ and $b^{(u)}$ are the $u$-th derivative of $\log\Pi(x)$ and $\log(1-\Pi(x))$. In particular:
\begin{equation*}
a^{(1)}(x)=\frac{\Pi'(x)}{\Pi(x)};
\quad \quad b^{(1)}(x)=-\frac{\Pi'(x)}{1-\Pi(x)}.
\end{equation*}

Therefore,
\begin{equation}\label{eq:ExpansionDeltaphi}
\delta \phi_k\pare{n,\omega} = \sum_{u=1}^{+\infty} \delta \phi^{(u)}_k\pare{n,\omega},
\end{equation}
where:
$$
\delta \phi^{(u)}_k\pare{n,\omega} =
\mathcal{H}_k^{(u)}(n,\omega) \,  \bra{\delta \Xk{n-1,\omega}}^u,
$$
with:
\begin{equation}\label{eq:Hku}
\mathcal{H}_k^{(u)}(n,\omega) = \frac{1}{u!} \bra{
\omega_{k}(n) \, a^{(u)} \, \pare{\Xksp{n-1,\omega}} \, 
+
\pare{1-\omega_{k}(n)} b^{(u)} \, \bra{\Xksp{n-1,\omega}}}.
\end{equation}

This expansion holds for any value of $\Xksp{n-1,\omega}$. However, when this quantity becomes large in absolute value, one has to consider more and more terms in the expansion to approach sufficiently well the function $\delta \phi_k\pare{n,\omega}$. This is well known property of the function $\Pi$ (which can be written in terms of the error function): the Taylor expansion converges very slowly near infinity and other expansions are more efficient (e.g. B\"urmann series \cite{schoepf:14}). Here, we want to consider the effect of a perturbation in a range where the function $\Pi$ does not saturate. In addition, we want to restrict ourselves to cases where the first order of the Taylor expansion is sufficient to characterize the response. This is ensured by conditions of the form (the same holds \textit{mutatis-mutandis} for $b$):

 $$
 \abs{\delta X_k} \ll \pare{u!}^{\frac{1}{u-1}} \abs{\frac{a^{(1)}\pare{X^{(sp)}_k}}{a^{(u)}\pare{X^{(sp)}_k}}}^{\frac{1}{u-1}}; \quad u > 1.
 $$
Applied to the second order this gives a condition:

\begin{equation}\label{eq:condLinRep}
\abs{\delta \Xk{n-1,\omega}} \ll \frac{2}{\abs{\Xksp{n-1,\omega}+a^{(1)}\pare{\Xksp{n-1,\omega}}}},
\end{equation}
which becomes more and more restrictive as one gets away from $\Xksp{n-1,\omega}=0$ i.e. away from the linear region of the sigmoid $\Pi$. Note that the condition $\Xksp{n-1,\omega} \sim 0$ corresponds to a voltage close to the firing threshold.
We insist that this constraint is not a limitation of our approach, but instead, a limitation of linear response applied in neuronal systems where the response of a neuron is characterized by a saturating function. Away from the linear part of the sigmoid, nonlinear effects dominate the response.\\

Under these conditions we obtain:
%
$$\delta \phi^{(1)}_k\pare{r,\omega} = - \frac{\mathcal{H}_k^{(1)}(r,\omega) }{C_k \, \sk{r-1,\omega}}\, \int_{\max(t_0,\nkrom)}^{r-1} S_k(t_1) \Gamma_k(t_1,r-1,\omega)  dt_1,$$
where we have replaced $n$ by $r$ in view of \eqref{eq:dmu}.

\bigskip
The function $\delta \phi^{(1)}_k\pare{r,\omega} $ is the \textit{first order variation of normalized potential}
when neurons are submitted to a weak time dependent stimulus, under the approximation \eqref{eq:Phi_GIF}. There are two contributions. The integral includes the effect of the stimulus on the dynamics flow; the term  $\mathcal{H}_k^{(1)}(r,\omega)$, given by eq.  \eqref{eq:Hku}, contains the effect of the network via the terms %
$a^{(1)}\pare{\frac{\theta-\Vkspont{r-1,\omega}}{\sk{n-1,\omega}}}$, $b^{(1)}\pare{\frac{\theta-\Vkspont{r-1,\omega}}{\sk{n-1,\omega}}}$ where
$\Vkspont{r-1,\omega}$ is given by eq. \eqref{eq:Vkspont}. Note that the dependence on synaptic weights is non-linear because $a^{(1)}$ and $b^{(1)}$ are non-linear.

\ssu{Linear response for the gIF model} \label{Sec:LinRepGenObs}

From equation (\ref{eq:dmu}) we obtain the linear response of the general observable $f(t)$ to the first order:
$$
\dmu{f(t)} \, = - \,\sum_{k=1}^N \,\frac{1}{C_k} \, \sum_{r=n_0+1}^{n= \ent{t}}  \Csp{f(t,\cdot)}{  
\frac{\mathcal{H}_k^{(1)}(r,\cdot) }{\sk{r-1,\cdot}}\, \int_{\max(t_0,\nkrm)}^{r-1} S_k(t_1) \Gamma_k(t_1,r-1,\cdot)  dt_1}.
$$
where we have used the linearity of correlations functions to get the sum $\sum_{k=1}^N$ out. The dots in the functions stand for $\omega$. This is to recall that the correlations $\cC^{(sp)}$ correspond to the average of the functions of $\omega$ over the spontaneous Gibbs distribution $\musp$. Thus, as expected from linear response theory, the first variation of the average of $f(t,\omega)$ reads as a series in correlation functions computed with respect to the spontaneous dynamics. \\

Let us discuss the main parts of equation \eqref{Sec:LinRepGenObs}. It decomposes as a sum over neurons (sum over $k$ from 1 to $N$), and as a sum over $r$ the time-correlation functions. Note that correlations decay exponentially fast in this model \cite{cessac:11b}. This allows, on one hand, to truncate the sum if necessary. On the other hand, it allows us to take the limit $n_0 \to -\infty$ corresponding to start the stimulation arbitrary far in the past. In this case, $t_0 \to -\infty$ as well and we may write:
\beq\label{eq:LinRepgIFtinf}
\dmu{f(t)} \, = - \,\sum_{k=1}^N \, \frac{1}{C_k} \, \sum_{r=-\infty}^{n= \ent{t}}  \Csp{f(t,\cdot)}{  
\frac{\mathcal{H}_k^{(1)}(r,\cdot) }{\sk{r-1,\cdot}}\, \int_{\nkrm}^{r-1} S_k(t_1) \Gamma_k(t_1,r-1,\cdot)  dt_1}.
\eeq

The correlation involves the time integral of the stimulus over the flow, with a weight $\Gamma_k(t_1,r-1,\cdot) $ depending on the spike history via the conductance. 
This integral is weighted by the term $\sk{r-1,\cdot}$ integrating the influence of noise, and $\mathcal{H}_k^{(1)}(r,\cdot)$ containing the network contribution. Thus, the linear response of an arbitrary observable $f(t)$, which may include one or more neurons, is not only a function of the external stimulus, but also depends on the network connectivity and on spike history.

\ssu{Time scales} \label{Sec:TimeScales}

We now introduce two time scales important for further computations.

\begin{enumerate}[(i)]
\item \textbf{The characteristic decay time of the flow}
$\Gamma_k(t_1,r-1,\cdot)=e^{-\frac{1}{C_k}\int_{t_1}^{r-1}\gk{u,\omega}}\, du$.

We define this characteristic time, $\tau_{d,k}$, by $\frac{1}{\tau_{d,k}}=\frac{\Expsp{\gk{u,\omega}}}{C_k}$ $=\frac{1}{C_k} \pare{g_L + \sum_{j=1}^N \Expsp{g_{kj}(u,.)}}$, where the subscript $d$ stands for ``dynamics", so that
 $\frac{1}{C_k} \Expsp{\int_{t_1}^{r-1}\gk{u,\omega} \, du} = \frac{1}{\tau_{d,k}}\pare{t_1-r-1}$.
From \eqref{eq:alpha}, \eqref{eq:gkj}, \eqref{eq:def_alpha_t_omega}, we have 
$$\Expsp{g_{kj}(u,.)} =G_{kj} \, \sum_{n=-\infty}^{\ent{u}} e^{-\frac{ u - n }{\tau_{kj}}} \, \nu_j.$$
$\nu_j =\Expsp{\omega_j(n)}$ is the firing rate of neuron $j$ in spontaneous dynamics. It is thus time-independent. Denoting $\Set{u}$ the fractional part of $u$, $u=\ent{u} + \Set{u}$, we have:
$$
\Expsp{g_{kj}(u,.)} =   \frac{G_{kj} \, \nu_j}{1-e^{-\frac{1}{\tau_{kj}}}} e^{-\frac{\Set{u}}{\tau_{kj}}}
$$	
and:

$$ \Expsp{\int_{t_1}^{r-1} g_{kj}(u,.) du}  =G_{kj} \, \nu_j \, \tau_{kj} \, \pare{r-1-\ent{t_1}- 
\frac{1-e^{-\frac{\Set{t_1}}{\tau_{kj}}}}{1-e^{-\frac{1}{\tau_{kj}}}}.} $$

We have therefore a piecewise continuous function, with jumps at integer values of $t_1$ and an exponential decay in between integer values, depending on the fractional part of $t_1$. This is obviously due to the time discretization of spike trains. Note that the fractional part term $1-e^{-\frac{\Set{t_1}}{\tau_{kj}}} \in [1,1-e^{-\frac{1}{\tau_{kj}}}[$ so that the expectation of the integrated conductance is mainly constrained by the first term. This gives us:
 \begin{equation*}
\tau_{d,k} \sim \frac{C_k}{g_L+ \sum_{j=1}^N G_{kj} \nu_j \tau_{kj}}
\end{equation*}
This time has the following physical interpretation. We can rewrite the characteristic time as:
$$
\tau_{d,k}= \frac{\tau_{L,k}}{1+\frac{G_k}{g_L}},
$$
where $G_k=\sum_{j=1}^N G_{kj} \nu_j \tau_{kj}$ is the total average synaptic conductance of neuron $k$. This allows for the appearance of the LIF characteristic time and the fact that the characteristic time of the flow gets smaller when the synaptic conductance increases or when the firing increases.

\item \textbf{Firing rate.} The second characteristic time is $\Expsp{\nkrm}$ which is equal to $\frac{1}{\nu_k}$, the inverse of the spontaneous firing rate of neuron $k$.
\end{enumerate}

\ssu{Mean-Field approximation} \label{Sec:Mean-Field}

In this section, we consider an approximation allowing us to simplify the linear response equation \eqref{eq:LinRepgIFtinf}. From this we obtain an explicit equation allowing us to make the link between the linear response in the gIF model and the formulation \eqref{eq:LinRepCorrMonom} in terms of correlations between monomials.

\sssu{Mean-Field limit}

We use two approximations:
\begin{enumerate}[(i)]
\item We  replace $\nkrm$ in \eqref{eq:LinRepgIFtinf} by $-\infty$; 
\item  We replace $\Gamma_k(t_1,r-1,\omega)=e^{-\frac{1}{C_k}\int_{t_1}^{r-1}\gk{u,\omega} \, du}$ by  $e^{-\frac{(r-1-t_1)}{\tau_{d,k}}}$.
\end{enumerate}
This way, we have removed the complex dependence in $\omega$ (appearing in $\nkrm$ and  $\gk{u,\omega}$) to keep a simpler dependence and making explicit the role played by correlations in the linear response, as shown in the remainder of this section.

These approximations, which are further commented in the discussion section, are qualitatively justified in the limit $\tau_{d,k} \ll \frac{1}{\nu_k}$ , which reads:
\begin{equation}\label{eq:CondMFLimit}
\frac{C_k \nu_k}{g_L+ \sum_{j=1}^N G_{kj} \nu_j \tau_{kj}} \ll 1.
\end{equation}
Indeed, in this limit, the characteristic time for the flow integration is faster than the mean-interspike interval and one may replace the time integral in the flow by the average of this function over $\musp$. 
In addition, we may replace $\tko$ by $-\infty$ in \eqref{eq:LinRepgIFtinf} as the difference in the integrals $\int_{\tko}$ and $\int_{-\infty}$ is negligible (because of the fast decay of the exponential).

Note that the flow for a LIF is $\Gamma_k(t_1,t,\omega)=e^{-\frac{(t-t_1)}{\tau_{L,k}}}$. Thus, our flow is close to a LIF model. What changes is the characteristic time. Yet, another big difference comes in the synaptic terms of the gIF model, analyzed below. 
\\
The mean-field approximation allows the following simplifications:
$$
\VkL{t,\omega} \, \sim E_L \frac{\tau_{d,k}}{\tLk}; 
\quad \sk{n-1,\omega} \, \sim \,  \sqrt{\frac{\tau_{d,k}}{2}} \, \frac{\sigma_B}{C_k}.
$$
\sssu{Linear response}
From the mean-field approximation we have:
$$
\dmu{f(t)} \sim - \,\sum_{k=1}^N \, \frac{1}{C_k} \, \sum_{r=-\infty}^{n= \ent{t}}  \Csp{f(t,\cdot)}{  
\frac{\mathcal{H}_k^{(1)}(r,\cdot) }{\sqrt{\frac{\tau_{d,k}}{2}} \, \frac{\sigma_B}{C_k}}\, \int_{-\infty}^{r-1} S_k(t_1) e^{-\frac{(r-1-t_1)}{\tau_{d,k}}}  dt_1}
$$
We can now get the integral of the stimulus out of the correlation as we have removed the dependence on $\omega$.

$$
\dmu{f(t)} \, \sim \,  - \frac{\sqrt{2}}{\sigma_B}  \,\sum_{k=1}^N \, \frac{1}{\sqrt{\tau_{d,k}}} \, \sum_{r=-\infty}^{n= \ent{t}}  \Csp{f(t,\cdot)}{  
\mathcal{H}_k^{(1)}(r,\cdot)} \,  \int_{-\infty}^{r-1} S_k(t_1) e^{-\frac{(r-1-t_1)}{\tau_{d,k}}}   dt_1.
$$
We note $e_{d,k}(t)=e^{-\frac{t}{\tau_{d,k}}} \, H(t)$, so that we may write:
\begin{equation}\label{eq:LinRepMF1}
\dmu{f(t)} \, =- \frac{\sqrt{2}}{\sigma_B}  \,\sum_{k=1}^N \, \frac{1}{\sqrt{\tau_{d,k}}} \, \sum_{r=-\infty}^{n= \ent{t}}  \Csp{f(t,\cdot)}{  
\mathcal{H}_k^{(1)}(r,\cdot)} \, \bra{S_k \ast e_{d,k}}(r-1).
\end{equation}
The term $\bra{S_k \ast e_{d,k}}(r-1)$ that we would have as the response of an isolated IF neuron is here weighted by correlations functions involving synaptic interactions. This becomes more explicit below.

From this approximation we have simplified the correlation term. We now have the correlation between $f$ and the first order expansion of $\delta \phi$. Note that this approximation could also be used to handle the higher order terms in the expansion \eqref{eq:ExpansionDeltaphi} of $\delta \phi$.

From \eqref{eq:Hku} we see that the computation of $\Csp{f(t,\cdot)}{\mathcal{H}_k^{(1)}(r,\cdot)}$

requires us to compute the term
$$
\Csp{f(t,\cdot)}{  
\omega_{k}(r) \, a^{(1)}\pare{\Xksp{r-1,\omega}}}
= \Csp{f(t,\cdot)}{  
\omega_{k}(r) \, \frac{e^{-\frac{1}{2}\Xksp{r-1,\omega}^2}}{\Pi\pare{\Xksp{r-1,\omega}}}},
$$
(and the corresponding term involving $b^{(1)}$), where $\Xksp{r-1,\omega}$ is given by \eqref{eq:Xsp}. In particular,  $\Xksp{r-1,\omega}$ depends on the synaptic term that can be simplified using the mean-field approximation as we will now show.\\

\sssu{The synaptic term}

In the mean-field approximation we have, from  \eqref{eq:Vksyn} and \eqref{eq:def_alpha_t_omega}
$$
\Vksyn{r-1,\omega} \sim \frac{1}{C_k} \sum_{j=1}^N   W_{kj} \, \int_{-\infty}^{r-1} e^{-\frac{(r-1-t_1)}{\tau_{d,k}}} \sum_{n=-\infty}^{\ent{t_1}} \, e^{-\frac{t_1 - n}{\tau_{kj}}} \, \omega_j(n) dt_1
$$
where:
 $$
 \int_{-\infty}^{r-1} e^{-\frac{(r-1-t_1)}{\tau_{d,k}}} \sum_{n=-\infty}^{\ent{t_1}} \, e^{-\frac{t_1 - n}{\tau_{kj}}} \, \omega_j(n) dt_1 = A_{kj} \, \sum_{s=-\infty}^{r-1} e^{-\frac{(r-1-s)}{\tau_{d,k}}} \,  \, \sum_{n=-\infty}^{s} \, e^{-\frac{(s - n)}{\tau_{kj}}} \, \omega_j(n),
$$
with $A_{kj}=
\int_{0}^{1}  e^{-\frac{u}{\eta_{k,j}}}  du = 
\eta_{k,j} \, 
\bra{1-e^{-\frac{1}{\eta_{k,j}}}}$ 
%
and $\frac{1}{\eta_{k,j}}=\frac{1}{\tau_{kj}}-\frac{1}{\tau_{d,k}}$.

Note that the integral is upper bounded by $\frac{A_{kj}}{
\pare{1-e^{-\frac{1}{\tau_{kj}}}}
 \pare{1-e^{-\frac{1}{\tau_{d,k}}}}}$.

We now simplify the series, making more explicit the dependence in $\omega_j$. After some algebra, we obtain:

 $$
 \sum_{s=-\infty}^{r-1} e^{-\frac{(r-1-s)}{\tau_{d,k}}} \,  \, \sum_{n=-\infty}^{s} \, e^{-\frac{(s - n)}{\tau_{kj}}} \, \omega_j(n)  = \sum_{p=1}^{+\infty} B^{(p)}_{kj} \, \omega_j(r-p) 
$$
with $B^{(p)}_{kj}=\displaystyle{
\sum_{
\tiny{
\begin{array}{lll}
q,s \geq 0;\\
q+s=p-1
\end{array}
}
}
}
e^{-\frac{q}{\tau_{kj}}-\frac{s}{\tau_{d,k}}}$.
It follows that:
\begin{equation*}
\Vksyn{r-1,\omega} \sim \frac{1}{C_k} \sum_{j=1}^N   W_{kj} \,A_{kj}  \, \sum_{p=1}^{+\infty} B^{(p)}_{kj} \, \omega_j(r-p).
\end{equation*}
and, finally: 
\begin{equation*}
\Xksp{r-1,\omega} \sim \theta_{L,k} -   \sum_{j=1}^N   \,  W_{kj}^{(\sigma)} \, \sum_{p=1}^{+\infty} B^{(p)}_{kj} \, \omega_j(r-p),
\end{equation*}
where $\theta_{L,k} =\frac{\theta- E_L \frac{\tau_{d,k}}{\tLk}}{\sqrt{\frac{\tau_{d,k}}{2}} \, \frac{\sigma_B}{C_k} }$  is the threshold term that one would obtain for an isolated LIF neuron, submitted to noise, with a characteristic decay time $\tau_{d,k}$ ($=\tLk$ for the LIF). Also, $W_{kj}^{(\sigma)}=\frac{1}{\sigma_B} \, \sqrt{\frac{2}{\tau_{d,k}}}  \,  W_{kj} \,A_{kj}$  are renormalized synaptic 
weights.

\sssu{Markovian approximation}
Even with the MF approximation, $\Xksp{r-1,\omega}$ depends on the whole history via the synaptic term. We now use a Markov approximation of order $D$ to further simplify this term. That is, we truncate the history dependence up to $D$ integer time steps in the past. This truncation is justified by the exponential correlation decay exposed in section \ref{Sec:CorrDecay}. This makes the link with the formal expansion proposed in section \ref{Sec:MonDecomp}.

We have
\begin{equation}\label{eq:XspMarkov}
\Xksp{r-1,\omega} \sim \theta_{L,k} -   \sum_{j=1}^N   \,  W_{kj}^{(\sigma)} \, \sum_{p=1}^{D} B^{(p)}_{kj} \, \omega_j(r-p),
\end{equation}
In this way, $\Xksp{r-1,\omega}$, $a^{(1)}\pare{\Xksp{r-1,\omega}}$ and $b^{(1)}\pare{\Xksp{r-1,\omega}}$ used in the linear response are random variables with finitely many values, whose law is constrained by the marginal law of $\musp$, on a past of depth $D$.
As we have an explicit value for $\Xksp{r-1,\omega}$, we also have the explicit value for $\mathcal{H}_k^{(1)}(r,\cdot)$. Thus, we can use the Hammersley-Clifford decomposition \eqref{eq:Hammersley-Clifford} to obtain the coefficients $\delta \phi_l(r)$.
However, an explicit, tractable computation requires additional approximations, as done in the next section. 

\ssu{Explicit computation of the linear response for the gIF model}
We consider different levels of approximations, in terms of the memory depth $D$.
The main objective is to instantiate the equation \eqref{eq:LinRepCorrMonom} of section \ref{Sec:MonDecomp}

\sssu{Markovian approximation of order $1$}
Here, 
$\Xksp{r-1,\omega} \sim \theta_{L,k} -   \sum_{j=1}^N   \,  W_{kj}^{(\sigma)} \, B^{(1)}_{kj} \, \omega_j(r-1)$.
This is a random variable in a space of cardinality $2^{N}$, constrained by the law of $\omega_j(r-1)$ under $\musp$. Likewise, $a^{(1)}\pare{\Xksp{r-1,\omega}}$ and $b^{(1)}\pare{\Xksp{r-1,\omega}}$ are random variables of the same type.
We can use the Hammersley-Clifford transformation \eqref{eq:Hammersley-Clifford} to convert these variables into a linear combination of monomials. These monomials have the form $m_l(\omega)=\prod_{k=1}^n \omega_{i_k}(r-1)$
where, in contrast to \eqref{eq:monomial_def}, all spikes occur at the same time ($r-1$). Still the degree $n$ of the monomial can be quite large, up to $N$, the number of neurons, corresponding to the spiking patterns where all neurons spike at time $r-1$.
Nevertheless it is possible to make a monomial expansion of these functions, truncating them to a certain monomial degree.
%
\begin{equation*}
\begin{split}
&a^{(1)}\pare{\Xksp{r-1,\omega}} \sim  \xi_k^{(0)} + \sum_{i=1}^N \xi_{k;i}^{(1)} \, \omega_i(r-1) \\
&+ \sum_{i,j=1}^N \xi_{k;ij}^{(2)} \, \omega_i(r-1) \omega_j(r-1)
 + \sum_{i,j,l=1}^N \xi_{k;ijl}^{(3)} \, \omega_i(r-1) \omega_j(r-1)  \omega_l(r-1) + \dots 
 \end{split} 
\end{equation*}
Note that this is a sum containing a finite number of terms, where the last one is proportional
to $\omega_1(r-1) \dots \omega_N(r-1)$. \\
The coefficients can be explicitly computed using the Hammersley-Clifford decomposition. For example we have: 
 $$\xi_k^{(0)}=a^{(1)}\pare{\Xksp{r-1,\vect{&0 \\& \cdots\\ &0}}},$$
 the block containing only $0$; 
 $$\xi_{k;i}^{(1)}= a^{(1)}\pare{\Xksp{r-1,\vect{&0 \\& \cdots \\ & 1_i \\ &\cdots\\ &0}}} - a^{(1)}\pare{\Xksp{r-1,\vect{&0 \\& \cdots\\ &0}}},$$
 where the block contains only a $1$ at position $i$; 
  $$
  \xi_{k;i,j}^{(2)}= a^{(1)}\pare{\Xksp{r-1,\vect{&0 \\& 1_i \\ &\cdots \\ & 1_j \\ &\cdots\\ &0}}}
 -a^{(1)}\pare{\Xksp{r-1,\vect{&0 \\& 1_i \\ &\cdots \\ & 0 \\ &\cdots\\ &0}}}
 $$
 $$
 -a^{(1)}\pare{\Xksp{r-1,\vect{&0 \\& 0 \\ &\cdots \\ & 1_j \\ &\cdots\\ &0}}}
 +a^{(1)}\pare{\Xksp{r-1,\vect{&0 \\& 0 \\ &\cdots \\ & 0 \\ &\cdots\\ &0}}}
 $$
%
and so on.\\

This finally gives:
\begin{equation}\label{eq:Hammersley-Clifford-gIF-synapses}
\begin{split}
& \xi_k^{(0)}=a^{(1)}\pare{\theta_{L,k}}; \\
& \xi_{k;i}^{(1)}=a^{(1)}\pare{\theta_{L,k}-W_{ki}^{(\sigma)}B^{(1)}_{ki}}-a^{(1)}\pare{\theta_{L,k}};\\
& \xi_{k;ij}^{(2)}=
\left(
\begin{split}
&a^{(1)}\pare{\theta_{L,k}-W_{ki}^{(\sigma)}B_{1;ki}-W_{kj}^{(\sigma)}B^{(1)}_{kj}} 
- a^{(1)}\pare{\theta_{L,k}-W_{ki}^{(\sigma)}B_{1;ki}}\\
-&a^{(1)}\pare{\theta_{L,k}-W_{kj}^{(\sigma)}B^{(1)}_{kj}}
+ a^{(1)}\pare{\theta_{L,k}};
 \end{split}
 \right)
 \\
&\vdots
 \end{split} 
\end{equation}
and so on. The same holds for $b^{(1)}\pare{\Xksp{r-1,\omega}}$ defining coefficients $\beta_k$. Thus, for $\mathcal{H}_k^{(1)}(r,\omega)$ we have:
\begin{equation*}
\begin{split}
&\mathcal{H}_k^{(1)}(r,\omega) \sim  \gamma_k^{(0)} +  \gamma_k^{(1)} \, \omega_k(r) +
\sum_{i=1}^N \gamma_{k;i}^{(2)} \, \omega_k(r) \,\omega_i(r-1) \\
&+ \sum_{i,j=1}^N \gamma_{k;ij}^{(3)} \, \omega_k(r) \,\omega_i(r-1) \omega_j(r-1)
 + \sum_{i,j,l=1}^N \gamma_{k;ijl}^{(4)} \,  \omega_k(r) \,\omega_i(r-1) \omega_j(r-1)  \omega_l(r-1) + \dots
 \end{split} 
\end{equation*}
\begin{equation*}
\begin{split}
& \gamma_k^{(0)}=\beta_k^{(0)}; \\
& \gamma_k^{(1)}=\xi_k^{(0)}-\beta_k^{(0)}; \\
& \gamma_{k;i}^{(2)}=\xi_{k;i}^{(1)} - \beta_{k;i}^{(1)};\\
& \gamma_{k;ij}^{(3)}=\xi_{k;ij}^{(2)} - \beta_{k;ij}^{(2)};\\
& \gamma_{k;ijl}^{(4)}=\xi_{k;ijl}^{(3)} - \beta_{k;ijl}^{(3)};\\
 \end{split} 
\end{equation*}
Finally, the linear response reads:
\begin{equation}\label{eq:LinRepMarkovD1}
\dmu{f(t)} \, =- \frac{2}{\sigma_B}  \,\sum_{k=1}^N \, \frac{1}{\sqrt{\tau_{d,k}}} \, \sum_{r=-\infty}^{n= \ent{t}}  
\bra{
\small{
\begin{array}{ll}
&\gamma_k^{(1)} \, \Csp{f(t,\cdot)}{\omega_k(r)}\\
+\sum_{i=1}^N &\gamma_{k;i}^{(2)} \, \Csp{f(t,\cdot)}{\omega_k(r) \,\omega_i(r-1) 
}\\
+\sum_{i,j=1}^N &\gamma_{k;ij}^{(3)} \, \Csp{f(t,\cdot)}{\omega_k(r) \,\omega_i(r-1) \,\omega_j(r-1)
}\\
+& \cdots
\end{array}
}
}
\, \bra{S_k \ast e_{d,k}}(r-1),
\end{equation}
where the sum into brackets contains $N$ terms, corresponding to correlations functions of increasing degree. If we now consider an observable of the form \eqref{eq:monomial_decomp_t} we may write the linear response in the form \eqref{eq:LinRepCorrMonom}:
$$
\dmu{f(t)}  \, = \, \sum_{r=-\infty}^{n=\ent{t}} \sum_{l,l'} f_l(t)  \,  C_{l,l'}(n-r) \, \delta \phi_{l'}(r),
$$
where, as in section \ref{Sec:LinRepFiniteRange} we have used the time-translation invariance of $\musp$ letting $n-r$ appearing in correlations. Here, $l'$ is the index of the monomial (definition \eqref{eq:DefBlockIndex}) corresponding to the monomial $m_l'(\omega)=\omega_k(1)  \prod_{u=1}^m \omega_{i_u}(0)$. Then,
\begin{equation*}
\delta \phi_{l'}(r)=\sum_{m=1}^N \sum_{i_1, \dots, i_m=1}^N \gamma_{k;i_1 \dots i_m}^{(m)} \bra{S_k \ast e_{d,k}}(r-1)
\end{equation*}
%
%
and:
$$
C_{l,l'}(n-r) = \Csp{m_l(n)}{m_{l'}(r)}
$$

This instantiates an example where the terms entering the general form of the linear response \eqref{eq:LinRepCorrMonom} are explicitly computed. In eq. \eqref{eq:LinRepMarkovD1}, one clearly sees the correlations between the observable $f$ and the spike events hierarchy. Each term is weighted by the coefficients $\gamma_{k;i_1 \dots i_m}^{(m)}$ integrating the effect of synaptic interactions with history-dependent conductances. 
Let us repeat that the term $\bra{S_k \ast e_{d,k}}(r-1)$ is what we would obtain for an isolated LIF neuron with a characteristic time $\tau_{d,k}$. 

We note that the response of a neuron $k$ to a stimulus applied to neuron $i$ does not only depend on the synaptic weight $W_{ki}$ but, in general, on all synaptic weights because dynamics creates complex causality loops which build up the response of neuron $k$. 

\sssu{Markovian approximation of order $D$}
The procedure generalizes to a finite memory depth $D>1$. Using \eqref{eq:XspMarkov} we write the synaptic contribution in terms of spike events and using the Hammersley-Clifford decomposition \eqref{eq:Hammersley-Clifford} we  write the function $\cH^{(1)}$ in \eqref{eq:Hku} as a monomial expansion where coefficients can be explicitly computed. They involve pre-synaptic spikes events, as in \eqref{eq:Hammersley-Clifford-gIF-synapses}. The linear response is a series of correlation functions, which converges thanks to the spectral gap property. Clearly, the terms in the series involve monomials of increasing order whose probability is expected to decrease fast as the order increases. Thus, a Markovian approximation of memory depth $1$ or $2$, involving only pairwise or triplets might be sufficient to characterize the statistics of the gIF model, especially when dealing with finite samples obtained from numerical simulations. This remark deserves more thorough investigations, a subject for future studies.

\section{Numerical simulations} \label{sec:Numerics}

In this section, we illustrate our results using a simple example. We use a discrete-time integrate and fire neuronal network derived from the gIF model. The main reason to choose here a ``simple'' model is numeric.
Indeed, performing convincing numerical simulations illustrating our results is a work \textit{per se}, of the same level of difficulty as the mathematical results we have shown up to now. 
In order to study the linear response theory in a model like the gIF, one  has to properly handle numerically:
\begin{enumerate}[(1)]
\item The mixed dynamics, continuous-time evolution of the membrane potential depending on the discrete-time spikes history; 
\item The spike train statistics, especially finding a numerical way to perform a suitable averaging, not only for the spontaneous probability measure where time-ergodic average can be used but also for the non-stationary response where \textit{ergodicity does not take place};
\item The long memory tail in the dynamics; 
\item Find an illustrative example with the good range of parameters showing convincing, original, non-trivial results while avoiding prohibitive computational times. \end{enumerate}
It is not evident to us that classical spiking neural network simulators such as BRIAN \cite{simberg-brette-etal:19}, although quite efficient, could easily handle (1) in conjunction with (2).
The model presented below, which can be viewed as a mean-field approximation of the gIF has been studied both from the mathematical side and the numerical side \cite{cessac:08, cessac-vieville:08}. Additionally, we have designed a simulation tool, PRANAS, devoted to the analysis of population spike train statistics and allowing to properly handle Gibbs distributions from numerical simulations or experimental data, with a specific module dedicated to this model \cite{cessac-kornprobst-etal:17}.
This software is freely downloadable at \url{https://team.inria.fr/biovision/pranas-software/} on simple demand. 
The linear response C++ codes and instructions required to reproduce these numerical results can be found at  \url{https://github.com/brincolab/Linear-response}.

\subsection{Discrete time leaky integrate and fire model}

The model is obtained from the gIF model \eqref{eq:DyngIF} by setting   
$g_{k} (t,\omega) = g_L$ (constant conductance), $\alpha_{kj}(t,\omega)=\omega_j(n)$, $n=\ent{t}$ (instantaneous synapse), 

We then discretize time introducing $\gamma=1 - \frac{dt}{\tau}$, with $0 < \gamma < 1$. Setting $dt=1$ the sub-threshold dynamics is given by:
\begin{equation}\label{eq:BMS}
V_k(n+1)=\gamma \, V_k(n) + \sum_j W_{kj} \omega_j(n) + I_0  + S_k(t) + \sigma_B \xi_k (n), \quad \mbox{if } V_k(n)< \theta.
\end{equation}
It corresponds therefore to a discrete-time Leaky Integrate and Fire model \cite{cessac:11a}, where the post-synaptic potential from $j$ to $k$, at time $n+1$, is triggered by the pre-synaptic spike at time $n$. The results below are therefore an illustration of the mean-field approximation developed in section \ref{Sec:Mean-Field}.

Note that $\gamma$, the decay term, is related to the leak characteristic time of the LIF, $\tau=\frac{C}{g_L}$ by:
\begin{equation*}
\tau=\frac{1}{1-\gamma}.
\end{equation*}
The condition $\gamma < 1$ somewhat define the exponential decay in the spike history dependence via the characteristic time:
\begin{equation}\label{eq:tau_gamma}
\tau_\gamma=-\ent{\frac{1}{\log\gamma}}.
\end{equation}

This characteristic time can be interpreted as follows. It is easy to integrate equation \eqref{eq:BMS} up to the last time where voltage was reset in the past, $\nko$:
\begin{equation}\label{eq:IntegBMS}
V_k(n+1)=\sum_{j=1}^N W_{kj} \, \eta_{kj}(n,\omega) + I_0 \, \frac{1-\gamma^{n+1-\nko}}{1-\gamma} + \sum_{l=\nko}^n \gamma^{n-l} S_k(l)+ \sigma_B \sum_{l=\nko}^n \gamma^{n-l} \xi_k(l) ,
\end{equation} 
where:
\begin{equation}\label{eq:eta}
 \eta_{kj}(n,\omega)=\sum_{l=\nko}^n \gamma^{n-l} \, \omega_j(l),
\end{equation}

The term $ \gamma^{n-l}$ thus plays the role of the flow \eqref{eq:fgm} with a characteristic decay time \eqref{eq:tau_gamma}. 
Indeed, with $\gk{u,\omega}=g_L$, $-\frac{1}{C_k} \, \int_{t_1}^{t}\gk{u,\omega} \, du$ $=-\frac{g_L}{C_k} \pare{t-t_1}=-(n-l) \frac{dt}{\tau}$, where we discretize time in units $dt$ so that $t=n dt$, $t_1=l dt$. When $\frac{dt}{\tau}$ is small $-\frac{dt}{\tau} \sim \log \pare{1-\frac{dt}{\tau}}=\log \gamma$, so that $\Gamma_k(t_1,t,\omega) \sim \gamma^{n-l}$.
Comparing \eqref{eq:IntegBMS}, \eqref{eq:eta} to \eqref{eq:Vk}, \eqref{eq:Vkspont},  \eqref{eq:Vksyn},\eqref{eq:Vkext} we see that
the first term in the right hand side of  \eqref{eq:IntegBMS} corresponds to $\Vksyn{t,\omega}$.

Likewise the second, third term and fourth correspond to the integration \eqref{eq:Vkext} of a time dependent stimulus $I_0+S_k(t)+\sigma_B \xi_k(t)$ with a constant term $I_0$, used to fix the baseline activity, and a noise term with intensity $\sigma_B$.

\subsection{Linear response}

The potential of the discrete-time model has the same form as  \eqref{eq:Phi_GIF} \cite{cessac:11b}. Thus, equation \eqref{eq:dmu} adapts directly to the discrete-time case.
In the numerical simulations we use the mean-field approximation of section \ref{Sec:Mean-Field} where we replace $\nkrm$ by its average value $\moy{\nkrm}=r-1-\frac{1}{\nu_k}$ where the inverse $\frac{1}{\nu_k}$ of the average firing rate $\nu_k$ of neuron $k$ is the mean time between two spikes.
Under these approximations we examine two forms of linear response proposed in the paper. 

\begin{enumerate}

\item The first order expansion of the potential, \eqref{eq:dmu}. In the present context it becomes:

$$
\delta \mu^{(1)}\bra{f(n)} \, \sim - \,\sum_{k=1}^N  \, \sum_{r=-\infty}^{n}  
\sum_{l=r-1-\frac{1}{\nu_k}}^{r-1} \gamma^{r-1-l} \Csp{f(n,\cdot)}{  
\zeta(r-1,.)}\,S_k(l),
$$
where we have set $\zeta_k(r-1,\omega)=\frac{\mathcal{H}_k^{(1)}(r,\omega) }{\sk{r-1,\omega}}$. In the numerical simulations we use a Markovian approximation with memory depth $D$ such that 
the sum $\sum_{r=-\infty}^{n}$ is replaced by $\sum_{r=n-D}^{n}$. $D$ is typically determined by exponential decay of the terms  $\gamma^{l-r+1} \Csp{f(n,\cdot)}{  
\zeta_k(r-1,.)}$, which is controlled, on one hand, by $\gamma$ (with a characteristic time $\tau_\gamma=-\ent{\frac{1}{\log\gamma}}$), and the correlation $\Csp{f(n,\cdot)}{  
\zeta_k(r-1,.)}$, which is controlled by the spectral gap in the Perron-Frobenius matrix. In this approximation we have:
$$
\delta \mu^{(1)}\bra{f(n)} \, \sim - \,\sum_{k=1}^N  \, \sum_{r=n-D}^{n}  
\sum_{l=r-1-\frac{1}{\nu_k}}^{r-1} \gamma^{r-1-l} \Csp{f(n,\cdot)}{  
\zeta_k(r-1,.)}\,S_k(l).
$$ 
Using the stationarity of $\musp$ we have $\Csp{f(n,\cdot)}{  
\zeta(r-1,.)}=\Csp{f(n-r+1,\cdot)}{  
\zeta_k(0,.)}$ where $r \in [n-D,n]$ so that $m=n-r+1 \in [1,D+1]$. Then, introducing the $N \times D$ matrix $\cK^{(1)}$ with entries :
\begin{equation*}
\cK^{(1)}_{k,m}=\Csp{f(m,\cdot)}{\zeta_k(0,.)},
\end{equation*}
we may write $\delta \mu^{(1)}\bra{f(n)}$ in the form:
\beq\label{eq:LinRepBMSmu2}
\delta \mu^{(1)}\bra{f(n)} \, \sim - \,\sum_{k=1}^N  \, \sum_{m=1}^{D+1}  
\sum_{l=0}^{\frac{1}{\nu_k}} \, \gamma^{l} \, \cK^{(1)}_{km} \, S_k(n-m-l).
\eeq

\item The first order Hammersley-Clifford expansion \eqref{eq:LinRepMarkovD1} of $\mathcal{H}_k^{(1)}$ which corresponds to expand $\zeta_k(0,\omega)$ to the lowest order in the Hammersley-Clifford expansion, $\zeta_k(0,\omega) \sim \gamma_k^{(1)} \omega_k(0)$ with:
$$
\gamma_k^{(1)}=a^{(1)}(\theta_{Lk})-b^{(1)}(\theta_{Lk}),
$$
with $a^{(1)}(x)=\frac{\Pi'(x)}{\Pi(x)}$, $b^{(1)}(x)=-\frac{\Pi'(x)}{1-\Pi(x)}$. This gives an approximation similar to the fluctuation-dissipation theorem where the linear response is a sum of pairwise correlation functions.
In the discrete time model it reads:
\beq\label{eq:LinRepBMSmu3}
\delta \mu^{(HC_1)}\bra{f(n)} \, \sim - \frac{2 \, \sqrt{1-\gamma}}{\sigma_B} \, \gamma^{(1)} \, \,\sum_{k=1}^N  \, \sum_{m=1}^{D+1}  
\sum_{l=0}^{\frac{1}{\nu_k}} \gamma^{l} \, \cK^{(HC_1)}_{km} \, S_k(n-m-l).
\eeq
where :
\begin{equation*}
\cK^{(HC_1)}_{k,m}=\Csp{f(m,\cdot)}{\omega_k(0,.)}.
\end{equation*}
We have used the superscript ``$HC_1$'' to differentiate this lowest-order Hammersley-Clifford approximation from \eqref{eq:LinRepBMSmu2}. The interest of this approximation (and more generally, of the Hammersley-Clifford expansion \eqref{eq:LinRepMarkovD1}) is that it can be obtained without knowing explicitly the potential $\phi$ (by a mere fit of the coefficients $\gamma_k^{(1)}$). We expect however \eqref{eq:LinRepBMSmu2} to give a better approximation of the linear response than \eqref{eq:LinRepBMSmu3}. 
Note that the effective threshold $\theta_{L,k} =\frac{\theta- E_L \frac{\tau_{d,k}}{\tLk}}{\sqrt{\frac{\tau_{d,k}}{2}} \, \frac{\sigma_B}{C_k} }$ in the gIF model corresponds, in the discrete-time model, to $\theta_{L,k} =\frac{\theta- E_L}{\sqrt{\frac{\tau}{2}} \, \frac{\sigma_B}{C}}$, where $I_0 = \frac{E_L}{\tau}$,  so that:
$
\theta_{L,k} =\frac{\theta- \frac{I_0}{1-\gamma}}{\sqrt{\frac{1}{2(1-\gamma)}} \, \sigma_B},
$
is independent of $k$ in this case. This explains why $\gamma^{(1)}$ gets out of the sum and lost its index $k$ in \eqref{eq:LinRepBMSmu3}.
\end{enumerate}

\subsection{Averaging method}\label{Sec:AveragingMethod}

We need to compute numerically averages with respect to the invariant probability $\musp$. As $\musp$ is ergodic it is in principle possible to get them by time averaging. However, we also want to compute averages in the presence of a stimulus, where ergodicity does not take place. In addition, a notation like $\Expsp{f(n,\cdot) \, g(r,\cdot)}$ 
appearing all over the paper involves a subtlety:  we are computing the average of time-dependent quantities (via the first argument in $f(n,\cdot), g(r,\cdot)$) with respect to an invariant probability on the second argument, $\omega$. 
The mathematical meaning is the following:
$$
\Expsp{f(n,\cdot) \, g(r,\cdot)} = \int_{\Omega} f(n,\omega) \, g(r,\omega) \musp(d \omega).
$$

We numerically compute such quantities by generating $M$ spike trains, denoted $\omega^{(m)}$, $m=1 \dots M$, of length $T$, with the spontaneous dynamics, thus distributed according to $\musp$. Then:
\begin{equation}\label{eq:EmpAverage}
\int_{\Omega} f(n,\omega) \, g(r,\omega) d\moysp{\omega} \sim \frac{1}{M}
\, \sum_{m=1}^M f(n,\omega^{(m)}) \, g(r,\omega^{(m)}).
\end{equation}
The equality holds in the limit $M \to \infty$. Here, we chose $M=10000$ or $M=100000$. Fluctuations about the mean are ruled by the central limit theorem, so they decrease like $\frac{1}{\sqrt{M}}$ with a proportionality factor depending on the observables $f$ and $g$.

\subsection{Results} \label{Results}

\begin{figure}
\centering
\includegraphics[width=9cm,height=10cm, angle=270]{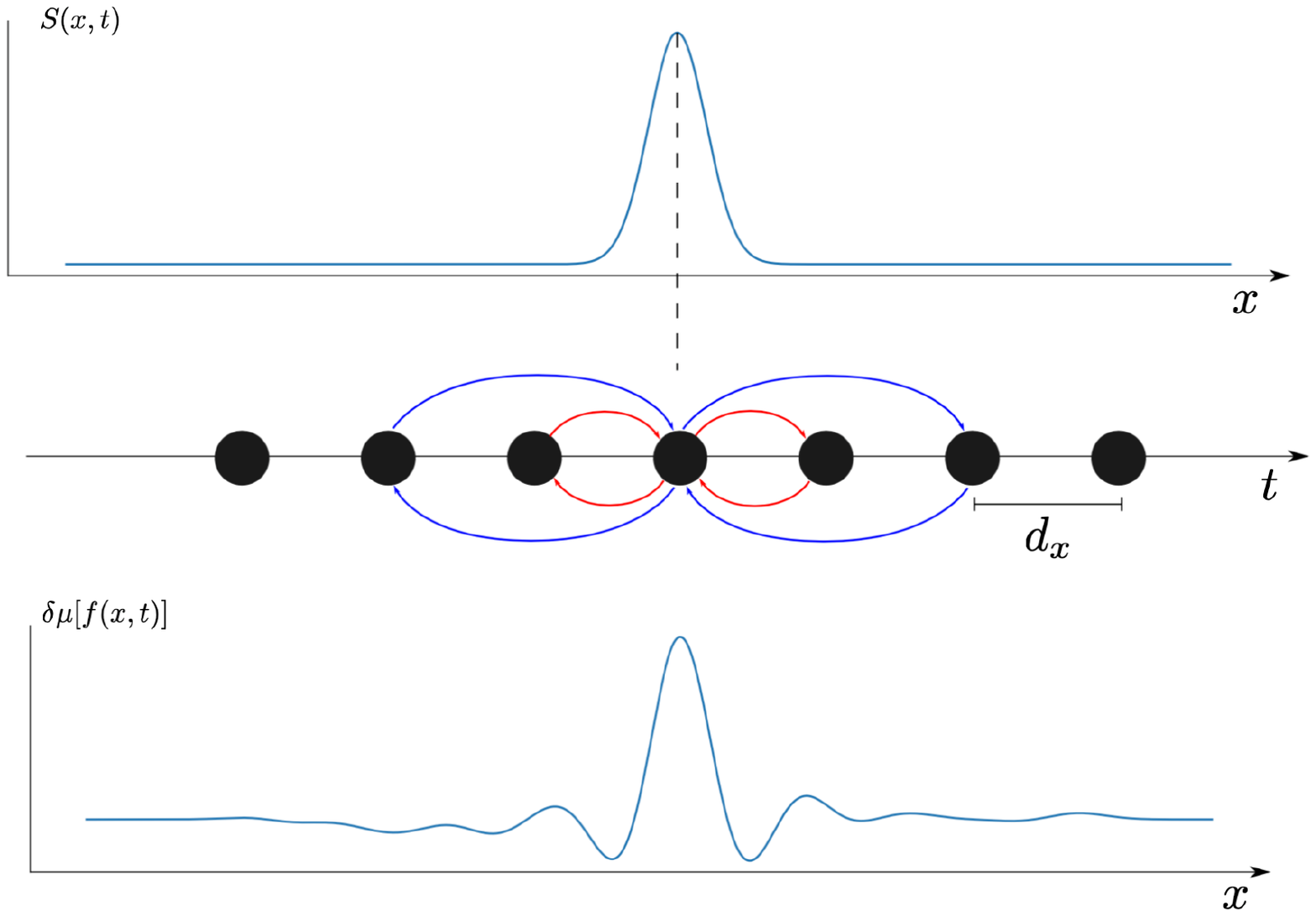}
\caption{Connectivity pattern of the model. Neurons (black filled circles) are located into a one-dimensional lattice with lattice spacing $d_x$. They are connected to nearest neighbours through excitatory connections (red) with weight $w^+$ and to second nearest neighbours through inhibitory connections (blue) with weight $w^-$. These neurons are submitted to a time dependent stimulus $S(x,t)$ (top line, blue trace) travelling at speed $v$. This modifies the average activity $\mu\bra{f(x,t)}$ by a variation $\delta \mu\bra{f(x,t)}$ (bottom trace).}
\label{Fig:Connectivity}
\end{figure}

We consider a one-dimensional lattice of $N=30$ neurons, separated by a lattice spacing $d_x$, with null boundary conditions. 
The connectivity is depicted in Fig. \ref{Fig:Connectivity}.
Each neuron excites its neighbours with a weight $W_{k\pm1,k}=w^+>0$ and inhibits its second neighbours with a weight $W_{k\pm2,k}=w^- < 0$. We compare the spontaneous activity (with a noise term and a constant term $I_0$ to fix the baseline activity, as in \eqref{eq:BMS}) to the dynamics in the presence of Gaussian pulse, with width $\Delta$, propagating at speed $v$ from left to right:
\begin{equation}\label{eq:StimGaussian}
S(x,t)=\frac{A}{\sqrt{2 \pi} \, \Delta} e^{-\frac{1}{2} \, \frac{\pare{x-v \,t}^2}{2 \, \Delta^2}},
\end{equation}
where $x,t$ are continuous space-time variables. The neuron $k$ is located at $x_k=k\,\delta$ and time is updated at each $t=n \, b$ where $b$ is a time bin. In the simulations $d_x=1 \, mm, v=2 \, mm/s, b=10 \, ms, \Delta=1 \, mm$. The term $A$ represent the amplitude and is variable to show the validity of the linear response theory as the amplitude of the stimulus increases (see Fig. \ref{Fig:Error}).
We fix $\theta=1$, $\gamma=0.6$, corresponding to a characteristic decay time $\tau_\gamma=-\ent{\frac{1}{\log\gamma}} \sim 2$. The memory $D$ appearing in the summations \eqref{eq:LinRepBMSmu2}, \eqref{eq:LinRepBMSmu3} was taken to be $D=10$ from the study of correlation decay rate (Fig. \ref{Fig:Correlation_23_A0.2} ). This is a good compromise between the convergence of these sums and the computational time. 

\begin{figure}
\centering
\includegraphics[width=7cm,height=7.4cm, angle=270]{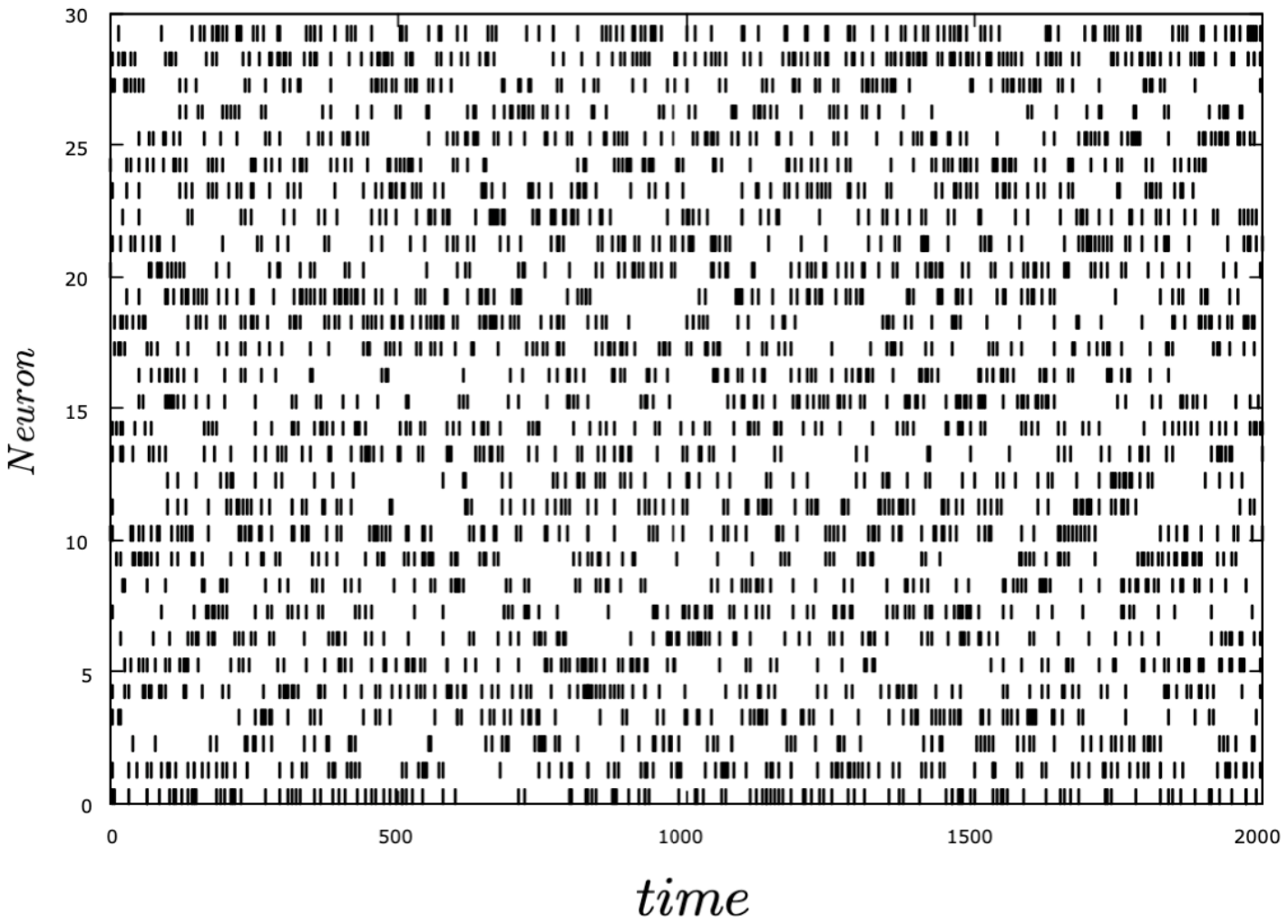}
\includegraphics[width=7cm,height=7.4cm, angle=270]{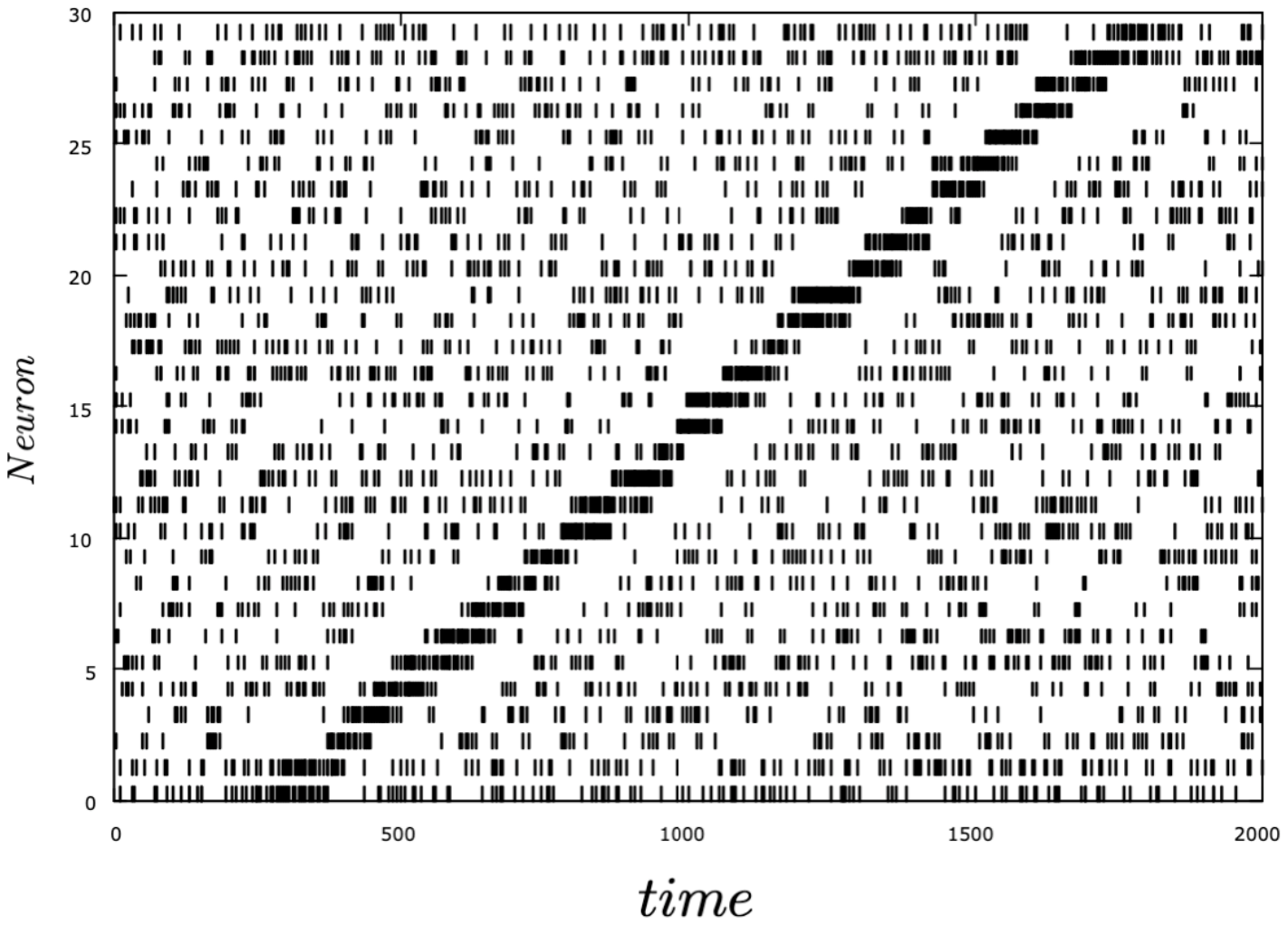}
\vspace{-0.5cm}
\caption{Spiking activity of the network. Left panel shows the spontaneous activity and right panel in the presence of a moving stimulus  \eqref{eq:StimGaussian}. }
\label{Fig:TraceMovingBar}
\end{figure}

The constant stimulus $I_0$ is fixed to have a good baseline activity (see Fig. \ref{Fig:TraceMovingBar}). We consider the activity without and with stimulus with moderate excitatory connectivity and strong inhibitory connectivity ($w^+=0.2$, $w^-=2$). In fig. \ref{Fig:TraceMovingBar} we show the spike network activity in spontaneous activity (left) and in the presence of a moving stimulus (right). The strong inhibition is particularly visible in the presence of the stimulus. 

\subsection{Linear response for firing rates}\label{Sec:rates}
\begin{figure}
\centering
\includegraphics[width=1.1\textwidth,height=0.75\textheight]{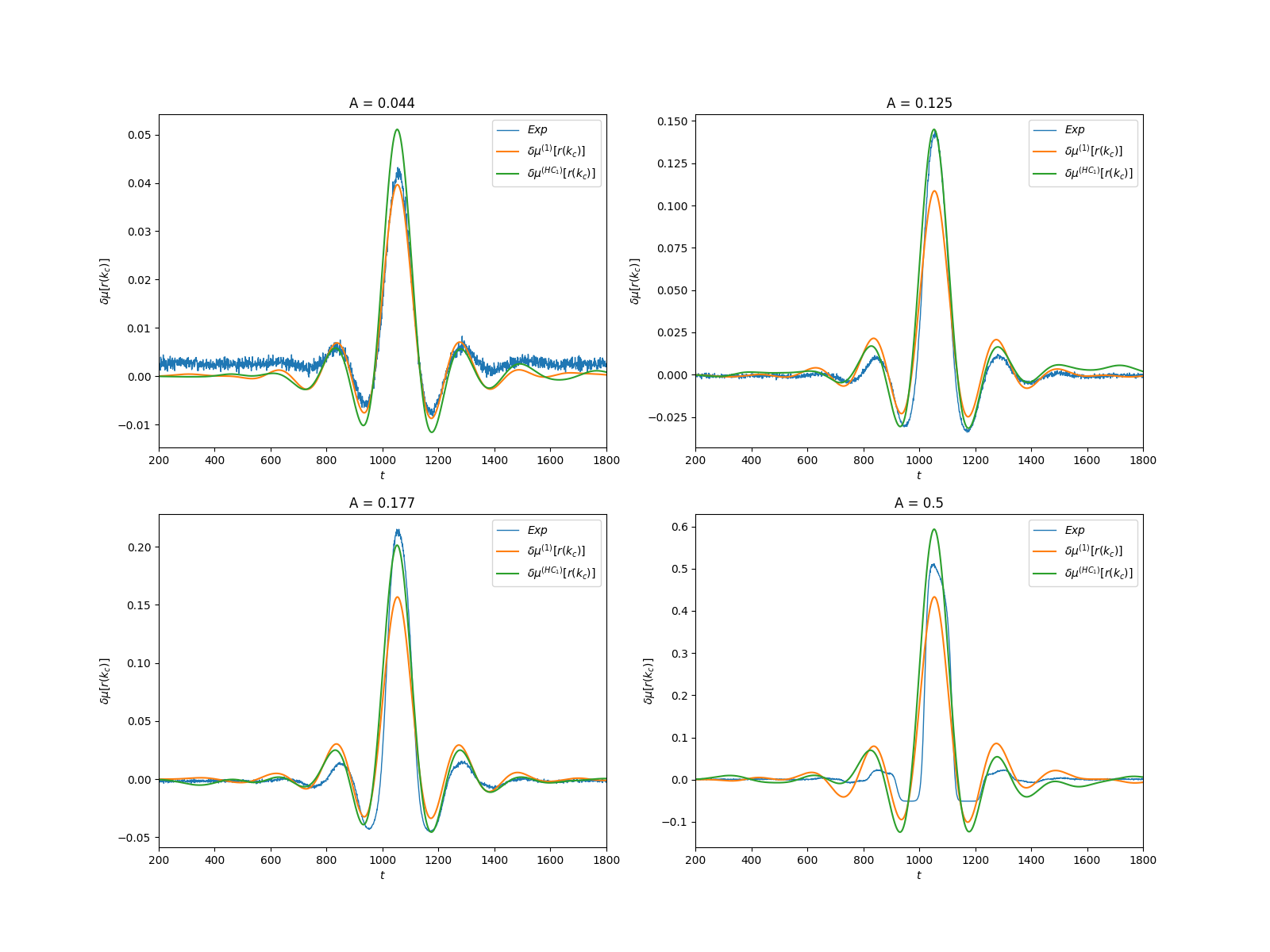}
\caption{
Linear response of $f(\omega,n)=\omega_{k_c}(n)$ for different values of stimulus amplitude $A$. Blue curve: Empirical average trace computed from \eqref{eq:EmpAverage}. Orange: Linear response computed from eq. \eqref{eq:LinRepBMSmu2}. Green: Linear response computed from eq. \eqref{eq:LinRepBMSmu3}.}
\label{Fig:LinRepRate}
\end{figure}

We first present the results of linear response for the observable $f(n,\omega)=\omega_{k_c}(n)$, where $k_c=\frac{N}{2}$ is the index of the neuron located at the center of the lattice. Thus, $\moy{f(n,\omega)} \equiv r(k_c,t)$, the firing rate of this neuron  as a function of time. In spontaneous activity it is a constant; under stimulation it depends on time. In fig. \ref{Fig:LinRepRate} we show the effect of the stimulus on the average value of this observable for different amplitudes values. One observes the combined effect of the stimulus and of the connectivity.

\begin{figure}
\centering
\includegraphics[width=12cm,height=15cm, angle=270]{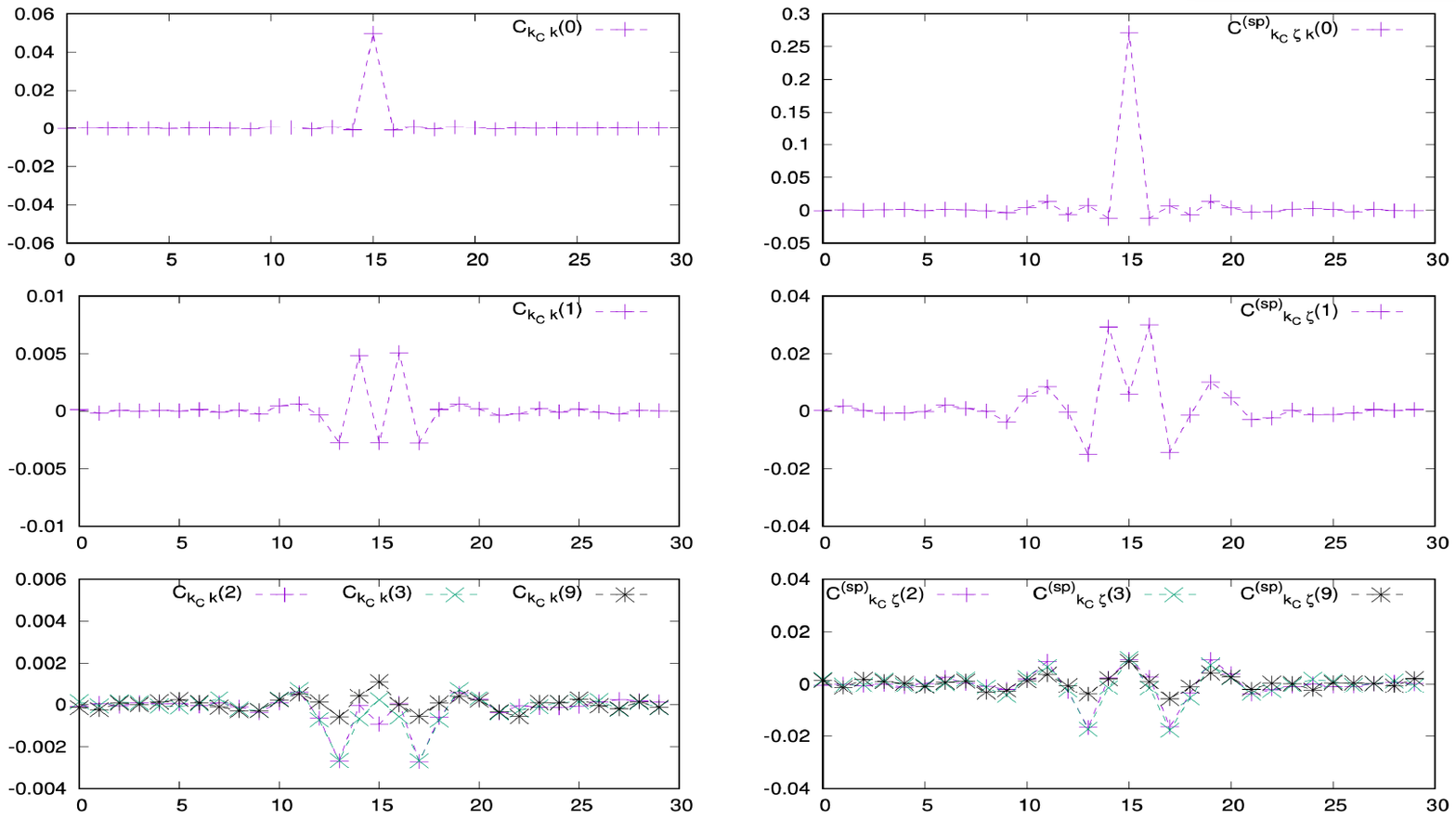}
\vspace{-1cm}
\caption{Correlation functions corresponding to 
 the firing rate of the neuron $k_c=\frac{N}{2}$ as a function of the neuron index $k$ (abscissa), for different values of the time delay $m$. Left) correlations with stimulus. Right) correlations in the spontaneous regime. Top. $m=0$, middle $m=1$, bottom $m=2,3, 9$. }
\label{Fig:Correlation_23_A0.2}
\end{figure}

We next studied how correlation functions in spontaneous activity depend on space and time. One observes that they decay relatively fast with the time delay of $m$ (Fig. \ref{Fig:Correlation_23_A0.2}). In addition, they are multiplied by $\gamma^m$ in \eqref{eq:LinRepBMSmu2}, \eqref{eq:LinRepBMSmu3}. Therefore the contribution to the linear response series decay exponentially fast and the series can be truncated to low order. Here we took a maximal order $D=10$.
 
Finally, as shown in Fig. \ref{Fig:LinRepRate}, we compute the linear response  $\delta \mu^{(1)}\bra{f(t)},\delta \mu^{(HC_1)}\bra{f(t)}$ and compare them to the  response obtained by empirical averages.

\subsection{Linear response for higher order observables}\label{Sec:HigherOrder}

Here, we consider the pairwise observable $f(\omega,n)=\omega_{k_c-2}(n-3) \omega_{k_c}(n)$
where $k_c=\frac{N}{2}$. This is an example of an observable with a time delay.
Neurons $k_c-2$ and $k_c$ mutually inhibit each other so we expect that the state of neuron $k_c-2$ before $n$ impact the state of neuron $k_c$ at time $n$. However, the correlation between those states depends as well on the state of the other neurons.

\begin{figure}[h!]
\includegraphics[width=1.1\textwidth,height=0.75\textheight]{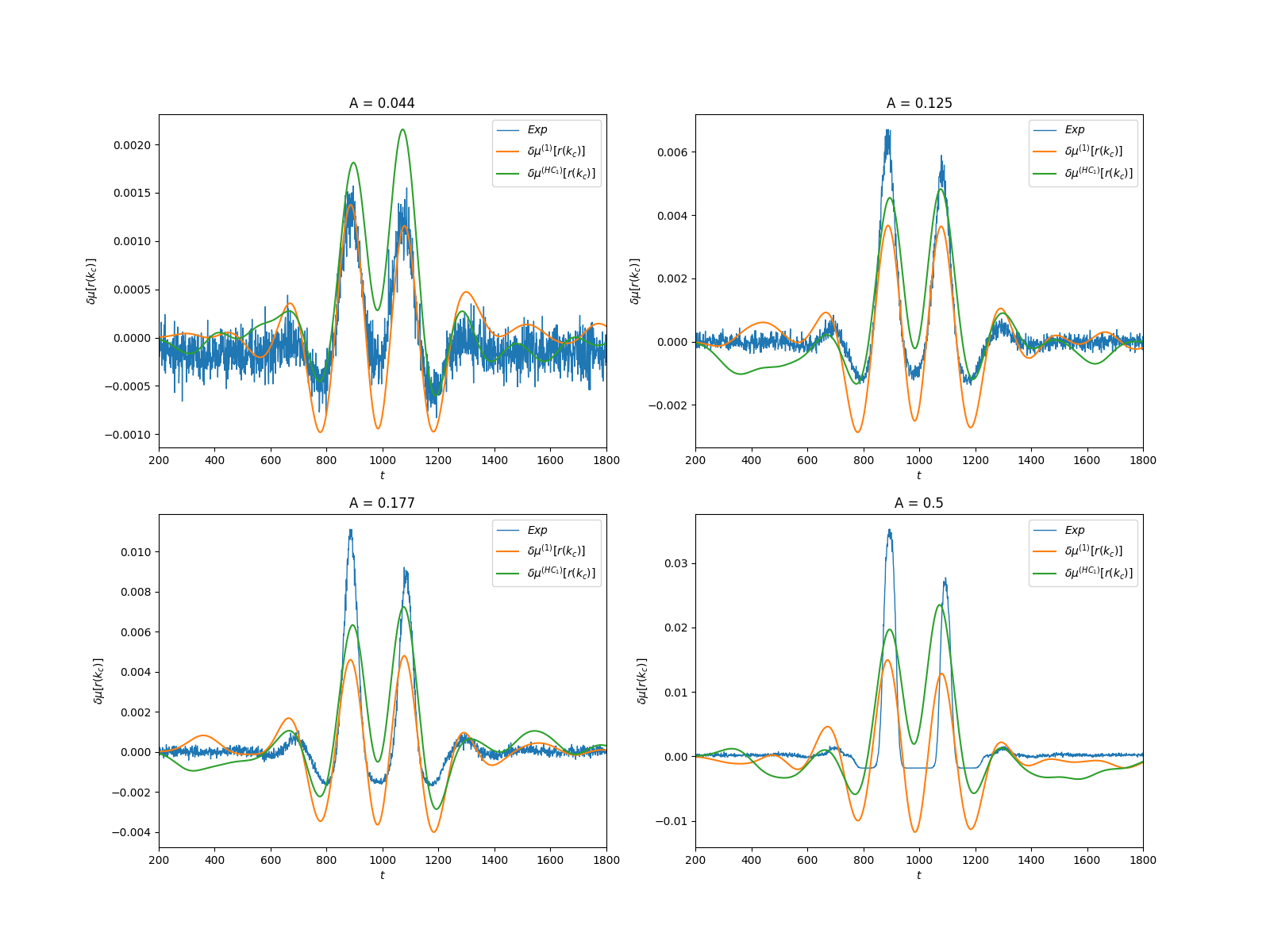}
\caption{Linear response of the observable  $f(n,\omega)=\omega_{k_c-2}(n-3) \omega_{k_c}(n)$. Here, we consider the same curves and amplitudes and as Fig. \ref{Fig:LinRepRate}. }
\label{Fig:LinRepPairwise}
\end{figure}

Similarly to the previous section we have plotted in Fig. \ref{Fig:LinRepPairwise} the empirical estimation of the linear response under the two approximations \eqref{eq:LinRepBMSmu2}, \eqref{eq:LinRepBMSmu3}. 

\subsection{Validity of the linear response}\label{Sec:ValidityLinRep}
\begin{figure}[h!]
\hspace*{-1.6cm}  
\includegraphics[width=1.2\textwidth,height=0.29\textheight]{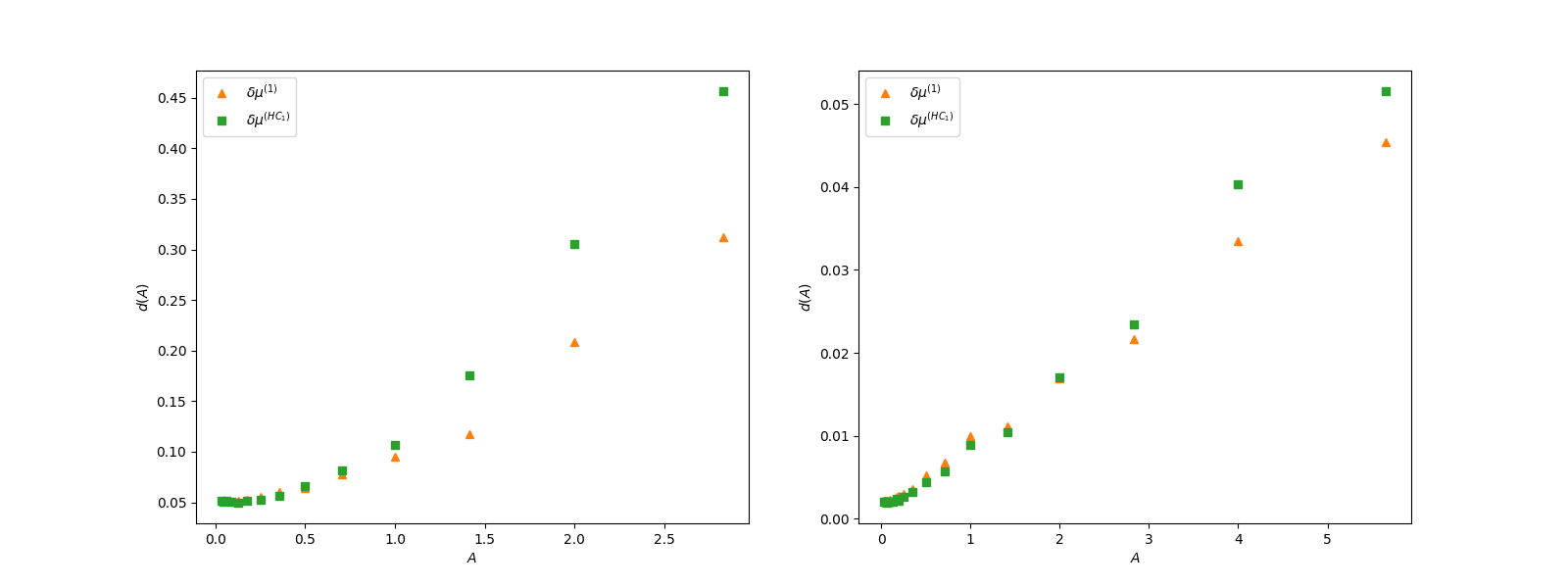}
\caption{$d^2$ distance (\ref{eq:dist}) between the curves $\delta \mu^{(1)}\bra{f(n)}$, $\delta \mu^{(HC_1)}\bra{f(n)}$ and the empirical curve, as a function of the stimulus amplitude $A$. Left panel show distance between rate curves (section \ref{Sec:rates}) and right panel distance between pairwise with delay (section \ref{Sec:HigherOrder}).}
\label{Fig:Error}
\end{figure}

The linear response is expected to hold when the stimulus amplitude is weak. What does that mean ? A mathematical answer is given by eq. \eqref{eq:condLinRep} although remaining at a rather abstract level. Here, we compute the distance:
\begin{equation}\label{eq:dist}
d^2(\delta \mu^{(th)}\bra{f(n)} ,  \delta \mu^{(exp)}\bra{f(n)})  =\sum_{n=1}^T \pare{\delta \mu^{(th)}\bra{f(n)} -  \delta \mu^{(exp)}\bra{f(n)}}^2,
\end{equation}
between the theoretical curves $\delta \mu^{(th)}\bra{f(n)}$ and the experimental curve $\delta \mu^{(exp)}\bra{f(n)}$, as a function of the stimulus amplitude $A$, in the two examples of observable investigated here. Note that distance here is not normalized: it does not take into account the amplitude of the response. This explains why the distance is larger in the case of firing rates than in the delayed pairwise case, as in the later the norm of the curve is quite smaller. The result is presented in Fig. \ref{Fig:Error}. As expected, the error in both cases increases with the amplitude of the stimulus, and it increases slower for  \eqref{eq:LinRepBMSmu2} than for the lowest order Hammersley-Clifford expansion \eqref{eq:LinRepBMSmu3}. It is interesting to see how the curves differ when $A$ increases (see Fig. \ref{Fig:LinRepRate} and \ref{Fig:LinRepPairwise}).  In the empirical average curves we clearly see a non-linear saturation (e.g. firing rate cannot exceed $1$) that is not reproduced by the linear response theory. This is further commented in the discussion section.

\subsection{Comments on numerical results}\label{Sec:cor}

With these simulations, we have been able to illustrate the predictions made by our theory. We are able to predict the time variation of an observable under a non-stationary stimulation, from the mere knowledge of the stimulus and spontaneous dynamics statistics. In particular, our theory works for observable with time delays, considerably enlarging the scope of linear response theories in neuronal networks. The comparison with classical fluctuation theory also emphasizes the role played by higher-order terms, taken into account in our approach. Linear response requires that the stimulus has a weak enough amplitude. We see very well this effect numerically. As the amplitude $A$ of the stimulus increases, we check the increasing discrepancy between the observation and the prediction. 

\su{Discussion}  

When an object moves across our visual field, it generates a transient spiking activity in the retina, conveyed to the thalamus and to the visual cortex \cite{berry:99,marre-etal:15}. The trajectory of this object - which is, in general, quite more complex than a moving bar with constant speed - involves long-range correlations in space and in time. Local information about this motion is encoded by retinal ganglion cells. Decoders based on the firing rates of these cells can extract some of the motion features \cite{palmer:15,palmer:16,deny:17,sederberg:18,srinivasan:82,hosoya:05,kastner:13}. Yet, lateral connectivity in the retina - especially via amacrine cells connecting bipolar cells - plays a central role in motion processing (see e.g.  \cite{gollisch-meister:10}). In addition, ganglion cells are directly connected through electric synapses \cite{gollisch-meister:10}. What is the role of this lateral connectivity in motion processing? Clearly, one may expect it to induce spatial and temporal correlations in spiking activity, as an echo, a trace, of the object's  trajectory. These correlations cannot be read in the variations of firing rate; they also cannot be read in synchronous pairwise correlations as the propagation of information due to lateral connectivity necessarily involves \textit{delays}. This example raises the question about what  information can be extracted from spatio-temporal correlations in a 
network of connected neurons submitted to a transient stimulus. 
What is the effect of the stimulus on these correlations? How can one handle this information from data where one has to measure \textit{transient} correlations? 

In this paper, we have addressed the first of these questions in a theoretical setting, using the linear response theory and probability distributions with unbounded memory generalizing the basic definition \eqref{eq:GibbsStatPhys} of Gibbs distributions in statistical physics courses. Our goal was to show the most general mathematical formalism allowing one to handle spike correlations as a result of a neuronal network activity in response to a stimulus. The most salient result of this work is that the difference of an observable average in response to an time dependent external stimulus of weak amplitude can be computed from the knowledge of the spontaneous correlations, i.e., from the dynamics without the stimulus. This result is not surprising from a non-equilibrium statistical physics perspective (Kubo relations, fluctuation-dissipation relation \cite{kubo:66, ruelle:99}). However, to the best of our knowledge, this is the first time it has been established for spiking neuronal networks. The novelty of our approach is that it provides a consistent treatment of the expected perturbation of higher-order correlations, going in this way, beyond the known linear perturbation of firing rates and instantaneous pairwise correlations; in particular, it extends to time-dependent correlations. 

In addition, we wanted to make explicit the linear response kernel in terms of the parameters determining individual networks dynamics and neurons connectivity. We have provided an explicit example of this for a well-known class of models, the Integrate and Fire model, where, in our example, synaptic conductances depend on the spike history. This makes explicit the role of the neuronal network structure (especially synaptic weights) in the spiking response. As we show, and as expected, the stimulus-response and dynamics are entangled in a complex manner. For example, the response of a neuron $k$ to a stimulus applied on neuron $i$ does not only depends on the synaptic weight $W_{ki}$ but, in general, on all synaptic weights, because the dynamics create complex causality loops which build up the response of neuron $k$ \cite{cofre-cessac:14, cessac-sepulchre:04,cessac-sepulchre:06}. We formally obtained a linear response function in terms of the parameters of a spiking neuronal network model and the spike history of the network. 
Although a linear treatment may seem a strong simplification, our results suggest that already, in this case, the connectivity architecture should not be neglected. In the presence of stimuli, the whole architecture of synaptic connectivity, history and the dynamical properties of the networks are playing a role in the correlations through the perturbed potential. This agrees well with results from a recent study exhibiting an exact analytical mapping between neuronal network models and maximum-entropy models, showing that, in order to accurately describe the statistical behavior of any observable in the Maximum Entropy model, all the synaptic weights are needed, even to predict firing rates of single neurons \cite{cofre-cessac:14}. 

We have also introduced the monomial expansion by providing a canonical way of decomposing the potential, describing stationary dynamics in a similar way that statistical physics does. Moreover, the Hammersley-Clifford decomposition allows us to obtain the coefficients weighting the monomials in terms of the parameters constraining dynamics.
In the case of the gIF model, this allowed us to show the explicit dependence of the coefficients in terms of synaptic weights.
Although the basis of monomials is quite huge, standard results in ergodic theory and transfer matrices/operators state that we can neglect high order terms because of the exponential correlation decay. Yet, the decay rate is controlled by the spectral gap in the transfer matrix, which itself depends on the network parameters. In this setting, one cannot exclude situations where the spectral gap is very tiny, leading to very slow correlations decay, reminiscent of second-order phase transitions (critical phenomena).

Beyond the models, we also wanted to characterize the linear response of biological neurons from the knowledge of the spontaneous activity. As we have shown, this raises the question of which spontaneous correlations are relevant. Obviously, it is natural to start from the lowest orders (firing rate and pairwise interactions). Nevertheless, higher order terms can not only play a significant role in spatial terms, as shown in \cite{ganmor:11a,ganmor:11b}, but also in temporal terms. Indeed, as argued throughout this paper, neuronal interactions involve delays that have to be integrated in a model attempting to explain spike statistics \cite{marre-etal:12}. Note however, that contrary to what is usually believed, detailed balance is absolutely unnecessary in order to properly handle time correlations \cite{cofre:18a, cofre:18b, cofre:19}. Also note that binning - which can be convenient to remove short-range time-correlations from the analysis - dramatically changes the nature of the process under investigation, rendering it non-Markovian \cite{cessac-etal:17} \\

Yet, our paper raises several issues and perspectives that we briefly discuss here.

\paragraph{Mean field assumption.} The Mean-Field assumption allowed us to obtain the explicit form of the response \eqref{eq:LinRepMF1} in terms of correlations and network parameters. This approximation, which holds under the condition \eqref{eq:CondMFLimit}, and more generally  when neurons are more often
silent than active allowed us to simplify the more general equation \eqref{eq:LinRepgIFtinf} by, (i) replacing the history dependent flow with a history independent one; (ii) Replacing the last firing time before $t$, $\tko$, by $-\infty$. At the moment, we have no idea this would work  without (i), i.e. handling a flow which depends on the spike trajectory. This point seems quite hard to handle, even numerically, and we have no idea yet how a deviation from hypothesis (i) could be diagnosed. Concerning (ii) the opposite limit is when the firing rate is quite faster than the characteristic time for the flow. This essentially means that the neurons are firing at a high rate and are thus reset so fast that the dynamics \eqref{eq:DyngIF} has no time to settle. We did not consider this case. 
In the most general case, $\tko$ in \eqref{eq:LinRepgIFtinf} is a random variable whose law is constrained by the stationary probability $\musp$. We have not yet been able to handle this case.

\paragraph{Beyond linear response.} In this paper, we have focused our analysis on the linear response because one of our main goals was to establish a mathematical setting allowing us to derive the kernel in the convolution equation \eqref{eq:KernelRF} in terms of spontaneous spatio-temporal correlations. Yet, two questions are pending: (i) What about higher order terms ? and, (ii) Is this type of Volterra-like expansion the most adapted to handle the non-stationary, collective, spike response to a stimulus? Concerning (i) higher order terms appear in two places. First, in the expansion \eqref{eq:ExpansionExpPhi}, where one can easily access higher order terms from the expansion of the exponential. The difficulty is to express these terms as correlations with respect to the stationary probability. A systematic
expansion for smooth dynamical systems has been done by D. Ruelle in \cite{ruelle:98}. 
Second, higher order terms come from the Taylor expansion of the potential $\phi$ (eq. \eqref{eq:ExpansionDeltaphi}). 
Concerning (ii) there exist other ways of handling the non-linearity in spike response. A typical case is the so-called Linear-Non-Linear (LN) model \cite{chichilnisky:01} or the Generalized Linear Model (GLM) where the linear response term \eqref{eq:KernelRF} is corrected by a static non-linearity \cite{ahmadian:11}. The relation with the LN-GLM form and the potential form \eqref{eq:Phi_GIF} has been studied in \cite{cessac-cofre:13}. Actually, the potential \eqref{eq:Phi_GIF} is a GLM where the static non-linearity and the parameters constraining the model are explicitly known. LN-GLM are built to characterize firing rates, but, to infer correlations from them requires considerable work. LN-GLM defines, in fact, a family of transition probabilities, possibly non-stationary, from which, as developed in section \ref{Sec:LinRepGen}, one can construct a Gibbs distribution giving access to spike statistics. The stationary case is handled by the Perron-Frobenius theorem, but we do not know about a general way to handle the non-stationary response, except with the Volterra expansion. 

\paragraph{Monomial expansion.} The monomials expansion results in a plethora of terms, where many of them can be irrelevant. As mentioned in the text, this expansion has the advantage of being generic; it has the drawback that no general principle tells us which monomials are significant. In fact, this is even worse. As shown, in \cite{cofre-cessac:14}, starting from a GLM model, or a potential like \eqref{eq:Phi_GIF}, constrained by $O(N^2)$ parameters, a monomial expansion with memory depth $D>1$ will generically have $O(2^{ND})$ terms, most of them being therefore related by hidden non-linear relations. This is a serious criticism of the Maximum Entropy related modeling approach, except if something unexpected happens in biological neural networks, somewhat pruning a large part of the monomials. There is, up to now, no evidence of this. But this point is related to another question: is there a method to remove irrelevant monomials from the observation of data? In addition to classical Akaike and Bayesian information criterion \cite{claeskens:08}, Occam's factors \cite{balasubramanian:97}, we have investigated a method based on information geometry \cite{amari-nagaoka:00,amari:10}. Exponential measures like Gibbs constitute a suitable space of probability measures where the Fisher metric is closely related to pairwise spatio-temporal correlations. In this setting, pruning methods for monomials can be proposed \cite{herzog:18}. 

\paragraph{Effective connectivity.} The network architecture together with the stimuli influences the statistics of spike correlations produced. In recent years, considerable efforts have been  dedicated to the construction of detailed connection maps of neurons on multiple scales \cite{helmstaedter:13,marc:13}. In this paper we have introduced a general formalism allowing for the computation of how a neuron responds to an excitation submitted to another neuron, yet defining a notion of effective connectivity based on stimulus-response causality. It follows from our analysis that this effective connectivity depends on stimulus, as already observed for the effective connectivity defined through the Ising model \cite{cocco:09}; it depends also on synaptic weights, in a complex manner. It was observed in \cite{cessac-sepulchre:06} that the existence of resonances in the power spectrum (with a similar origin as the resonances discussed in section \ref{Sec:resonances}) generates stimuli dependent graphs e.g. controlled by
the frequency of an harmonic stimulus. It would be interesting to check whether such a property holds here as well.

\paragraph{Data.} Another possible application of our results comes from the fact that correlations are taken with respect to the statistics in  spontaneous activity, which can be approximated from experimental recordings of a neuronal network spiking, in absence of external stimuli, using the Maximum Entropy Principle. As correlations between monomials can also be computed from data, assuming that the neuronal tissue from which the spikes have been recorded can be modeled by the gIF, only the values of the parameters of the model are needed to compute $\delta \phi$ and to predict the linear response. A similar approach has been used in \cite{cocco:09} with quite interesting results. Our work extends this to interactions with delays, allowing, as a future work, to take a step further 
in answering the important questions raised by these authors: ``First, how do the effective couplings depend
on the (visual) stimulus? Second, to what extent are the inferred
couplings affected by the incomplete sampling of the activity,
both from temporal (finite duration of the recordings) and
spatial (small area of the retina covered by the electrode array)
points of views? Third, do the couplings strongly depend on the
model used for the inference?"
Note  that the spontaneous Gibbs distribution with pairwise interactions (including time delays) can be well approximated from data as shown in \cite{vasquez-palacios-etal:12,cessac-etal:17b}.

\section*{List of Abbreviations}

\textbf{LIF} \quad Leaky Integrate-and-Fire\\

\noindent
\textbf{gIF} \quad generalized Integrate-and-Fire\\

\noindent
\textbf{MEA} \quad Multi-Electrode Arrays\\

\noindent
\textbf{GLM} \quad Generalized Linear Model\\

\noindent
\textbf{LN} \quad Linear non-Linear Model\\

\noindent
\textbf{Acknowledgments:} We thank Arnaud Le Ny and Roberto Fernandez for helpful comments on theoretical aspects of generalized Gibbs measures and chains with complete connections.\\

\noindent
\textbf{Funding:}  All authors have been supported by MAGMA-EQA-041903 INRIA associated team. RC was supported by CONICYT-PAI Inserci\'{o}n \# 79160120, Fondecyt Iniciaci\'{o}n 2018 Proyecto 11181072. BC was supported by the French government through the UCA-Jedi project managed by the National Research Agency (ANR-15- IDEX-01) and, in particular, by the interdisciplinary Institute for Modeling in Neuroscience and Cognition (NeuroMod) of the Universit\'{e} C\^{o}te d'Azur.\\

\noindent
\textbf{Ethics approval and consent to participate:}  Not applicable.\\

\noindent
\textbf{Competing interests:} The authors declare that they have no competing interests.\\

\noindent
\textbf{Consent for publication:} Not applicable.\\

\noindent
\textbf{Availability of data and materials:} Data sharing not applicable to this article as no datasets were generated or analysed during the current study.\\

\noindent
\textbf{Author's contributions:} BC and RC contributed equally. IA and BC run the simulations. All authors read and approved the final manuscript.\\

\end{document}